\documentstyle[12pt,aasms]{article}

\newcommand{\go}{\gtrsim}
\newcommand{\lo}{\lesssim}

\def\simless{\mathbin{\lower 3pt\hbox
   {$\rlap{\raise 5pt\hbox{$\char'074$}}\mathchar"7218$}}} 
\def\simgreat{\mathbin{\lower 3pt\hbox
   {$\rlap{\raise 5pt\hbox{$\char'076$}}\mathchar"7218$}}} 

\tighten

\begin{document}

\title{
Tests of Spurious Transport in Smoothed Particle Hydrodynamics
}

\author{James C.~Lombardi, Jr.\altaffilmark{1}}
\affil{Department of Astronomy and Center for Radiophysics and Space
Research, Cornell University, Ithaca, NY 14853; lombardi@spacenet.tn.cornell.edu or lombardi@vassar.edu}
\altaffiltext{1}{Current Address: Sanders Physics Laboratory, Vassar College, Poughkeepsie, NY 12604.}
\author{Alison Sills\altaffilmark{2}}
\affil{Department of Astronomy, Yale University, P.O.\ Box 208101,
New Haven, CT 06520; asills@astronomy.ohio-state.edu}
\altaffiltext{2}{Current Address: Department of Astronomy, The Ohio State University, 174 West 18th Avenue, Columbus, Ohio 43210.}
\author{Frederic A.~Rasio}
\affil{Department of Physics, MIT 6-201, Cambridge, MA 02139; rasio@mit.edu}
\and
\author{Stuart L.~Shapiro}
\affil{Departments of Physics and Astronomy and National Center for
Supercomputing Applications, University of Illinois at Urbana
Champaign, 1110 West Green Street, Urbana, IL 61801; shapiro@astro.physics.uiuc.edu}


%
%
%

\begin{abstract}
We have performed a series of systematic tests to evaluate
quantitatively the effects of spurious transport in three-dimensional
smoothed particle hydrodynamics (SPH) calculations.  Our tests
investigate (i) particle diffusion, (ii) shock heating, (iii) numerical
viscosity, and (iv) angular momentum transport.  The effects of various
program parameters on spurious mixing and on viscosity are
investigated.  The results are useful for quantifying the accuracy of
the SPH scheme, especially for problems where shear flows or shocks are
present, as well as for problems where true hydrodynamic mixing is
relevant.

We examine the different forms of artificial viscosity (AV) which have
been proposed by Monaghan, by Hernquist \& Katz, and by Balsara.  Our
tests suggest a single set of values for the AV parameters $\alpha$ and
$\beta$ (coefficients of the linear and quadratic terms) which are
appropriate in a large number of situations: $\alpha\approx 0.5$,
$\beta\approx 1$ for the classical AV of Monaghan,
$\alpha\approx\beta\approx0.5$ for the Hernquist \& Katz AV, and
$\alpha\approx\beta\approx\gamma/2$ for the Balsara AV (where $\gamma$
is the adiabatic index).  However, we also discuss how these choices
should be modified depending on the goals of the particular
application.  For instance, if spurious particle mixing is not a
concern and only weak shocks (Mach number ${\cal M}\lo 2$) are expected
during a calculation, then a smaller value of $\alpha$ is appropriate.
Somewhat larger values for $\alpha$ and $\beta$ may be preferable if an
accurate treatment of high Mach number shocks (${\cal M}\go 10$) is
required.  We find that both the Hernquist \& Katz and Balsara forms
introduce relatively small amounts of numerical viscosity.
Furthermore, both Monaghan's and Balsara's AV do well at treating
shocks and at limiting the amount of spurious mixing.  For these
reasons, we endorse the Balsara AV for use in a broad range of
applications.
\end{abstract}

\clearpage
\section{Introduction}

Smoothed particle hydrodynamics (SPH) is a Lagrangian method introduced
specifically to deal with astrophysical problems involving
self-gravitating fluids moving freely in three dimensions.
Pressure-gradient forces are calculated by kernel estimation, directly
from the particle positions, rather than by finite differencing on a
grid as in other particle methods such as PIC (the particle-in-cell
method; see, e.g.\ [\cite{Harlow1988}]) or grid-based methods like PPM (the
piecewise parabolic method; see, e.g.\ [\cite{Porter-Woodward1994}]).  This
idea was originally introduced by Lucy [\cite{Lucy1977}] and Gingold \& Monaghan
[\cite{Gingold-Monaghan1977}], who applied it to the calculation of dynamical fission
instabilities in rapidly rotating stars. Since then, a wide variety of
astrophysical fluid dynamics problems have been tackled using SPH (see
[\cite{Monaghan1992,Rasio-Lombardi1998}] for reviews).  In recent
years, these have included
planet and star formation
[\cite{Monaghan-Lattanzio1991,Burket-etal1997,Nelson-etal1998}], solar
system formation [\cite{Boss-etal1992}], supernova explosions
[\cite{Herant-Benz1992,Garcia-Senz-etal1998}], tidal disruption of
stars by massive black holes [\cite{Laguna-etal1993}], large-scale cosmological structure formation [\cite{Katz-etal1996,Shapiro-etal1996}], galaxy formation [\cite{Katz1992,Steinmetz1996}], stellar collisions [\cite{Lombardi-Rasio-Shapiro1995,Lombardi-Rasio-Shapiro1996}] and
binary coalescence [\cite{Rasio-Shapiro1992,Rasio-Shapiro1995,Davies-etal1994,Centrella-McMillan1993,Zhuge-etal1994,Zhuge-etal1996}].
The SPH method itself has also undergone
major advances.  Most notably, artificial viscosity (AV) has been
incorporated [\cite{Gingold-Monaghan1983,Hernquist-Katz1989,Monaghan1989,Balsara1995,Morris-Monaghan1997}], as well as powerful
algorithms for the calculation of self-gravity including particle-mesh
methods [\cite{Evrard1988}] and tree algorithms [\cite{Hernquist1987,Hernquist-Katz1989,Benz-etal1990}].

We have performed systematic tests of the SPH method.  In particular,
we concentrate on the examination of spurious transport, including the
motion of SPH particles introduced as a numerical artifact of the SPH
scheme.  Many applications require a careful tracing of particle
positions, and in these cases it is essential that the spurious
diffusion of SPH particles is small.  For example, SPH calculations can
be used to establish the amount of composition mixing during stellar
collisions [\cite{Benz-Hills1987,Lombardi-Rasio-Shapiro1995,Lombardi-Rasio-Shapiro1996}], which is of primary importance in determining the subsequent
stellar evolution of the merger remnant (see, e.g., [\cite{Sills-etal1997}]).
Since
some of the mixing observed in a SPH calculation is always spurious,
the observed amount of mixing is an upper limit to the actual amount.
Low-resolution SPH calculations in particular tend to be very noisy, and
this noise can lead to spurious diffusion of SPH particles, independent
of any real physical mixing of fluid elements.

SPH particles are coupled via a smoothing kernel, and there are
therefore strong
interactions among neighboring particles.  Regardless of the
type of fluid being simulated, the physical analogue of a system of SPH
particles is a crystal, liquid, or very imperfect gas (depending on the
noise level in the calculation) but never an ideal gas of
noninteracting particles.  These particle interactions introduce a
numerical viscosity into the SPH scheme and allow for the spurious
exchange of momentum and angular momentum among shear layers.  We have
studied and measured this viscosity, both in the context of a pure
shear flow constructed in a periodic box with slipping boundary
conditions, and in a rapidly, differentially rotating, self-gravitating
system.

One main goal of this paper is to present a thorough comparison of
three different AV forms, namely those of Monaghan
[\cite{Monaghan1989}], Hernquist \& Katz [\cite{Hernquist-Katz1989}], and Balsara [\cite{Balsara1995}].  Our tests of
these forms include a modified version of the Riemann shock-tube
problem in which periodic boundary conditions are imposed.  For each of
the AV forms, we investigate how the AV parameters can be adjusted to
achieve an accurate description of shocks, while still controlling
spurious mixing and shear viscosity.

In addition, we test the effects of varying a number of SPH-specific
parameters and schemes, including:  the number of neighbors $N_N$, the
choice of evolution equation (energy vs.~entropy), and the type of
advection algorithm.  Other tests of SPH include those by Hernquist \&
Katz [\cite{Hernquist-Katz1989}] and Steinmetz \& M\"uller [\cite{Steinmetz-Muller1993}].  In addition, comparisons
between SPH and Eulerian codes have been presented in the literature in
a variety of contexts: stellar collisions [\cite{Davies-etal1993}],
cosmology [\cite{Kang-etal1994}], rotating stars [\cite{Smith-etal1996}],
coalescing neutron stars [\cite{Ruffert-etal1997}], and shock-tube tests
[\cite{Morris-Monaghan1997}].

Many different implementations of SPH exist (e.g. [\cite{Evrard1988,Hernquist-Katz1989,Monaghan-Lattanzio1985}]), and in \S 2 we
give a brief description of the more popular schemes.  The degree of
spurious diffusion of SPH particles is quantified by diffusion
coefficients which we measure in \S 3.  In \S 4 we examine particle
diffusion in simple self-gravitating system.  In \S 5 we examine how
well various strengths and forms of AV handle shocks
in a modified version of the Riemann shock-tube test with periodic
boundary conditions.  In \S 6, we measure shear viscosity and examine
the spurious effects of AV on the transfer of angular
momentum.  It is the tests of \S5 and \S6 upon which we base our
comparison of the various AV forms.  Finally in \S 7 we summarize and
discuss our major results.

\section{Numerical Method \label{method}}
\subsection{Density, Pressure, and Entropy}

An SPH particle can be thought of as a Lagrangian fluid element.
Associated with particle $i$ is its position ${\bf r}_i$, velocity
${\bf v}_i$ and mass $m_i$.  In addition, each particle carries
SPH-specific parameters including a purely numerical ``smoothing
length'' $h_i$, specifying the local spatial resolution. An estimate of
the fluid density at ${\bf r}_i$ is calculated from the masses,
positions, and smoothing lengths of neighboring particles as a local
weighted average,
\begin{equation}
\rho_i=\sum_j m_j W_{ij}, \label{rho}
\end{equation}
where the symmetric weights $W_{ij}=W_{ji}$ can be calculated from the
method of Hernquist \& Katz [\cite{Hernquist-Katz1989}], as
\begin{equation}
W_{ij}={1\over2}\left[W(|{\bf r}_i-{\bf r}_j|,h_i)+W(|{\bf r}_i-{\bf
                      r}_j|,h_j)\right].
\end{equation}
Here $W(r,h)$ is a smoothing (or interpolation) kernel, for which we
use the
second-order accurate form of Monaghan \& Lattanzio [\cite{Monaghan-Lattanzio1985}],
\begin{equation}
W(r,h)={1\over\pi h^3}\cases{1-{3\over2}\left({r\over h}\right)^2
      +{3\over4}\left({r\over h}\right)^3,& $0\le{r\over h}<1$,\cr
 {1\over4}\left[2-\left({r\over h}\right)\right]^3,& $1\le{r\over
 h}<2$,\cr
      0,& ${r\over h}\ge2$.\cr}
\end{equation}

Depending on which evolution equation is integrated (see
equations~(\ref{udot}) and (\ref{adot}) below), particle $i$ also
carries either the physical parameter $u_i$, the internal energy per
unit mass in the fluid at ${\bf r}_i$, or $A_i$, the entropy variable,
a function of the specific entropy in the fluid at ${\bf r}_i$.
Arbitrary equations of state (e.g.~adiabatic, isothermal, even
equations of state for metals and rocky materials; cf.\ [\cite{Benz-etal1986}]) are permitted in SPH.  The calculations presented in this
paper use, unless otherwise noted, polytropic equations of state with
$\gamma=5/3$, appropriate for an ideal monatomic gas.
The pressure at ${\bf r}_i$ is therefore calculated
either as
\begin{equation}
p_i=(\gamma-1)\,\rho_i\, u_i,
\end{equation}
or
\begin{equation}
p_i=A_i\,\rho_i^\gamma.
\end{equation}
We define the specific entropy of particle $i$ to be
\begin{equation}
s_i\equiv
{1\over \gamma-1} \ln \left({p_i \over \rho_i^\gamma (\gamma-1)}\right),
\label{entropy}
\end{equation}
and consequently the total entropy of the system $S=\sum_i m_i s_i$.
Equation (\ref{entropy}) is a definition of convenience: we refer to
the quantity $s_i$ as entropy, even though it differs from the true
thermodynamic entropy (which depends on the composition of the fluid
being represented).  Although both $s_i$ and the true thermodynamic
entropy are conserved in adiabatic processes, it is $s_i$ which arises
naturally when studying the dynamical stability of self-gravitating
fluids.

\subsection{Dynamic Equations and Gravity}
\label{subsection:dynamic}

Particle positions are updated either
by
\begin{equation}
\dot{\bf r}_i = {\bf v}_i, \label{rdot}
\end{equation}
or the more general XSPH method
\begin{equation}
\dot{\bf r}_i = {\bf v}_i+\epsilon \sum_j m_j{{\bf v}_j-{\bf v}_i\over
\rho_{ij}}W_{ij} \label{XSPH}
\end{equation}
where $\rho_{ij}=(\rho_i+\rho_j)/2$ and $\epsilon$ is a constant
parameter in the range $0 < \epsilon < 1$ [\cite{Monaghan1989}].  Equation
(\ref{XSPH}), as compared to equation~(\ref{rdot}), changes particle
positions at a rate closer to the local smoothed velocity.  The XSPH
method was originally proposed as a means of decreasing spurious
interparticle
penetration across the interface of two colliding fluids.

The velocity of particle $i$ is updated according to
\begin{equation}
           \dot{\bf v}_i = {\bf a}^{(Grav)}_i+{\bf a}^{(SPH)}_i
\end{equation}
where ${\bf a}^{(Grav)}_i$ is the gravitational acceleration and
\begin{equation}
{\bf a}^{(SPH)}_i=-\sum_j m_j \left[\left({p_i\over\rho_i^2}+
    {p_j\over\rho_j^2}\right)+\Pi_{ij}\right]{\bf \nabla}_i W_{ij}.
    \label{fsph}
\end{equation}
Various forms for the AV term $\Pi_{ij}$ are discussed below.
The AV term ensures that correct jump
conditions are satisfied across (smoothed) shock fronts, while the rest of
equation~(\ref{fsph}) represents one of many possible SPH-estimators
for the acceleration due to the local pressure gradient (see, e.g.,
[\cite{Monaghan1985}]).

To provide reasonable accuracy, an SPH code must solve the equations of
motion of a large number of particles (typically $N>>1000$). This rules
out a direct summation method for calculating the gravitational field
of the system, unless special purpose hardware such as the GRAPE is
used [\cite{Steinmetz1996,Klessen1997}].  In most implementations of SPH,
particle-mesh algorithms [\cite{Evrard1988,Rasio-Shapiro1992,Couchman-etal1995}] or tree-based algorithms [\cite{Hernquist-Katz1989,Dave-etal1997}] are used to calculate the gravitational accelerations ${\bf
a}^{(Grav)}_i$.  Tree-based algorithms perform better for problems
involving large dynamic ranges in density, such as star formation and
large-scale cosmological calculations. For problems such as stellar
interactions, where density contrasts rarely exceed a factor
$\sim10^2-10^3$, grid-based algorithms and direct solvers are generally
faster.  Tree-based and grid-based algorithms are also used to
calculate lists of nearest neighbors for each particle exactly as in
gravitational $N$-body calculations (see, e.g., [\cite{Hockney-Eastwood1988}]).

Our SPH codes are
slightly modified versions of codes originally developed by Rasio [\cite{Rasio1991}], with implementations similar to those adopted by Hernquist \&
Katz [\cite{Hernquist-Katz1989}].  Our 3D code has the option of including gravity, and
calculates the gravitational field by a particle-mesh convolution
algorithm which uses a grid-based FFT solver [\cite{Hockney-Eastwood1988,Wells-etal1990}].  More specifically, the smoothed density
sets the values of the source term for Poisson's equation at grid
points.  The FFT-based convolution algorithm then solves for the
gravitational potential on that grid.  Forces at grid points are
obtained by finite differencing, and then interpolated onto the
particle positions.
We have found that, for our tests involving self-gravitating fluids, it
is relatively easy to make the gravity accurate enough
that it is not a significant source of error.  Therefore, the results
of this paper can be applied to any SPH code regardless of its
gravitational scheme.


\subsection{Artificial Viscosity}

We now present three commonly used AV forms which are tested in
this paper.  In \S 7.2 and \S 7.3 we will discuss the results of these
tests, while in \S 7.4 we discuss which of the AV forms performs best
in which circumstances.

For the AV, a symmetrized version of the form
proposed by Monaghan [\cite{Monaghan1989}] is often adopted,
\begin{equation}
\Pi_{ij}={-\alpha\mu_{ij}c_{ij}+\beta\mu_{ij}^2\over\rho_{ij}},
\label{pi}
\end{equation}
where $\alpha$ and $\beta$ are constant parameters,
$c_{ij}=(c_i+c_j)/2$ (with $c_i=(\gamma
p_i/\rho_i)^{1/2}$ being the speed of sound in the fluid at ${\bf
r}_i$), and
\begin{equation}
\mu_{ij}=\cases{ {({\bf v}_i-{\bf v}_j)\cdot({\bf r}_i-{\bf r}_j)\over
h_{ij}(|{\bf r}_i -{\bf r}_j|^2/h_{ij}^2+\eta^2)}& if $({\bf v}_i-{\bf
v}_j)\cdot({\bf r}_i-{\bf r}_j)<0$\cr
	     0& if $({\bf v}_i-{\bf v}_j)\cdot({\bf r}_i-{\bf
	     r}_j)\ge0$\cr}
		\label{mu}
\end{equation}
with $h_{ij}=(h_i+h_j)/2$.  We will refer to viscosities of this form
as the ``classical'' AV.  This form represents a combination of a bulk
viscosity (linear in $\mu_{ij}$) and a von~Neumann-Richtmyer viscosity
(quadratic in $\mu_{ij}$).  The von~Neumann-Richtmyer
AV was initially introduced to suppress particle
interpenetration in the presence of strong shocks.
Morris \& Monaghan [\cite{Morris-Monaghan1997}] have recently implemented equation (\ref{mu})
with a {\it time varying} coefficient $\alpha$, and with $\beta=2\alpha$.
Our tests will demonstrate that, for constant $\alpha$ and $\beta$, equation (\ref{pi}) performs
best when $\alpha\approx0.5$, $\beta\approx 1$, and
$\eta^2\sim10^{-2}$, although, as discussed in
\S 7.4, these choices should be adjusted to fit
the particular goals of an application.

Another form for the AV, introduced by Hernquist \& Katz [\cite{Hernquist-Katz1989}]
calculates $\Pi_{ij}$ directly from the SPH estimate of the divergence
of the velocity field:
\begin{equation}
\Pi_{ij}=\cases{ {q_i\over\rho_{i}^2}+{q_j\over\rho_{j}^2}& if $({\bf
v}_i-{\bf v}_j)\cdot({\bf r}_i-{\bf r}_j)<0$\cr
	     0& if $({\bf v}_i-{\bf v}_j)\cdot({\bf r}_i-{\bf
	     r}_j)\ge0$\cr},
		\label{pi2}
\end{equation}
where
\begin{equation}
q_i=\cases{ \alpha \rho_i c_i h_i |{\bf \nabla}\cdot {\bf v}|_i+
	       \beta \rho_i h_i^2 |{\bf \nabla}\cdot {\bf v}|_i^2
		 & if $\left({\bf \nabla}\cdot {\bf v}\right)_i<0$\cr
             	0& if $\left({\bf \nabla}\cdot {\bf v}\right)_i\ge0$\cr}
			\label{q}
\end{equation}
and
\begin{equation}
({\bf \nabla}\cdot {\bf v})_i={1 \over \rho_i}\sum_j m_j
	({\bf v}_j-{\bf v}_i)\cdot{\bf \nabla}_i W_{ij}. \label{divv}
\end{equation}
We will refer to this form as the HK AV.
Although this form provides a slightly less accurate description of 
shocks than equation (\ref{pi}), it does
exhibit less shear viscosity.  Our tests show that
$\alpha\approx\beta\approx0.5$ is an appropriate choice for the HK AV
for a broad range of circumstances (see \S 7.4).

More recently, Balsara [\cite{Balsara1995}] has proposed the AV
\begin{equation}
\Pi_{ij}=
\left({p_i\over\rho_i^2}+{p_j\over\rho_j^2}\right)
	\left(-\alpha \mu_{ij} + \beta \mu_{ij}^2\right),
	\label{piDB}
\end{equation}
where
\begin{equation}
\mu_{ij}=\cases{ {({\bf v}_i-{\bf v}_j)\cdot({\bf r}_i-{\bf r}_j)\over
h_{ij}(|{\bf r}_i -{\bf r}_j|^2/h_{ij}^2+\eta^2)}{f_i+f_j \over 2
c_{ij}}& if $({\bf v}_i-{\bf v}_j)\cdot({\bf r}_i-{\bf r}_j)<0$\cr
	     0& if $({\bf v}_i-{\bf v}_j)\cdot({\bf r}_i-{\bf
	     r}_j)\ge0$\cr}.
		\label{muDB}
\end{equation}
Here $f_i$ is the form function for particle $i$, defined by
\begin{equation}
f_i={|{\bf \nabla}\cdot {\bf v}|_i \over |{\bf \nabla}\cdot {\bf v}|_i
+|{\bf \nabla}\times {\bf v}|_i + \eta' c_i/h_i,
} \label{fi}
\end{equation}
where the factor $\eta'\sim 10^{-4}-10^{-5}$ prevents numerical
divergences, $({\bf \nabla}\cdot {\bf v})_i$ is given by equation
(\ref{divv}), and
\begin{equation}
({\bf \nabla}\times {\bf v})_i={1 \over \rho_i}\sum_j m_j
	({\bf v}_i-{\bf v}_j)\times{\bf \nabla}_i W_{ij}. \label{curlv}
\end{equation}
The function $f_i$ acts as a switch, approaching unity in regions of
strong compression ($|{\bf \nabla}\cdot {\bf v}|_i >>|{\bf
\nabla}\times {\bf v}|_i$) and vanishing in regions of large vorticity
($|{\bf \nabla}\times {\bf v}|_i >>|{\bf \nabla}\cdot {\bf v}|_i$).
Consequently, this AV has the advantage that it is suppressed in shear
layers.  Throughout this paper we use $\eta'=10^{-5}$, a choice which
does not significantly affect our results.  Note that since
$(p_i/\rho_i^2+p_j/\rho_j^2)\approx 2c_{ij}^2/(\gamma\rho_{ij})$,
equation~(\ref{piDB}) resembles equation (\ref{pi}) when $|{\bf
\nabla}\cdot {\bf v}|_i >> |{\bf \nabla}\times {\bf v}|_i$, provided
one rescales the $\alpha$ and $\beta$ in equation (\ref{piDB}) to be a
factor of $\gamma/2$ times the $\alpha$ and $\beta$ in equation
(\ref{pi}).  We will show that $\alpha\approx\beta\approx\gamma/2$ is
often an appropriate choice for the Balsara AV.

\subsection{Thermodynamics}

To complete the description of the fluid, either $u_i$ or $A_i$ is
evolved according to a discretized version of the first law of
thermodynamics:
\begin{equation}
{d\,u_i\over d\,t}= {1\over 2} \sum_j m_j \left({p_i\over\rho_i^2}+{p_j\over\rho_j^2}+
    \Pi_{ij}\right)\,({\bf v}_i-{\bf v}_j)\cdot{\bf \nabla}_i
    W_{ij},
	\label{udot}
\end{equation}
or
\begin{equation}
{dA_i\over dt}={\gamma-1\over 2\rho_i^{\gamma-1}}\,
     \sum_jm_j\,\Pi_{ij}\,\,({\bf v}_i-{\bf v}_j)\cdot{\bf \nabla}_i
     W_{ij}.
	\label{adot}
\end{equation}
We call equation~(\ref{udot}) the ``energy equation,'' while equation
(\ref{adot}) is the ``entropy equation.''  Which equation one should
integrate depends upon the problem being treated.  Each has its own
advantages and disadvantages.
For instance, thermodynamic processes
such as heating and cooling [\cite{Katz-etal1996}] and nuclear burning
[\cite{Garcia-Senz-etal1998}] can be incorporated more easily into the
energy equation.

The derivations of
equations~(\ref{udot}) and (\ref{adot}) neglect 
the time variation of $h_i$.  Therefore if we integrate the energy
equation, even in the absence of AV, the total
entropy of the system will not be strictly conserved if the particle
smoothing lengths are allowed to vary in time; if the entropy equation
is used to evolve the system, the total entropy would then be strictly
conserved when $\Pi_{ij}=0$, but not the total energy [\cite{Rasio1991,Hernquist1993}]. For more accurate treatments involving
time-dependent smoothing lengths, see Nelson \& Papaloizou [\cite{Nelson-Papaloizou1993,Nelson-Papaloizou1994}]
and Serna et al.\ [\cite{Serna-etal1996}].

There are many other equivalent forms of the basic SPH equations which
reduce to the correct fluid equations in the limit
$N\rightarrow\infty$, $h_i\rightarrow0$. However, most of them will
satisfy their associated conservation equations only approximately,
i.e., up to errors which tend to zero only in this limit. In contrast,
the above equations have the virtue of conserving energy and momentum
exactly, independent of the number of particles used, as long as the
smoothing lengths are held fixed (eg., [\cite{Rasio1991}]). Of course, in the
numerical solution, errors will still be introduced by the
time-integration scheme.

\subsection{Integration in Time}

For a stable time integration scheme, the timestep must satisfy a
Courant-like condition with $h_i$ replacing the usual grid separation. For
accuracy, the timestep must be a small enough fraction of the system's
dynamical time.  We calculate the timestep as
\begin{equation}
\Delta t=C_N\,{\rm Min}(\Delta t_1,\Delta t_2), \label{dt}
\end{equation}
where the constant dimensionless Courant number $C_N$ typically
satisfies $0.1\simless C_N \simless 0.8$, where
\begin{equation}
\Delta t_1={\rm Min}_i\,(h_i/\dot v_i)^{1/2}, \label{dt1}
\end{equation}
and where for $\Delta t_2$ we use one of two types of expressions,
the simplest being
\begin{equation}
\Delta t_2={\rm Min}_i\left({h_i\over (c_i^2+v_i^2)^{1/2}}\right).
\label{simple.dt}
\end{equation}
In the presence of strong shocks, equations such as (\ref{simple.dt})
can allow for fairly large entropy changes in a single timestep when
$C_N$ is large.  This problem can be eliminated by using smaller $C_N$,
or by adopting a more sophisticated expression introduced by
Monaghan [\cite{Monaghan1989}]:
\begin{equation}
\Delta t_2={\rm Min}_i\left(
{h_i \over c_i+1.2\alpha c_i+1.2\beta {\rm Max_j}|\mu_{ij}|}
\right).  \label{good.dt}
\end{equation}
If the Hernquist \& Katz AV [eq.~(\ref{pi2})] is used, the quantity
Max$_j|\mu_{ij}|$ in equation (\ref{good.dt}) can be replaced by
$h_i|{\bf \nabla}\cdot {\bf v}|_i$ if $({\bf \nabla}\cdot
{\bf v})_i<0$, and by $0$ otherwise.  By accounting for AV-induced
diffusion, the $\alpha$ and $\beta$ terms in the denominator of
equation (\ref{good.dt}) allow for a more efficient use of
computational resources than simply using a smaller value of $C_N$.  In
this paper, we will label the timestep routine by an S (for ``simple'')
when we implement equations (\ref{dt}), (\ref{dt1}), and
(\ref{simple.dt}), and by an M (for Monaghan) when we implement
(\ref{dt}), (\ref{dt1}), and (\ref{good.dt}).

The evolution equations are integrated using a second-order explicit
leap-frog scheme. Such a low order scheme is appropriate because
the dominate source of error for the evolution is the noise in
particle interactions due to numerical discreteness effects.  Other details
of our implementation, as well as a number of test-bed calculations
using our SPH code, are presented in Rasio \& Shapiro [\cite{Rasio-Shapiro1991,Rasio-Shapiro1992}].

\subsection{Smoothing Lengths and Accuracy}

The size of the smoothing lengths is often chosen such that particles
roughly maintain some predetermined number of neighbors $N_N$.  Typical
values of $N_N$ range from about 20 to 100.  If a particle
interacts with too few neighbors, then the forces on it are sporadic, a
poor approximation to the forces on a true fluid element. In general, one
 finds that, for given physical conditions, 
the noise level in a calculation always decreases when $N_N$ is increased.  

At the other
extreme, large neighbor numbers degrade the resolution by requiring
unreasonably large smoothing lengths.  
However,  higher accuracy is obtained in SPH calculations only when 
{\em both\/}
the number of particles $N$ {\em and\/} the number of neighbors $N_N$ are
increased, with $N$ increasing faster than $N_N$ so that the smoothing
lengths $h_i$ decrease. Otherwise (e.g., if $N$ is increased while maintaining
$N_N$ constant) the SPH method is {\it inconsistent}, i.e., it converges to an
unphysical limit [\cite{Rasio1991}]. The choice of $N_N$ for a given
calculation is therefore dictated by a compromise between an acceptable
level of numerical noise and the desired spatial resolution (which is
$\approx h\propto 1/N_N^{1/d}$ in $d$ dimensions) and level of accuracy.

\section{Simple Box Tests \label{boxtests}}
\subsection{Measuring SPH Particle Diffusion}
\label{measure.diffusion}

Simulations of a homogeneous volume of gas, at rest and in the absence of
gravity, provide a natural environment to examine spurious diffusion of
SPH particles.   In the ideal simulation of a motionless fluid, no
SPH particles would move, and the  
thermodynamic variables would remain constant.  However, there is always
some level of noise in an SPH system, and this leads to the
spurious motion of particles even in the absence of any bulk flow.

In order to model such a system, we introduce periodic boundary
conditions in a cubical box, adopting the standard technique of
molecular dynamics (cf.\ [\cite{Allen-Tildesley1989}]): whenever an SPH
particle leaves the box, it is reintroduced with the same velocity
vector on the opposing face, directly across from where it exited.
Particles with smoothing kernels extending beyond a side of the box can
have neighbors near the opposing side, once periodicity is taken into
account.  More precisely, particle $j$ has particle $i$ as a neighbor
if there exists integers $k$, $l$ and $m$ such that the position
($x_i+kL,y_i+lL,z_i+mL$) is within a distance $2h_j$ of
($x_j,y_j,z_j$), where $L$ is the length of the box.  Unless otherwise
noted, the calculations presented in this section employ equal mass
particles, all with the same time-independent smoothing length $h$
chosen such that the average number of neighbors $N_N$ is 20, 32, 48 or
64.  The total number of particles $N$ in the box is unimportant, as
long as it is large enough that surface effects can be neglected.  To
ensure this, we always choose $N$ such that $L/h \simgreat 16$.

For the diffusion tests of this section, the natural units are given by
$n=c_s=1$, where $n$ is the number density of SPH particles and $c_s$
is the local sound speed.  With this choice, velocities are in units of
$c_s$, distances are in units of $n^{-1/3}$, and times are in units of
$n^{-1/3} c_s^{-1}$.  In practice, we implement $c_s=1$ by choosing the
entropy variable $A=\rho^{1-\gamma}/\gamma$.  Furthermore, the mass of
the particles is chosen such that the cubical box contains unit mass:
$M=Nm=1$.  Since the local number density and sound speed are known in
any SPH calculation, these units make our results applicable to many
contexts.

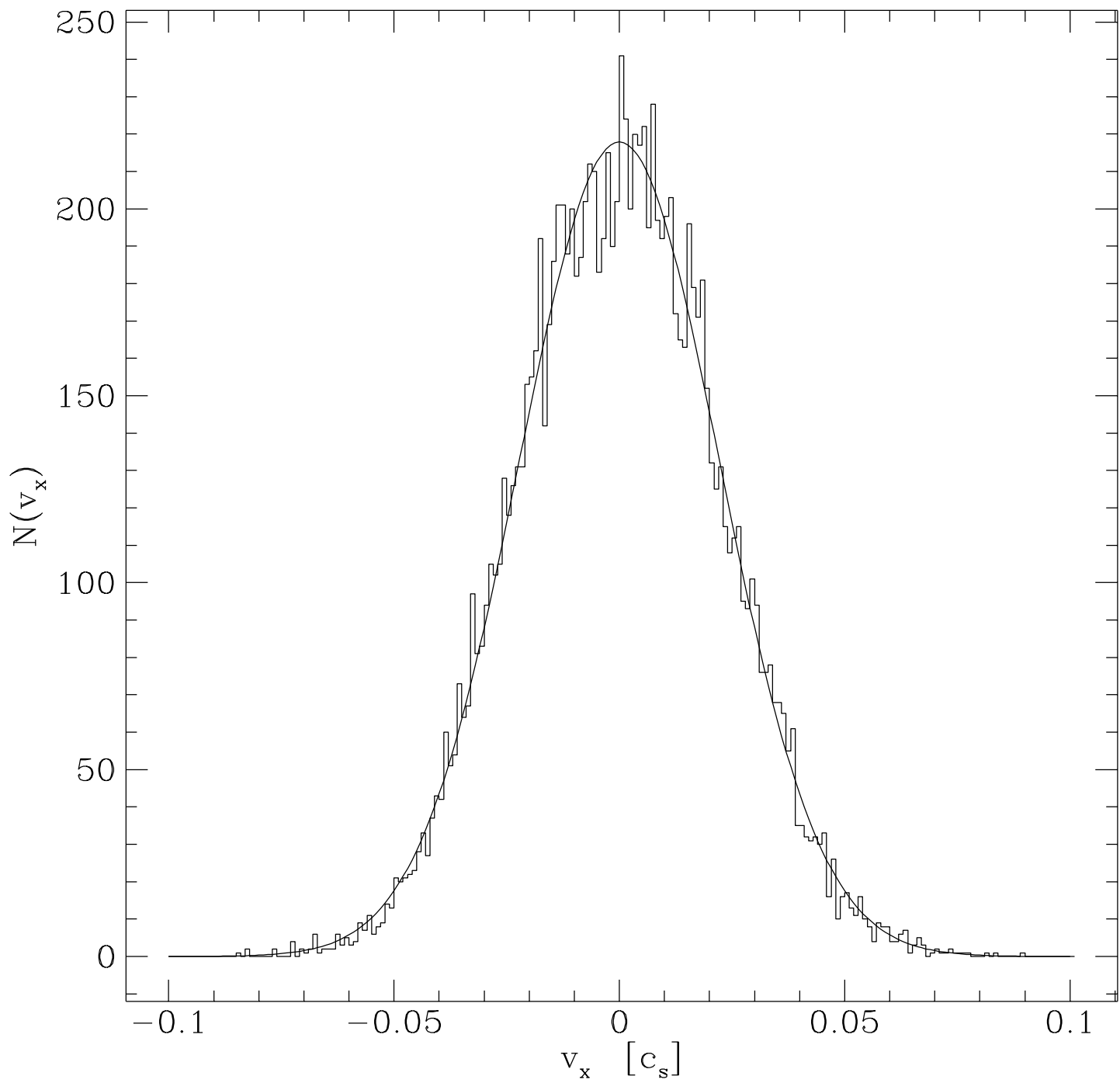
\begin{figure}
\caption{
The number of particles $N(v_x)$ in velocity bins of width $0.001 c_s$
for the equilibrium state in a typical simple box test, where $c_s$ is
the sound speed.  The $N=23^3$ particles interacted with $N_N\approx
64$ neighbors and began in a simple cubic lattice configuration with
noise artificially introduced at $t=0$.  The solid line shows the best
fit Maxwellian, corresponding to $v_{rms}=0.404$, once the system has
reached equilibrium.  Deviations from this best fit are consistent with
statistical fluctuations.
} \label{mbhist}
\end{figure}

After positioning the particles on a regular lattice and assigning
their velocities (with zero net momentum), we allow the system to
evolve, without AV.  Although each SPH particle
represents a physical fluid element with a certain temperature and
density, the SPH particles themselves have their own numerical
``temperature'' (due to the particle velocity dispersion) and number
density.  While there is an obvious correlation between the number
density of the SPH particles and the density of the gas being
represented, no such correlation exists between the numerical
temperature of the SPH particles and the physical temperature of the
gas being simulated.  Regardless of the initial velocity distribution
chosen, the velocities ultimately settle into an equilibrium
Maxwell-Boltzmann distribution (see Figure~\ref{mbhist}) with some root
mean square particle velocity $v_{rms}$ which quantifies the noise
level, or numerical temperature, of the system.  The energy exchange
which causes this thermalization is due to the strong coupling between
neighboring particles through equation~(\ref{fsph}).  We have also
found that the velocity distribution in real calculations tends to be
roughly a Maxwellian centered on the local smoothed velocity.

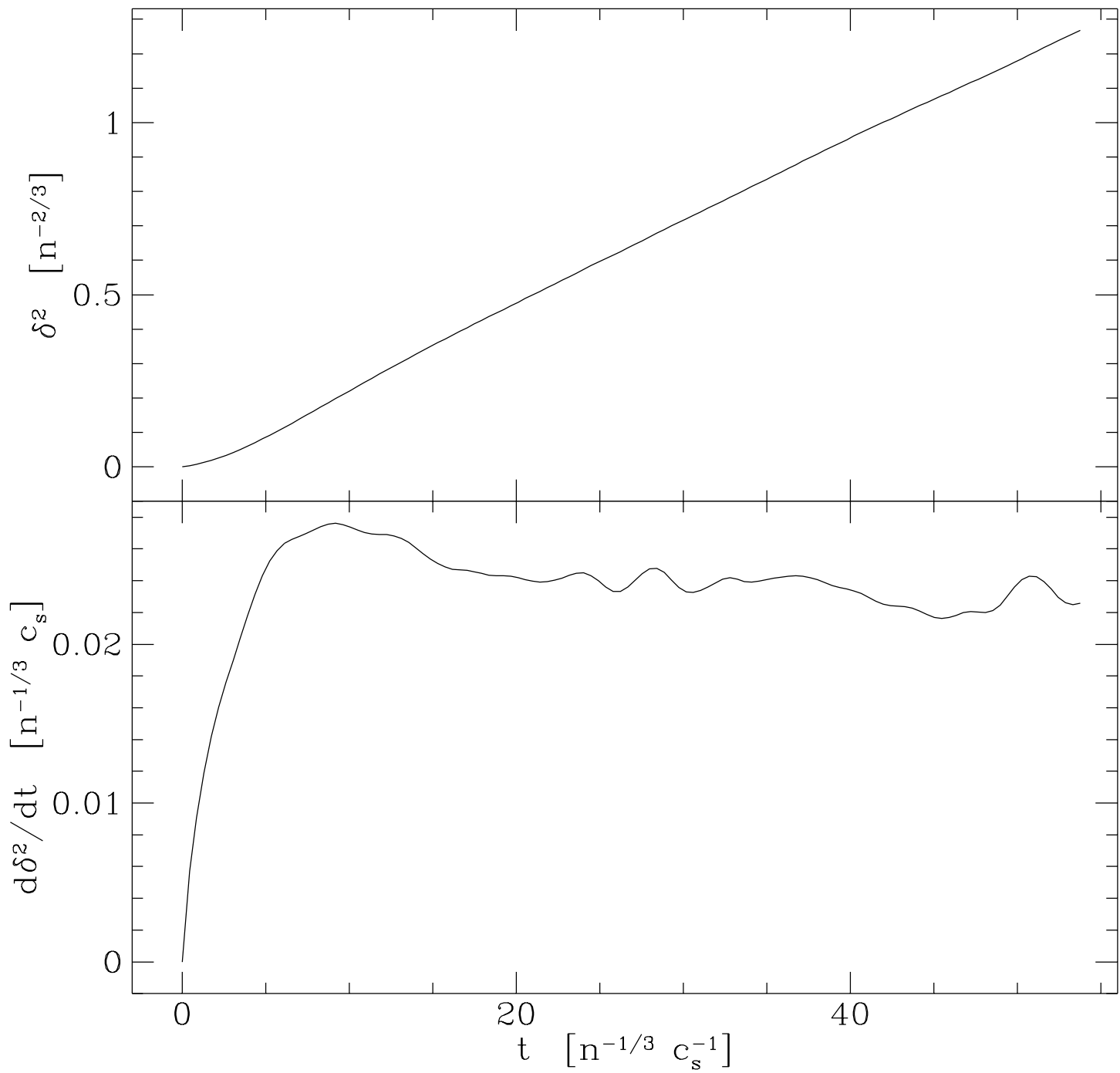
\begin{figure}
\caption{
The mean square deviation $\delta^2$ and slope $d\delta^2/dt$ as a
function of time after an equilibrium particle velocity dispersion
$v_{rms}=0.069 c_s$ has been reached
in a typical simple box test with $N_N=48$ and no AV.  At late times,
the mean square deviation $\delta^2$
increases approximately linearly with time, and we define the diffusion
coefficient $D$ as the slope of this line.  Units are discussed in \S
3.1.
} \label{d2vt}
\end{figure}

The level of diffusion is quantified as follows.  Once the velocity
distribution has settled into an equilibrium Maxwellian, we record the
positions of all particles.  Since ideally the particles would not move
far from their initial positions, it is then easy to monitor the mean
square spurious diffusion distance $\delta^2$ as a function of time $t$
(properly accounting for particles which cross the faces of the box).
At late times the mean square deviation $\delta^2$ increases at an
nearly constant rate, so that the system obeys the usual diffusion
equation $\delta^2=D t$, and the diffusion coefficient $D\equiv
d\delta^2/dt$, evaluated at late times, is easily measured. (In
molecular dynamics, the diffusion coefficient $D$ is sometimes defined
to be a factor of six smaller than in our definition.) As an
example, Figure~\ref{d2vt} shows $\delta^2$ and $d\delta^2/dt$ for a
system with an equilibrium $v_{rms}=0.069$; it is clear that
$d\delta^2/dt$ is essentially constant at late times, and we measure
$D\approx 0.024$.

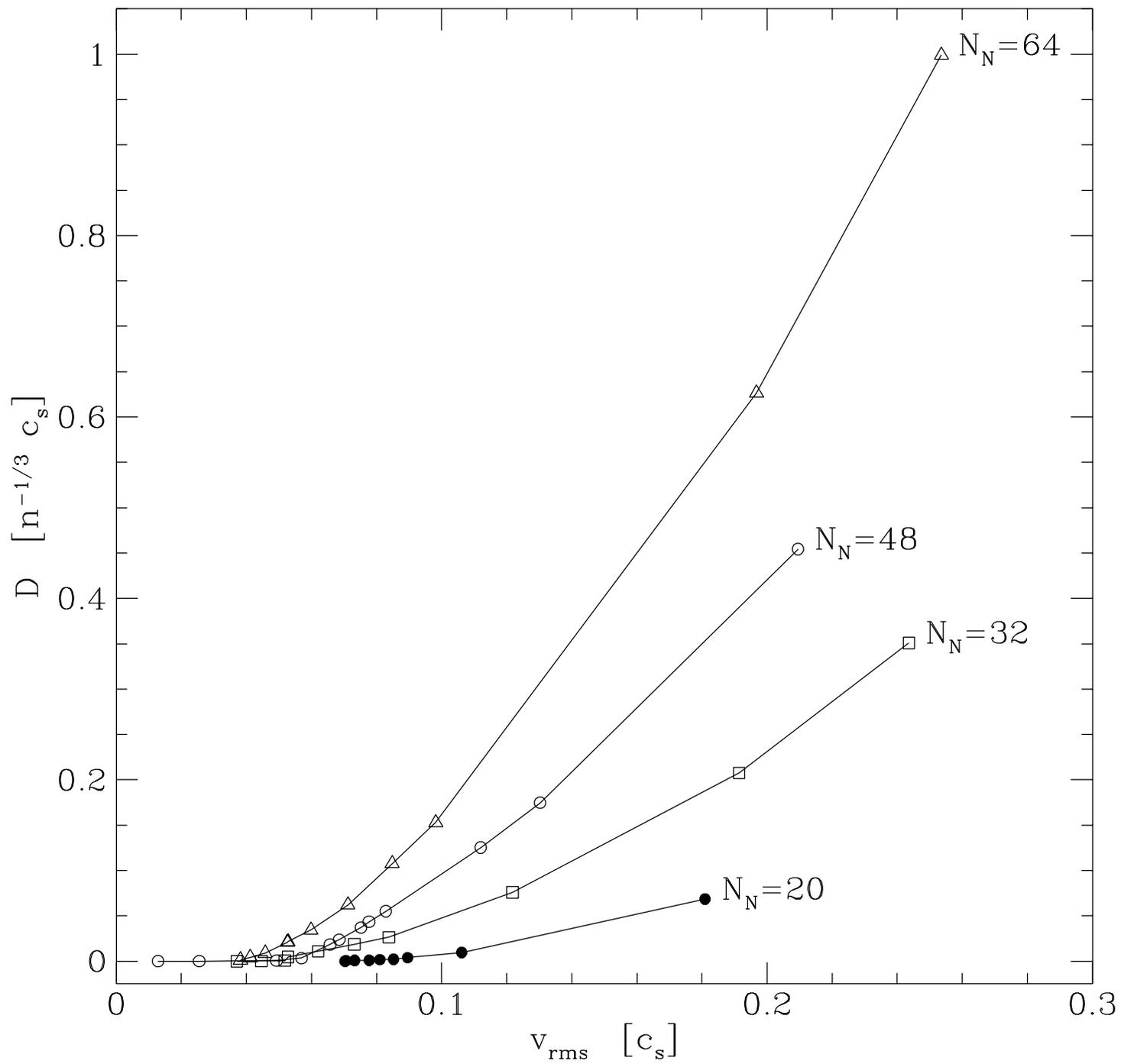
\begin{figure}
\caption{
The diffusion coefficient $D$ as a
function of the root mean square velocity dispersion $v_{rms}$ for
various neighbor numbers $N_N$, as measured by simple box tests in
which the SPH particles began on a simple cubic lattice.
} \label{chall}
\end{figure}

Figure~\ref{chall} shows the diffusion coefficients $D$ for various
$v_{rms}$ and for $N_N=$20, 32, 48 and 64.  Not surprisingly, spurious
diffusion increases as $v_{rms}$ increases.  Note that, for a given
$N_N$, there is a critical noise level below which the diffusion
coefficient $D$ is essentially zero.  In this regime, the SPH particles
settle into a regular lattice and oscillate around their equilibrium
positions, and we say the system has ``crystallized'' (see \S 3.2).
There seems to be a crystallization point for all the curves at some
critical velocity dispersion $v_{cr}>0$.  The trend is for $v_{cr}$ to
decrease as $N_N$ increases.  During the dynamical phase of real
applications, AV typically keeps the noise level low
enough that the numerical temperature is at most slightly above that
required for crystallization.

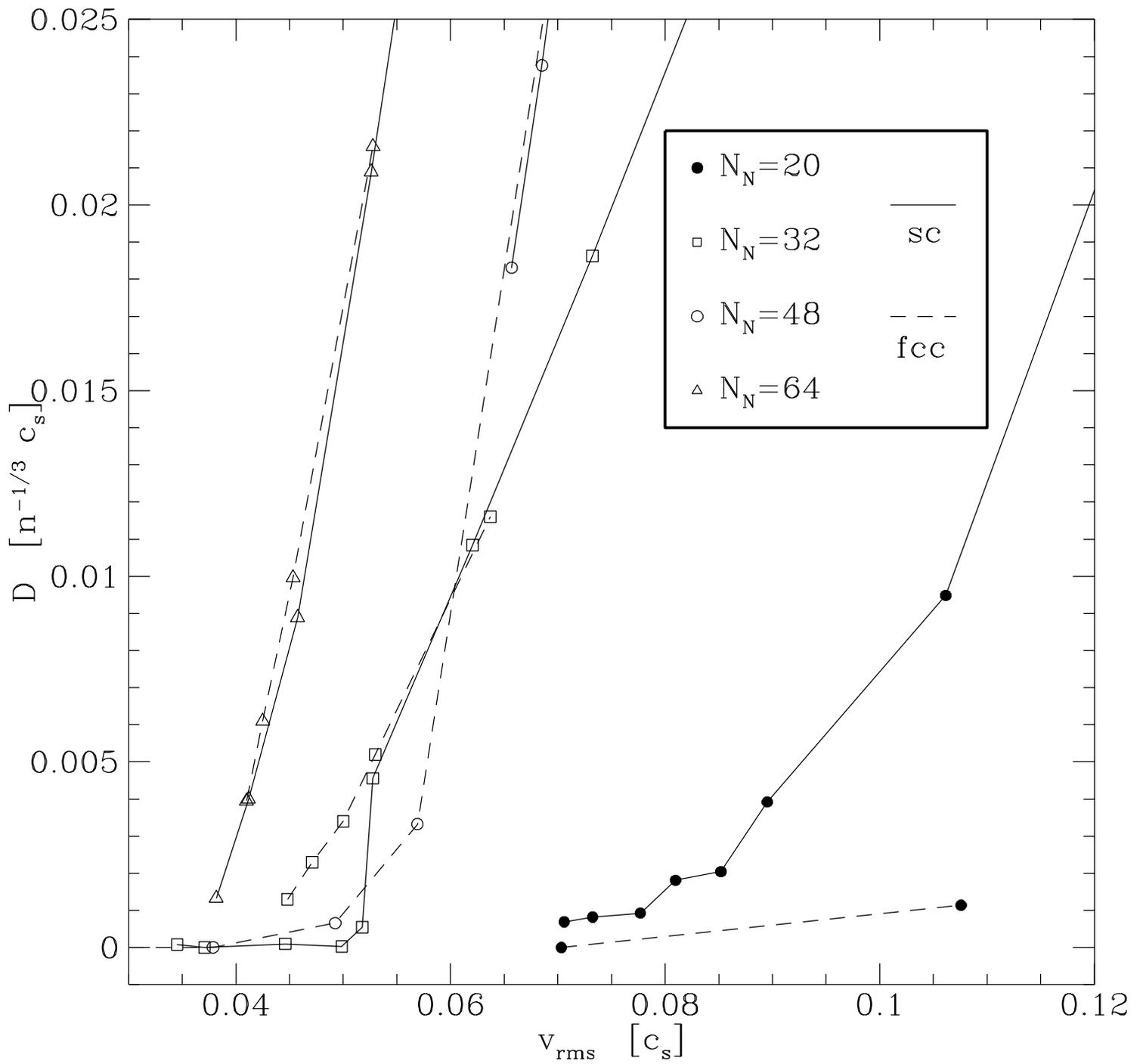
\begin{figure}
\caption{
The diffusion coefficient $D$ near
crystallization.  Conventions are as in Fig.~\ref{chall}. At $t=0$, the
SPH particles began on either a simple cubic lattice (data points
connected by solid lines) or a face centered cubic lattice (data points
connected by dashed lines).  In this regime, $D$ has an obvious
dependence on this system's history.
} \label{chball}
\end{figure}

The diffusion coefficient is not always a unique function of $N_N$ and
$v_{rms}$, but can also depend on the history of the SPH particles.  To
demonstrate this we started the particles on various types of
lattices.  Figure~\ref{chball} shows the measured values of the
diffusion coefficient $D$ in the crystallization regime for systems of
particles which began in either face centered cubic (dashed lines) or a
simple cubic (solid lines) configurations.  There is a clear dependence
on the system's history in this regime, making it impossible to define
a precise crystallization velocity dispersion.  Note that all of the
data points in Figure~\ref{chball} have a small diffusion coefficient,
$D<0.025$.  Well above the crystallization noise level (that is,
outside of the region displayed in Figure 4) the diffusion coefficient
is largely independent of initial conditions; that is, there is
negligible history dependence for sufficiently large $v_{rms}$.

The diffusion coefficients shown in Figures~\ref{chall} and
\ref{chball} are measured while integrating the entropy
equation~(\ref{adot}) with a Courant number $C_N=0.4$ and with the S
timestep algorithm [see eqs. (\ref{dt}), (\ref{dt1}), and
(\ref{simple.dt})].  However, measurements which use the energy
equation~(\ref{udot}) or different Courant numbers, or both, give
similar coefficients, provided only that the Courant number is small
enough that the integration routine is stable.

\subsection{Lattices of SPH particles}

By experimenting with various lattice types as initial conditions in
the simple box tests, we have found that not all equilibrium
configurations of SPH particles are stable.  For example, simple cubic
lattice configurations are unstable to perturbations, while other
lattice types, such as hexagonal close-packed, are stable.  If the
particles begin motionless and slightly perturbed from equilibrium
simple cubic lattice sites, they achieve a non-zero noise level and
readjust their positions to a different, preferred lattice type (see
Figure~\ref{crystal}).  Although the instability develops more slowly
for smaller $C_N$, it cannot be avoided altogether.

For a few of our simple box tests, we allowed the smoothing lengths
$h_i$ to vary both in time and in space, {\it without} including the
corrections in the evolution equations described by Nelson \&
Papaloizou [\cite{Nelson-Papaloizou1993,Nelson-Papaloizou1994}] and Serna et al.\ [\cite{Serna-etal1996}].  The system's behavior is
greatly affected: there is a secular, spurious increase in the total
energy $E$.  Almost all of this spurious energy is kinetic.  If the
AV is active during such runs, energy conservation
is much better; however, the error then emerges as a spurious entropy
increase (see Figure~\ref{utes}).  The AV run in Figure~\ref{utes} used
$\alpha=1$, $\beta=2$, $\eta^2=0.01$ and the classical form of AV; both
runs use $C_N=0.8$ and an initial Maxwell-Boltzmann velocity
distribution with a velocity dispersion $v_{rms}=0.107$.

In many SPH applications, shocks play an important role in the
dynamics.  Therefore, understanding how various AV
schemes affect the level of spurious diffusion is essential.  A uniform
SPH gas is {\it not\/} an appropriate arena to study this effect, since
the AV quickly solidifies the particles into a
lattice structure.  In a calculation with AV but
without shocks or shear, the diffusion coefficient $D$ is always essentially
zero (see Figures~\ref{d2vtAV}~and~\ref{quench}), since diffusion
occurs only as a transient.  

We can derive approximate analytic expressions for the artificial
viscous dissipation timescale by dimensional analysis on the AV term in
equation~(\ref{fsph}).  Here we focus on the classical AV
[eq.~\ref{pi}]; in \S 6.2 we will analyze all three AV forms in a
different context.  Beginning with equation~(\ref{mu}), we note that
since $|{\bf r_i}-{\bf r_j}|\sim h_{ij}$ we have $\mu_{ij}\sim \Delta
v$, where $\Delta v$ is a typical relative velocity of neighboring
particles.  If, in the vicinity of particles $i$ and $j$, the sound
speed is $c_s$ and the density is $\rho$, then equation~(\ref{pi})
gives us $\Pi_{ij} \sim -\alpha \,\Delta v c_s/\rho$ if $\beta \Delta v
<< \alpha c_s$ (as is typically the case in the absence of shocks).  If
the local number density of particles is $n$, then a typical particle
mass $m_j\sim \rho/n$, and $|{\bf \nabla}_i W_{ij}|\sim n/(h N_N)$.
Combining these expressions, we find that the acceleration of particle
$i$ due to the AV is
\begin{equation}
\dot v_i^{AV}\equiv|-\sum_j m_j \Pi_{ij}{\bf \nabla}_i W_{ij}| \sim {\alpha
c_s \,\Delta v \over h N_N^{1/2}}, \label{vdotAV}
\end{equation}
where we have assumed that the sum over $N_N$ terms in
equation~(\ref{fsph}) scales as $N_N^{1/2}$ since there is no
preferred direction for ${\bf \nabla}_i W_{ij}$.

The artificial viscous dissipation timescale $\tau$ is then just $v/\dot
v^{AV}$, where $v$ is a typical particle velocity.  For the simple box
tests we have $v\sim \Delta v \sim v_{rms}$, so that the viscous
timescale is
\begin{equation}
\tau\sim{h N_N^{1/2} \over\alpha c_s}=\left({3 \over 32
\pi}\right)^{1/3} {N_N^{5/6}\over \alpha} n^{-1/3} c_s^{-1}.
\label{tauAV}
\end{equation}
Our numerical results agree well with this simple expression.  For
$\alpha=1$ and $N_N=32$, equation~(\ref{tauAV}) gives a timescale
$\tau\sim 6 n^{-1/3} c_s^{-1}$, which is in reasonable agreement with the
time it takes to form a lattice ({\it i.e.\/} the timescale on which
the kinetic energy drops to zero) in the case presented in
Figure~\ref{d2vtAV}.  Although the timescale depends on both $N_N$ and
the AV, it is always quite short: typically just a
few sound crossing times between neighboring SPH particles.

\section{Polytrope Tests \label{polytrope}}
Applications of SPH often involve self-gravitating systems with
significant density gradients.  The results of our simple box tests can
be applied to such calculations, which we will demonstrate by
considering a set of equilibrium $n=1.5$ polytropes (spherical
hydrostatic equilibrium configurations with
$p=\hbox{const}\times\rho^{1+1/n}$) all with mass $M$ and radius $R$,
but modeled with various total numbers of particles $N$ and neighbor
numbers $N_N$.  In this section, all calculations implement the simple
timestep routine given by equations (\ref{dt})--(\ref{simple.dt}) and
have no AV.  The natural units are given by $G=M=R=1$, so that
consequently the unit of time is $(R^3/GM)^{1/2}$.

We relax the polytrope to equilibrium by applying an artificial drag
force which opposes motion for 20 time units.  We then remove the drag
force and record the particle positions.
Ideally, the particles would remain stationary.  However,
as expected from the results of
\S 3.1, these particles spuriously diffuse from
their starting positions, and this diffusion is easy to monitor.  By
periodically noting the particle velocity dispersion $v_{rms}$, we
can apply the simple box test results to get an ``instantaneous'' value
for the diffusion coefficient $D$ by interpolating between data points
in Figure~\ref{chall}.  In this way, we ``predict'' the mean
square displacement $\delta^2$ by a simple, numerically evaluated
integral,
\begin{equation}
\delta^2=\int D(t) dt, \label{predicted.delta2}
\end{equation}
and then compare this prediction to the actual, measured mean square
displacement.

Figure~\ref{d2vtpoly} shows, as a function of time, the mean square
spurious displacement for the innermost 6400 particles in an $n=1.5$
polytrope modeled with $N=13949$ equal mass particles, each with
$N_N=48$ neighbors on average.  We do not track the particles of the
outer layers here, since they are subject to an effect which we do not
attempt to model: when such a particle diffuses outward beyond the
surface, gravity pulls it back, making the actual diffusion distance
somewhat smaller than predicted.  For those particles which always
remain inside the surface, gravity is everywhere balanced by pressure
gradient forces, so that the rate of diffusion is essentially the same
as in our simple box tests.  The usual advection scheme
equation~(\ref{rdot}) was used for the calculation presented in the top
frame of Figure~\ref{d2vtpoly}, while the XSPH equation~(\ref{XSPH})
with $\epsilon=0.5$ was used in the bottom frame.  The ``predicted''
mean square displacement (dashed curve), as calculated from equation
(\ref{predicted.delta2}), agrees well with the actual square
displacement (solid curve).  To obtain the predicted curve in the XSPH
calculation, the root mean square of the right hand side of
equation~(\ref{XSPH}) was used in place of $v_{rms}$ when determining
the diffusion coefficient $D$.  The Courant number $C_N=0.8$ and the
simple timestep routine determine the integration timesteps for both
cases.

The slight differences between predicted and actual displacements arise
because of our interpolating to obtain $D$ and because our diffusion
coefficients are only approximate in the crystallization regime (due to
history dependence).  Since the SPH particles are melting out of their
crystalline phase around $t\approx10$, our values for $D$ are
overestimated then.  The XSPH advection method does indeed diminish the
amount of spurious diffusion: the final ($t=30$) mean square
displacement for the XSPH calculation is nearly one fourth of the value
from the simple advection scheme.  However, one must be careful when
using XSPH: using too large of an $\epsilon$ can cause certain modes to
become numerically unstable.  For instance, for the extreme case of
$\epsilon=1$ we are not able to evolve an equilibrium $n=1.5$ polytrope
without the integration becoming unstable.

In order to test the importance of the Courant number $C_N$, we evolved
a set of equilibrium $n=1.5$ polytropes using several values of $C_N$
between $0.1$ and $1.6$.  In all cases we turned off the AV, used
$N=13949$ equal mass particles each with $N_N=48$ neighbors on average,
allowed the smoothing lengths to vary in space and time, used the
``simple'' timestep routine, and monitored three measures of error: the
fractional (spurious) change in total energy $\Delta E/E$, the velocity
dispersion $(v/c_s)_{rms}$, and the mean square diffusion distance
$\delta^2$.  As $C_N$ increased from $0.1$ to $1.1$, these errors,
evaluated at $t=25$, increased only very slightly:  $\Delta E/E$
increased from $0.014$ to $0.017$, $(v/c_s)_{rms}$ from $0.13$ to
$0.14$, and $\delta^2/R^2$ from $0.02$ to $0.03$.  For $C_N=1.2$, the
integration becomes unstable, with the errors at $t=25$ then being
$\Delta E/E=0.7$, $(v/c_s)_{rms}=0.3$, and $\delta^2/R^2=0.2$.  This
result suggests that in certain cases, for fixed computational
resources, it may be more efficient to use a relatively large Courant
number like $C_N=0.8$ and more particles, rather than a small Courant
number like $C_N=0.3$ and fewer particles.

Figure~\ref{nn} shows $\Delta E/E$, $(v/c_s)_{rms}$, and $\delta^2/R^2$
at $t=25$ for a set of calculations with $C_N=0.8$ and various $N_N$.
Here the $n=1.5$ polytropes are modeled by either $N=30000$ particles
(circular data points) or $N=13949$ particles (square data points).
For a given $N_N$, the $N=30000$ models always have larger
accumulated errors: as $N$ is increased, one must also increase $N_N$
in order for the SPH scheme to remain accurate.  Although increasingly
larger $N_N$ results in increasingly smaller errors, this does not mean
one should strive to use as large a value for $N_N$ as possible.  Large
$N_N$ yields large smoothing lengths and hence poor spatial
resolution.  The optimal $N_N$ must be determined by a compromise
between the competing factors of accuracy and resolution, and depends
on the particular application.  Nevertheless, we can place very loose
constraints on how fast the optimal $N_N$ should be increased as $N$ is
increased.  From Figure~\ref{nn} we see that in going from $N=13949$ to
$N=30000$ we need to increase $N_N$ by at least (very roughly) 15\% in
order to prevent the errors from increasing.  This corresponds to a
scaling $N_N\propto N^q$ with $0.2\lo q < 1$, assuming a power-law
relation.  The upper limit of 1 on $q$ stems from the requirement that
the smoothing lengths must decrease as $N$ and $N_N$ increase.

We have performed our diffusion tests using equal mass particles.
Sometimes, however, SPH calculations use particles of unequal mass so
that less dense regions can still be highly resolved.  Unfortunately,
the more massive particles tend to diffuse to the bottom of the
gravitational potential more so than less massive ones.  In other
words, each particle has a preferred direction to diffuse, and in a
dynamical application this direction can be continually changing.  As
an example, we evolved an equilibrium $n=1.5$  polytrope in which the
SPH particles initially in the envelope were, on average, heavier than
those in the core.  Over the course of the calculation, the heavier
particles settled to the core while the lighter particles tended to the
envelope (see Figure~\ref{bhist}).  Such behavior makes spurious
diffusion more difficult to predict for calculations which use unequal
mass particles.

\section{Periodic Shock-Tube Tests \label{shock-tube}}
Since the simple box tests of \S 3 are helpful only for calculations
without AV, we turn
now to a periodic version of the 1D Riemann shock-tube
problem of Sod [\cite{Sod1978}], a standard test of hydrodynamic codes and AV
schemes containing
many of the same qualitative features as real applications
which involve shocks.  The physical setup is as follows.  

Initially, fluid slabs with constant (and alternating) density $\rho$
and pressure $p$ are separated by an infinite number of planar,
parallel, and equally spaced interfaces. If we define the unit of
length to be twice the distance between adjacent interfaces, and if we
let the $x=0$ plane coincide with one of these interfaces, then

\begin{equation}
\left. \begin{array}{ll}
	\rho=\rho_l,~p=p_l & \mbox{if $-{1 \over 2}<x\le 0$} \\
	\rho=\rho_r,~p=p_r & \mbox{if $0<x\le {1 \over 2}$},
	\end{array}
\right.	\label{thermo}
\end{equation}
where $\rho_l$, $p_l$, $\rho_r$ and $p_r$ are constants specifying the
density and pressure of the slabs to the ``left'' and ``right'' of
$x=0$.  Pressures and densities for $|x|>{1 \over 2}$ are given by
repeatedly stacking the thermodynamic slabs described by
equation~(\ref{thermo}) along the x-axis to infinity, hence the name
{\it periodic} shock-tube tests.  At $t=0$ the interfaces are removed
and, if $p_l \neq p_r$, a shock wave moves from the high pressure
material into the low.  A rarefaction wave also originates at each
interface, propagating in the direction opposite to its corresponding
shock.  Before the initial collision of shock waves from adjacent
interfaces, regions of five different thermodynamic states coexist and
the entropy of the fluid increases linearly with time.  A
quasi-analytic solution can be constructed for these early times using
standard methods (see, e.g., [\cite{Courant-Friedrichs1976}]) and is
presented in detail by Rasio \& Shapiro [\cite{Rasio-Shapiro1991}].

\subsection{Low Mach Number Cases \label{lowMach}}

For the first set of shock-tube calculations we consider, the fluid slab
to the left of the interface at $x=0$ initially has density
$\rho_l=1.0$ and pressure $p_l=1.0$, while on the right $\rho_r=0.25$
and $p_r=5/2^{16/3}=0.12402$.  Consequently this box contains $0.625$
units of mass: $0.5$ on the left and $0.125$ on the right.  An
adiabatic equation of state is used with $\gamma=5/3$, so that the
entropy variable $A$ equals $1.0$ on the left and $1.25$ on the
right.  From equation~(\ref{entropy}), the initial entropy of each of
the periodic cells is thus $S=1.5[0.5\ln (1.5)+0.125\ln
(1.5\times1.25)]=0.4220$.  For these initial conditions, the initial
shock waves have a relatively low Mach number ${\cal M}\approx 1.6$.
In these units, the speed of sound in the initial left hand slab is
$c_s^l=(\gamma p_l/\rho_l)^{1/2}=\gamma^{1/2}$, and the unit of time is
therefore $\gamma^{1/2}L/c_s^l$, where $L$ is the length of a periodic
cell (our unit of length).

Employing AV with the form of equation~(\ref{pi}), we
obtained a good representation of the shock with our 1D code by using
the AV parameters $\alpha=\beta=1$ and $\eta^2=0.05$ in the classical
AV.   The smoothing length $h$ of the $N=2500$ equal mass particles was
constant and chosen such that the particles would have $N_N=16$
neighbors on average.  Our 1D code integrates the energy equation, and
uses the Monaghan timestep routine with $C_N=0.2$.  Figure~\ref{rhox}
shows the density and velocity profiles as given by the quasi-analytic
solution (solid curve) and our 1-dimensional code (dotted curve) at a
time $t=0.15$.  As expected, discontinuities are smoothed over a few
smoothing lengths.  Figure~\ref{svst} shows the entropy [see
eq.~(\ref{entropy})] given by our 1D SPH code (dotted curve), which
nearly matches the quasi-analytic solution (solid curve).

The above calculation helps establish the accuracy of our 1D code, but
does not realistically assess the accuracy of a 3D calculation, where
the much smaller number of particles per dimension leads to a reduced
spatial resolution.  Furthermore, many sources of numerical errors,
including spurious mixing, are artificially reduced when motion with
only one degree of freedom is allowed.  We test our 3D code by using it to
simulate exactly the same physical problem: at $t=0$, slabs of fluid
with alternating thermodynamic states are separated by equally spaced
planar interfaces perpendicular to the $x$-axis.  Periodic boundary
conditions are imposed on all six sides of a cube with faces at $x=\pm
{1 \over 2}$, $y=\pm {1 \over 2}$, and $z=\pm {1 \over 2}$.  We
considered cases only with a constant smoothing length $h<<1$, and,
unless otherwise stated, we integrate the entropy equation.

Our calculations with the 3D code use $N=10^4$ equal-mass particles.
All the particles initially in the left hand slab have the same
smoothing length, smaller than the smoothing length common to particles
initially in the right hand slab.  These smoothing lengths are not
allowed to vary with time, and are chosen such that particles which are
farther than $2h$ from an interface have $N_N=64$ neighbors on
average.  Within each constant density slab, the SPH particles start in
a stable lattice with a randomly chosen orientation (choosing the
lattice face to be parallel to the interface would be too artificial of
a setup).  The initial conditions for each slab are constructed by
randomly distributing particles in a periodic box of dimensions
${1\over 2}\times 1\times 1$ and then slowly relaxing the system with
an artificial drag force.  The resulting lattices are preferred to
initially randomly distributed particles, since a random distribution
would introduce a high noise level not representative of real
applications.

We determine the accuracy of our calculations with the 3D code by
comparing its results against those of the much more accurate 1D code.
Such 3D calculations are a useful and realistic way to calibrate spurious
transport in simulations with AV, since the test
problem, which includes shocks and some large fluid motions, has many
of the same properties as real astrophysical problems.  In fact, the
recoil shocks in stellar collisions do tend to be nearly planar, so
that even the 1D geometry of the shock fronts is realistic.  The
periodic boundary conditions play the role of gravity in the sense that
they prevent the gas from expanding to infinity.

Figure~\ref{latetmost} shows the pressure $P$, entropy variable $A$,
density $\rho$ and velocity $v_x$ as given by our 1D code (solid curve)
and by our 3D code (dots) at the relatively late time $t=1$.  Here the
3D calculation
implements the classical AV with $\alpha=0.5$ and $\beta=1$.  The bar
in the lower left corner of the uppermost frame displays the average
region of influence (i.e.~the mean diameter of the smoothing kernels)
for the particles in the 3D calculation: the total length of this bar
is $4\langle h\rangle$, where $\langle h\rangle=0.058$ is the average
smoothing length.  The 3D calculation does well at reproducing the
major features in the thermodynamic profiles, but, not surprisingly,
smoothes out any small scale structure which occur on lengths scales
shorter than a few smoothing lengths.  In the regions near $x=0.1$ and
$x=0.4$, where the fluid is being shock-heated, the pressure, entropy
variable and density in the 3D calculation are double-valued due to the
shock front not remaining perfectly planar throughout the calculation.

Since the fluid motion in these calculations should be solely in the
$x$-direction, spurious motion in the $y$- and $z$-directions is easy
to measure.  Spurious motion in the $x$-direction can be studied by the
following method, based on the idea that planes of fluid should not
cross in one-dimension.  That is, the shape of a composition profile
should remain unchanged throughout a calculation.  Once the shock-tube
system has reached a steady state, we examine the distribution of the
Lagrangian labels $x_i(t=0)$ as a function of $m(x)$, the amount of
mass between the interface (contact discontinuity) and $x$.  Deviations
from the initial profile must be spurious, so we can immediately
calculate spurious displacements in the $x$-direction for individual
particles.  Diffusion measurements in each of the three directions give
similar results.

We have studied the quality of the 3D code's results for various
AV parameters and forms.  We have completed a number
of shock-tube tests which began with the same initial conditions
described above, but with different values of the AV
parameters.  Varying $\eta^2$ by a factor of $25$ between $0.002$ and
$0.050$ makes little difference in the results, and we therefore
concentrate on the effects of $\alpha$ and $\beta$.  All calculations
described in this section have $\eta^2=0.01$.

Figure~\ref{varyAV} shows the dependence of the solution on $\alpha$
and $\beta$ for the classical AV by plotting, as a function of time,
the mean square spurious displacement in the directions perpendicular
to the bulk fluid motion (in units of $n^{-1/3}$, where $n$ is the SPH
particle number density), the internal energy $U$ and the entropy $S$.
The solid line results from our accurate calculation of the shock-tube
problem with the 1D code.  In frame (a) of Figure~\ref{varyAV},
$\alpha=0$ while $\beta$ is varied.  In (b) $\beta=0$ and $\alpha$ is
varied.  Finally in (c) $\beta=1$ and $\alpha$ is varied.  Runs with
$\alpha=0$ or $\beta=0$ are interesting since they represent an
AV which is either purely quadratic (von
Neumann-Richtmyer viscosity) or linear (bulk viscosity) in $\mu_{ij}$,
respectively, and these two types of AV generate different
numerical viscosities (see \S 6).

Table I
summarizes all of our low Mach number 3D shock-tube calculations and
reports how well each does matching the 1D solution.  All the
calculations in Table I employed $10^4$ particles and a fixed smoothing
length chosen such that the number of neighbors $N_N=64$ on average.
In Column 1, we identify the type of AV used: C for the classical AV
[eq.~(\ref{pi})], HK for the Hernquist \& Katz AV [eq.~(\ref{pi2})],
and B for the Balsara AV [eq.~(\ref{piDB})].  Columns 2 and list the
AV parameters $\alpha$ and $\beta$ (unless otherwise
noted $\eta^2=0.01$).  Column 4 gives the type of timestep routine
used: S for simple [eq.~(\ref{simple.dt})] and M for Monaghan
[eq.~(\ref{good.dt})].  Column 5 gives the Courant number $C_N$.
Columns 6 and 7 give the number of iterations required to reach $t=1$
and $t=4$, respectively.  Column 8 gives the fractional deviation in
the total energy away from its initial value: $\Delta
E/E=|E(t=4)-E(t=0)|/E(t=0)$.  The $t=4$ value of
$\delta_y^2+\delta_z^2$, the spurious displacement squared in the
direction perpendicular to the bulk fluid flow, averaged over all
particles, is listed in Column 9.  Columns 10 and 11 give the maximum
deviation in $U/E$ and $S$, respectively, from that of the 1D code:
$\Delta(U/E)_{max}\equiv \hbox{Max}|U_{3D}/E_{3D}-U_{1D}/E_{1D}|$ and
$\Delta S_{max}\equiv \hbox{Max}|S_{3D}-S_{1D}|$.

Figure~\ref{compareav} shows, as a function of time, the average square
displacement perpendicular to the bulk fluid flow
$\delta_y^2+\delta_z^2$, the ratio of internal to total energy
$U/E$, and the entropy $S$ for three calculations with different forms
of AV:  the classical AV with $\alpha=0.5$, $\beta=1$ (long dashed
curve), the HK AV with $\alpha=\beta=0.5$ (short dashed curve), and the
Balsara AV with $\alpha=\beta=\gamma/2$ (dotted curve).  In the bottom two
frames, the solid curve corresponds to our 1D SPH code.  As we will discuss
in \S 7.4, these choices for $\alpha$ and
$\beta$ are our recommended values.  We see that all three AV forms can
handle the shocks with roughly the same degree of accuracy, although
the HK AV does allow slightly more spurious mixing and does not match
the 1D code's $U/E$ curve quite as well.

We also ran several low Mach number calculations with the energy
equation being integrated.  Table II compares these runs against the
corresponding calculations in which the entropy equation was
integrated.  For given values of $\alpha$, $\beta$ and $\eta^2$, the
two schemes do equally well at conserving energy, at controlling
particle diffusion, and at matching the time evolution of $U/E$ from
the 1D calculation.  However, integrating the energy equation does allow
slightly larger errors in the evolution of entropy, with $\Delta
S_{max}$ being 0.005 to 0.007 larger than when the entropy equation is
integrated.  This larger error in the entropy accumulates mostly at
early times when the shocks are strongest.

\subsection{High Mach Number Cases}

Since many astrophysical situations involve shocks which are stronger
than the low Mach number situation described in the previous section,
we repeated shock-tube tests with a larger difference in pressure
between the alternating fluid slabs.  In particular, we initially set
$p_l=1.0$ , $\rho_l=1.0$ and $\rho_r=0.25$. but reduced the pressure of
the right-hand fluid slab to $p_r=1.2402\times 10^{-3}$, a factor of
100 less than in the low Mach number cases of \S 5.1. This
increases the Mach number of the initial shock waves to ${\cal M}
\approx 13.2$.  The initial entropy of each of the periodic cells is
$S=1.5[0.5\ln (1.5)]+0.125\ln (1.5\times0.0125)]=-0.4415$.

For our 1D code, we continued to use the classical AV [see 
eq.~(\ref{pi})] with parameters $\alpha = \beta = 1$ and $\eta^2 =
0.05$.  We used 2500 particles and constant (in time) smoothing lengths $h_i$,
chosen such that the particles have 16 neighbors initially.  Figure
\ref{1dhighMach} shows a comparison between our 1D SPH code (dotted
curve) and the quasi-analytic solution (solid curve) at a time $t=0.15$. As expected, the
1D code does smooth out discontinuities in the density over
a width of a few smoothing lengths.  However, the agreement between the 1D
code and the quasi-analytic solution is still very good.

As in the low Mach number case, we can compare the results from the 3D
code to that of the 1D code, in order to evaluate the amount of
spurious mixing and to determine the acceptable range of values for
the AV parameters for our 3D calculations. Table III
is the high Mach number equivalent of Table I.
%
%
These 3D calculations employ $N=10^4$ particles each
with $N_N=64$ neighbors, as in the 3D low Mach number calculations.

In Figure \ref{highMachs} we present the results of 3D high Mach number
shock-tube calculations for various $\alpha$ and $\beta$
with the classical AV.  For all
the 3D calculations in this figure, we chose $\eta^2=0.01$ and used the
Monaghan timestep routine with $C_N=0.8$.  The solid line is the result
of the 1D calculation.  It is apparent that the spurious
displacement is smaller for stronger AV, as expected
and as in the low Mach number tests.  As also seen in the low Mach
number tests, the case with the lowest spurious mixing ($\alpha=5,
\beta=0$) has the worst fit to the energy curve of the 1D
calculation.  The entropy curve from the 1D case lies
between the cases with the high values of $\alpha$ or $\beta$, and
those with the low values. Therefore, the best choice of AV parameters
will depend on the particular situation which is
to be modeled. If spurious mixing is important to control, then a
strong viscosity is favorable. On the other hand, if spurious mixing is
not an issue, one could use a weaker AV to more
accurately determine the evolution of the system.

Figure~\ref{compareavhighm} shows, as a function of time, the average
square displacement perpendicular to the bulk fluid flow
$\delta_y^2+\delta_z^2$, the ratio of internal to total energy $U/E$,
and the entropy $S$ for three calculations with different forms of AV:
the classical AV with $\alpha=0.5$, $\beta=1$ (long dashed curve), the
HK AV with $\alpha=\beta=0.5$ (short dashed curve), and the Balsara AV
with $\alpha=\beta=\gamma/2$ (dotted curve).  In the bottom two frames, the
solid curve corresponds to our 1D SPH code.  As will be discussed in
\S 7.4, these choices for $\alpha$ and $\beta$
are our recommended values.  We see that the HK AV does allow slightly more
spurious mixing and does not quite match the 1D code's $U/E$ curve as
well.  Nevertheless, all three AV forms can
adequately treat the strong shocks of this system.

\subsection{High Mach Number Cases with $\gamma=3$}

Of course, not all fluids are well-described by the ideal gas
($\gamma=5/3$) approximation.  For example, neutron star matter is best
represented by a stiff equation of state with $\gamma\approx2$--3,
while an isothermal gas can be described with $\gamma=1$. Changing the
value of $\gamma$ changes the thermodynamic
properties of the material we model with SPH, which
in turn affects the way the AV behaves. Therefore, to
investigate the dependence on $\gamma$ of the `optimal' AV
parameters, we have performed some shock-tube calculations with $\gamma
=3$. The fluid slabs were set up to have the same Mach number as the
previous high Mach number ideal gas tests (${\cal M}=13.2$):
$\rho_l=1, p_l=1, \rho_r=0.25,$ and $p_r=8.78\times10^{-7}$. The
initial entropy of each periodic cell is $S=0.5[0.5
\ln(0.5)+0.125\ln(0.0562/2)]=-0.3965$.

For the corresponding calculation with the 1-D code, we used the
classical AV scheme with $\alpha=\beta=1$ and $\eta^2=0.05$, 2500
particles and 16 initial neighbors, as in the previous high Mach number
case. For our 3-D calculations, we used $10^4$ particles with 64 initial
neighbors, and a variety of AV parameters with all three AV schemes. We
used the Monaghan timestep routine with $C_N=0.3$. The different 3-D
calculations and their comparison to the 1D runs are given in Table IV,
and a selection of the results are shown in Figure
\ref{compareavgam3}.

As in the ideal gas case, spurious diffusion is smaller for stronger
artificial viscosities.  The calculations with small $\alpha$ show
additional ``wiggles'' in the energy curve (see Fig.
\ref{compareavgam3}) and larger errors in energy conservation
(see Table IV), suggesting the appearance of numerical instabilities for
strong shocks treated by weak AV forms.  In general, we find that the level
of energy conservation is worse in our $\gamma=3$ calculations than in
our $\gamma=5/3$ calculations (compare Tables 3 and 4).

\section{Shear Flows \label{sheartests}}
\subsection{Periodic Box Tests}

In order to model a shear flow of infinite extent, we return to a
cubical box with a side length $L=1$ and periodic boundary conditions.
The boundary conditions on the $x=\pm{L\over 2}$ and $z=\pm{L \over 2}$
faces are identical to the periodic boundary conditions in the simple
box tests of \S 3: when a particle crosses one of these faces it is
reinserted with the same velocity at the corresponding position on the
opposing face.  On the $y=\pm{L\over 2}$ faces, however, we implement
``slipping'' boundary conditions in order to maintain a velocity field
with a shear flow: if a particle crosses a face with a velocity
$(v_x,v_y,v_z)$ at a position $(x,\pm{L\over 2},z)$, it is reinserted
with a new velocity $(v_x\mp v_0,v_y,v_z)$ at the position $(x\mp
v_0t+kL,\mp{L\over 2},z)$, where $t$ is the time elapsed since the
beginning of the calculation and $k$ is the integer which places the
particle in the central periodic cell.  The resulting ``stationary
Couette flow'' has a velocity field close to $(v_0y/L,0,0)$ (see
Fig.~\ref{couette}).

Neighbor searching across the $x=\pm{L\over 2}$ and $z=\pm{L\over 2}$
faces is done exactly as in \S 3.1.  Across
the $y=\pm{L\over 2}$ faces, the slipping boundary conditions are taken
into account: the criterion for particle $j$ having particle $i$ as a
neighbor is that there exists integers $k$, $l$ and $m$ such that the
position ($x_i+kL+lv_0t,y_i+lL,z_i+mL$) is within a distance $2h_j$ of
($x_j,y_j,z_j$).  In addition, the relative velocity of particles
interacting across the $y=\pm{L\over 2}$ boundaries is adjusted by
$v_0$ when computing the AV term $\Pi_{ij}$.  In this way,
particle interactions across the boundaries behave identically to
interactions within the box.


Our units of distance and mass are the length of the box and the total
mass in the box: $L=1$ and $M=Nm\equiv1$, where $N$ is the
number of particles.  We set the entropy variable $A=1$ for all
the particles initially.  Consequently the unit of velocity in our
calculations is $\gamma^{-1/2} c_s$, where $c_s$ is the initial sound
speed, and the unit of time is $\gamma^{1/2}L/c_s$.

Figure \ref{shear} shows the spurious square displacement, energies, and
entropy as a function of time in three calculations with $N=1000$, $N_N=64$,
$v_0=0.1 \gamma^{-1/2} c_s$,
and various forms of AV.
The system was relaxed for the first 10 time units (without AV) towards
a situation with $(v_x,v_y,v_z)=(v_0y/L,0,0)$, while from $t=10$ to 50
the system evolves freely with the slipping boundary conditions and AV.

Notice the increase in energy once the relaxational damping is turned
off: roughly a 1\% increase in $E$ per time unit.  This increase
results from the slipping boundary conditions and, for a given AV form
and AV parameters, is nearly independent of the resolution.  Since we
are moving the boundary surfaces by hand and since
there is viscosity, there is a shear stress at the boundaries and work
is being done on the system.  This behavior is analogous to that of a
truly viscous fluid forced to undergo shear flow between close moving
boundaries (as in a viscosimeter): the added energy goes into heating
the fluid.

Since the faces of our cubical box have surface area $L^2$, the viscous
force $F_x$ acting on the fluid inside of the $y=\pm L/2$ faces is
\begin{equation}
F_x=\pm\eta{\partial v_x \over \partial y} L^2=\pm\eta v_0 L,
\end{equation}
where $\eta$ is the dynamic viscosity (not to be confused with the AV
parameter $\eta^2$).  The rate of energy change of the system is
therefore
\begin{eqnarray}
{dE\over dt}&=&\left[F_x v_x\right]_{y=-L/2}+\left[F_x v_x\right]_{y=+L/2}\\
& = & \eta v_0^2 L, \label{dedt}
\end{eqnarray}
Measuring the rate of energy increase therefore allows us to
numerically determine the dynamic viscosity.  This procedure for
measuring viscosity is also implemented in molecular dynamics (eg.\ [\cite{Naitoh-Ono1976}]).  To calculate the kinematic viscosity $\nu$ from
the dynamic viscosity $\eta$, one simple uses $\nu\equiv\eta/\rho=\eta
N/(Mn)$, where $n$ is the number density of particles.

In the absence of any spurious motion, SPH particles should maintain
the same spatial coordinates $y$ and $z$ throughout the calculation.
By monitoring motion in these two dimensions, we can therefore easily
quantify the extent of spurious diffusion.  As in \S 3, the square
displacement increases linearly with time at late times.  Here we
measure the diffusion coefficient $D$ by fitting the relation
$3(\delta_y^2+\delta_z^2)/2=Dt$.  In practice, we determine $\eta$ and
$D$ from the average slope of the energy and square displacement
curves, respectively, between times $t=12$ and $t=50$.  Tables V and VI
list the results of a set of runs at two different shear velocities
with $N_N=64$.  We vary the AV scheme and the AV parameters, and
monitor the time averaged velocity dispersion
$\langle(v_y^2+v_z^2)/c_s^2\rangle$ between $t=12$ and 50.  We also
list the viscosity $\eta$ [as determined from eq.\ (\ref{dedt})], the
diffusion coefficient $D$, and the product $\eta D$ for each case (all
converted into units $M=c_s=n=1$ to keep our results applicable to
general situations).  In the last three columns, the number in
parentheses ``()'' is the error in the last digit, or last two digits,
that is quoted.  The uncertainties for the viscosity $\eta$ and the
diffusion coefficient $D$ are determined from the root mean square
deviation of $E(t)$ and of $\delta_y^2(t)+\delta_z^2(t)$ from the
best-fit linear curve.  In Tables VII we present results from a handful of
calculations with various neighbor numbers $N_N$.  All of the
calculations use constant smoothing lengths, as well as a constant
timestep $dt=0.01$ so that fixed computational resources are
available.

\subsection{Rapidly rotating, self-gravitating fluids}
\label{subsection:rapidly}

Rotation plays an important role in many hydrodynamic processes in
astrophysics.  For
instance, the collision of two stars typically results in a rapidly and
differentially rotating merger remnant.  Even in the absence of shocks,
AV tends to damp away differential rotation due to
the relative velocity of neighboring particles at slightly different
radii.
Many systems are best modeled as a perfect fluid, ideally with a
viscous timescale $\tau=\infty$.  In such cases, any viscosity
introduced by the SPH scheme is spurious.  In this section, we
consider a differentially rotating, self-gravitating fluid and
analytically estimate the viscous timescale for each of the three AV forms
examined in this paper.  Our analytic estimates are then compared
against numerical determinations of the viscous timescale.  The larger
the viscous timescale, the more closely the calculation is treating
the gas as a perfect fluid.

As our concrete example, we consider an axisymmetric equilibrium
configuration rotating with an angular velocity $\Omega \propto
\varpi^{-\lambda}$, where the cylindrical radius $\varpi$ is the
distance to the rotation axis.  In this case, the magnitude of the
quantity $({\bf v}_i-{\bf v}_j)\cdot ({\bf r}_i-{\bf r}_j)$ which
appears in equation~(\ref{mu}) is $\sim h \,\Delta v$ for two
neighboring particles separated by $\sim h$, a typical smoothing
length, where the shear velocity $\Delta v \equiv \lambda \Omega h$.
Note that $\Delta v=0$ for the special case of rigid rotation
($\lambda=0$).

If the AV is of the form of
equation~(\ref{pi}) with $\beta=0$, equation~(\ref{vdotAV}) gives $\dot
v^{AV}$ and the viscous dissipation timescale $\tau\equiv v/\dot
v^{AV}=\Omega\varpi/\dot v^{AV} \sim \varpi N_N^{1/2}/(\alpha \lambda
c_s)$.  Note that this timescale $\tau$ is not directly dependent on
$N$: increasing $N$ while keeping $N_N$ fixed would not therefore
affect the viscous timescale in this case.  For general $\alpha$ and
$\beta$, we analytically estimate from equation~(\ref{pi}) that
\begin{equation}
\Pi_{ij}\approx -j_1{\alpha \Delta v c_s\over \rho} - j_2{\beta \Delta
v^2\over \rho}, ~~~~~\hbox{(Classical AV)}
\end{equation}
where $j_1$ and $j_2$ are dimensionless coefficients of order unity.
In this case equation~(\ref{vdotAV}) must be replaced by $\dot v^{AV}\approx
k_1\alpha c_s\,\Delta v/(h N_N^{1/2})+k_2 \beta \Delta v^2/(h
N_N^{1/2})$, and the viscous timescale $\tau=v/\dot v^{AV}$ is then given
by
\begin{equation}
\tau\equiv {v\over\dot v^{AV}}
\approx v \left(k_1{\alpha\Delta v c_s\over h N_N^{1/2}}+
		k_2{\beta\,\Delta v^2\over h
		N_N^{1/2}}\right)^{-1}
=\left(k_1{\alpha\lambda c_s\over \varpi N_N^{1/2}}+
		k_2{\beta\lambda\,\Delta v\over \varpi
		N_N^{1/2}}\right)^{-1},~~~~~\hbox{(Classical AV)}
			 \label{tau}
\end{equation}
where $k_1$ and $k_2$ are dimensionless coefficients of order unity.
The ratio of the two terms on the right hand side of equation
(\ref{tau}) tells us that the von Neumann-Richtmyer viscosity
(corresponding to the term with $\beta$) yields a timescale longer than
that of the bulk viscosity by a factor of $\sim \alpha
c_s/(\beta\,\Delta v)$.  The bulk viscosity therefore dominates the
shear for the classical AV, provided only that $\Delta v << c_s$.

If the AV is instead given by HK form [eq.~(\ref{pi2})],
dimensional analysis gives
\begin{equation}
\Pi_{ij}\approx -j_1'{\alpha \Delta v c_s\over \rho N_N^{1/2}} -
j_2'{\beta \Delta v^2\over \rho N_N}, ~~~~~\hbox{(HK AV}) \label{PIij2}
\end{equation}
if $({\bf \nabla}\cdot {\bf v})_i$ or $({\bf \nabla}\cdot {\bf v})_j$
is negative (otherwise $\Pi_{ij}=0$).  Although our idealized velocity
field satisfies $({\bf \nabla}\cdot{\bf v})_i=0$, the numerical
estimation of the velocity divergence, as computed by
equation~(\ref{divv}), gives small but non-zero results.  In deriving
equation (\ref{PIij2}) we have used $|({\bf v}_i-{\bf v}_j)\cdot{\bf
\nabla}_iW_{ij}|/n\sim \Delta v/(h N_N)$, which implies $|{\bf
\nabla}\cdot{\bf v}|_i \sim \Delta v/(h N_N^{1/2})$ from
eq.~(\ref{divv}).
Before we can estimate $\dot v_i^{AV}\equiv|-\sum_j m_j \Pi_{ij}{\bf
\nabla}_i W_{ij}|$ we must note that the summation $-\sum_j m_j
\Pi_{ij} {\bf \nabla}_i W_{ij}$ appearing in equation~(\ref{fsph})
scales like the number of terms $N_N$ in the summation ({\it not}
$N_N^{1/2}$ as with the classical AV): the condition $({\bf
\nabla}\cdot{\bf v})_i<0$ in equation
(\ref{q}) requires that the vectors ${\bf \nabla}_iW_{ij}$ for which
$\Pi_{ij}\ne 0$ are found preferentially in the direction of particle
$i$'s velocity deviation from the local fluid flow. 
Therefore, $\dot v^{AV}\approx
k_1'\alpha c_s\,\Delta v/(h N_N^{1/2})+k_2' \beta \Delta v^2/(h N_N)$, and
the timescale satisfies
\begin{equation}
\tau\equiv {v\over\dot v^{AV}} \approx
v\left(k_1'{\alpha c_s\,\Delta v \over h N_N^{1/2}}+k_2' {\beta \Delta v^2 \over h N_N}\right)^{-1}
=\left(k_1'{\alpha\lambda c_s\over \varpi N_N^{1/2}}+
		k_2'{\beta\lambda\,\Delta v\over
		\varpi N_N}\right)^{-1},~~~~~\hbox{(HK AV})
			 \label{tau2}
\end{equation}
where $j_1'$, $j_2'$, $k_1'$ and $k_2'$ are coefficients of order
unity.

Comparing
equations~(\ref{tau}) and (\ref{tau2}) we see that the timescale due to
the bulk viscosity is of the same order of magnitude for the classical and HK
artificial viscosities; however, the timescale associated with the von
Neumann-Richtmyer term is longer in the HK AV by
a factor $N_N^{1/2}$.  Since typical 3D calculations have
$N_N\sim 50--100$, the increase in the viscous dissipation
timescale is substantial whenever von Neumann-Richtmyer viscosity
terms are significant.

If the AV is given by Balsara's form
[eq.~(\ref{piDB})], we need to estimate the size of $f_i$
[eq.~(\ref{fi})] before we can estimate $\Pi_{ij}$.  For our assumed
velocity field $|{\bf \nabla}\times{\bf v}|=(2-\lambda)\Omega$.
Therefore, provided that $\lambda$ is far enough from 2 that the curl
of the velocity dominates over the other terms in the denominator on
the right hand side of equation (\ref{fi}), an SPH evaluation of $f_i$
gives
\begin{equation}
f_i\approx {|{\bf \nabla}\cdot{\bf v}|_i\over |{\bf \nabla}\times{\bf v}|_i} \sim {\lambda \over N_N^{1/2} (2-\lambda)}\equiv f.
\end{equation}
Recalling that $(p_i/\rho_i^2+p_j/\rho_j^2)\approx
2c_s^2/(\gamma\rho)$, we estimate from equation (\ref{piDB}) that
\begin{equation}
\Pi_{ij} \approx -j_1^{\prime\prime}{\alpha \Delta v c_s \over \rho}\left({2\over \gamma}f\right) - j_2^{\prime\prime}{\beta \Delta
v^2\over \rho} \left({2\over \gamma} f^2\right), ~~~~~\hbox{(Balsara AV)}
\end{equation}
where $j_1^{\prime\prime}$ and $j_2^{\prime\prime}$ are coefficients of order unity.
Therefore, $\dot v^{AV}\approx
2k_1^{\prime\prime}\alpha c_s\,\Delta v f/(\gamma h N_N^{1/2})+2k_2^{\prime\prime} \beta \Delta v^2 f^2/(\gamma h
N_N^{1/2})$, and the viscous timescale is given
by
\begin{eqnarray}
\tau\equiv {v\over\dot v^{AV}}
&\approx &v \left[k_1^{\prime\prime}{\alpha\Delta v c_s\over h N_N^{1/2}} \left({2\over \gamma}f\right)+
		k_2^{\prime\prime}{\beta\,\Delta v^2\over h
		N_N^{1/2}} \left({2\over \gamma}f^2\right)\right]^{-1} \nonumber \\
&\approx&\left[k_1^{\prime\prime}{\alpha\lambda^2 c_s\over \varpi N_N (2-\lambda)}{2\over \gamma}+
		k_2^{\prime\prime}{\beta\lambda^3\,\Delta v\over \varpi
		N_N^{3/2} (2-\lambda)^2}{2\over \gamma}\right]^{-1},~~~~~\hbox{(Balsara AV)}
			 \label{tauDB}
\end{eqnarray}
where $k_1^{\prime\prime}$ and $k_2^{\prime\prime}$ are also
coefficients of order unity. 

To test these simple analytic estimates we computed
$\tau_i=v_i/|-\sum_j m_j \Pi_{ij}{\bf \nabla}_i W_{ij}|$ for a rapidly
and differentially rotating equilibrium configuration.  This
configuration was constructed in three steps: (1) we created an $n=3,
\Gamma_1=5/3$ polytrope (pressure profile $p=A \rho^{5/3}\propto
\rho^{4/3}$, and consequently $A\propto\rho^{-1/3}$) of radius $R$ and
mass $M$, (2) assigned a velocity
$v_0=0.5$ (in units where $G=M=R=1$) in the azimuthal direction $\hat
{\bf \phi}$ to all particles, and (3) relaxed to a rotating equilibrium
state by means of an artificial ``drag'' force $\propto v_0 \hat {\bf
\phi}-{\bf v}_i$ on the particles.  The resulting rapidly rotating
configuration ($T/|W|\approx 0.11$) is in virial equilibrium with a
rotation profile close to $\Omega \propto \varpi^{-1}$.  At small
$\varpi$, when the particle smoothing kernels overlap with the rotation
axis, the finite resolution of the SPH scheme cause deviations from the
$\Omega \propto \varpi^{-1}$, cutting off the divergence of $\Omega$ at
$\varpi=0$.  The centrifugal force near $\varpi=0$ nevertheless is
strong enough to make the density a local minimum there; in the
equatorial plane the maximum density actually occurs at $\varpi\approx
0.14$.

For such a configuration modeled using $N=10^4$ and $N_N\approx 64$,
Figure~\ref{tsall} compares the actual timescale $\tau_i=v_i/|-\sum_j
m_j \Pi_{ij}{\bf \nabla}_i W_{ij}|$ computed directly from the SPH code
(left frame) against our analytic estimates (right frame):
(a) classical AV with $\alpha=1$, $\beta=0$,
(b) classical AV with $\alpha=0$, $\beta=1.5$,
(c) HK AV with $\alpha=0.5$, $\beta=0$,
(d) HK AV with $\alpha=0$, $\beta=0.5$,
(e) Balsara AV with $\alpha=\gamma/2$, $\beta=0$, and
(f) Balsara AV with $\alpha=0$, $\beta=1.5\times \gamma/2$.
For all six cases, the same set of particles are analyzed, with the only
difference being the way $\dot v_i^{AV}$ is calculated.  It is
clear that our analytic estimates do a good job of reproducing the
overall trend in position and magnitude of the actual timescale $\tau$.
The estimates for cases (a) and (c) are identical, while the average
measured timescale in case (a) is slightly less than that of case (c),
which implies $k_1'<k_1$.  For each of the AV forms, the timescale due to
the bulk viscosity is significantly less than that due to the von
Neumann-Richtmyer viscosity.

Our analytic estimates of $\Pi_{ij}$ and the viscous dissipation
timescale $\tau$ have neglected the effects of additional velocity
contributions due to particle noise.  For this reason, the numerical
coefficients in equations (\ref{tau}), (\ref{tau2}), and (\ref{tauDB})
are not strictly constant but instead have some complicated dependence
on the neighbor number $N_N$ and noise level in the system.
Consequently when the particle noise is comparable to the shear
velocity, our estimates tend to over estimate the timescale.
Figure~\ref{tsall_noisy} shows the timescales in 6 different
calculations which have evolved freely for 1 time unit from the relaxed
particle state of Figure~\ref{tsall}.  During this evolution, the
particle noise level grows large enough to make our analytic formulae
overestimate the timescale for cases (d), (e) and (f) by a factor of
$\sim 2$.  Furthermore, while both the HK and Balsara AVs continue to
have significantly longer timescales than the classical AV, the
timescale for the Balsara AV is now only slightly larger than for the
HK AV.

Figure~\ref{oall} shows the evolution of the angular momentum profile
$\Omega$ in seven different calculations which began with the same
initial conditions but implemented the different artificial viscosities:
equations~(\ref{pi}), (\ref{pi2}) and (\ref{piDB}).  The Balsara
AV best preserves the angular velocity profile.

One might worry that the spurious increase in the internal energy $u$
or entropy variable $A$ due to shear might also occur on as short a
timescale as the viscous dissipation.  However, dimensional analysis on
equation~(\ref{udot}) and (\ref{adot}) shows that the spurious increase
in $u$ and $A$ occurs on a timescale $\sim \tau
c_s^2/(\gamma(\gamma-1)v\,\Delta v)$.  In typical systems $v\Delta
v<<c_s^2$, so that the timescale for $u$ or $A$ to change is
considerably longer than the viscous dissipation timescale $\tau$.
Figure \ref{svstshear} shows the entropy $S$ as a function of time $t$
for various types of AV.  Although AVs with more
shear viscosity naturally produce more spurious increase in entropy, in
all cases the rate of entropy increase is rather small.

\section{Discussion and Summary}
\subsection{Particle Diffusion}
 
Many of our tests focus on spurious diffusion, the motion of SPH
particles introduced as an artifact of the numerical scheme.  Often
applications require a careful tracing of the particle positions, and
in these cases it is essential that spurious diffusion be small.  For
example, SPH calculations can be used to establish the degree of fluid
mixing during stellar collisions, which is of primary importance in
determining the subsequent stellar evolution of the merger remnants
(e.g.\ [\cite{Sills-etal1997}]).  It must be stressed that the amount of
mixing determined by SPH calculations is always an upper limit.  In
particular, low-resolution calculations tend to be noisy, and this
noise can lead to spurious diffusion of particles, independent of any
real physical mixing of fluid elements.

We have analyzed spurious diffusion by using SPH particles in a box
with periodic boundary conditions to model a stationary fluid of
infinite extent.  For various noise levels (particle velocity
dispersions) and neighbor numbers $N_N$, we measure the rate of
diffusion, which we quantify by calculating a diffusion coefficient
$D$.  Although strong shocks and AV in SPH calculations can lead to
additional particle mixing [\cite{Monaghan1989}], particle diffusion is the
dominant contribution to spurious mixing in weakly shocked fluids.

Once expressed in terms of the number density of SPH particles and the
sound speed, these diffusion coefficients can therefore be used to
estimate spurious deviations in particle positions in a wide variety of
applications, including self-gravitating systems.  For each particle in
some large-scale simulation, this spurious deviation is estimated
simply from equation (\ref{predicted.delta2}).  The coefficient $D$ in
the integrand of equation (\ref{predicted.delta2}) depends on the
particle's velocity deviation from the local flow, the local number
density $n$ of particles, and the local sound speed $c_s$, so that
these quantities need to be monitored for each particle during the
calculation.  Such a scheme is used in \S 4 to estimate
spurious mixing in an equilibrium polytrope, and has also been
successfully applied in the context of stellar collisions [\cite{Lombardi-Rasio-Shapiro1996}].

For sufficiently low noise levels, the diffusion coefficient
essentially vanishes, as the particles simply oscillate around
equilibrium lattice sites.  We say that such a system has
``crystallized.'' For a neighbor number $N_N\approx 64$, a system of
SPH particles will crystallize if the root mean square velocity
dispersion is less than about 3--4\% of the sound speed.  We find that
crystallized cubic lattices are unstable against perturbations, while
lattice types with large packing fractions, such as hexagonal
close-packed, are stable.  For this reason it may sometimes be better
to construct initial data by placing particles in a close-packed
lattice, rather than in a cubic lattice as is often done.  In practice,
initial particle data are typically constructed by first relaxing the
system with an artificial drag force, a procedure which automatically
produces a stable lattice structure but also spuriously removes small
amounts of internal energy.

The diffusion coefficients have been measured using equal mass
particles.  Sometimes, however, SPH calculations use particles of
unequal mass so that less dense regions can still be highly resolved.
To test the effects of unequal mass particles in a self-gravitating
system, we constructed an equilibrium $n=1.5$ polytrope, using particle
masses which increased with radius in the initial configuration.
Allowing the system to evolve, we observed that the heaviest particles
gradually migrated towards the center of the star, exchanging places
with less massive particles.  For a polytrope modeled with 
$N\approx 1.4\times 10^4$ particles and a neighbor number $N_N\approx 64$, 
the distribution of particle masses is reversed within roughly 
80 dynamical timescales. This is caused by the
interactions among neighboring particles via the smoothing kernel.
These interactions allow energy exchange, and equipartition of energy
then requires the heavier particles to sink into the gravitational
potential well.  Spurious mixing is therefore a more complicated
process in calculations which use unequal mass particles: each particle
has a preferred direction to migrate, and in a dynamical application
this direction can be continually changing.  For simulations in which
fluid mixing is important, equal mass particles are an appropriate
choice.

\subsection{Shock Tube Tests}

The diffusion tests just described are all done in the absence of
shocks and without AV.  To test the AV schemes described in
\S 2, we turn to a periodic version of the 1-D Riemann shock-tube
problem of Sod [\cite{Sod1978}].  Initially, fluid slabs with constant (and
alternating) density $\rho$ and pressure $p$ are separated by an
infinite number of planar, parallel, and equally spaced interfaces.  We
treat this inherently 1-D problem with both a 1-D and a 3-D SPH code.
The 1-D code is naturally more accurate, and provides a benchmark
against which we can compare the results of our 3-D code.
In both cases, periodic boundary conditions allow us to model 
the infinite number of slabs.

Using various values of $\alpha$ and $\beta$, we performed a number of
such shock-tube calculations with our 3-D code, at both Mach numbers
${\cal M}\approx1.6$ and ${\cal M}\approx 13.2$ for $\gamma=5/3$.  In
addition, we performed tests with $\gamma=3$ and ${\cal M}\approx
13.2$. For each 3-D calculation, we compare the time variation of the
internal energy and
entropy of the system against that of the 1-D calculation. Furthermore,
since any motion perpendicular to the bulk fluid flow is spurious, we
were also able to examine spurious mixing.  We
find that all three forms of AV can handle shocks well.  For example,
with $N=10^4$ and $N_N\approx 64$, there is better than 2\% agreement
with the 1-D code's internal energy vs.~time curve when ${\cal
M}\approx 1.6$, and agreement at about the 3\% level when ${\cal
M}\approx 13.2$.  We also find that both equations (\ref{pi}) and
(\ref{piDB}), as compared to equation (\ref{pi2}), allow less spurious
mixing and do somewhat better at reproducing the 1-D code's results.
For all three forms of AV, increasing the strength of the AV allows for
less spurious diffusion.

From Tables 1--4, which present results for numerous shock-tube tests,
we see that the level at which energy conservation is satisfied depends
only weakly on the AV parameters but strongly on the length of the
timesteps.  Energy is typically conserved to better than 0.1\% in the
$\gamma=5/3$ 3D calculations whenever the number of timesteps to reach
$t=4$ exceeded 1000.  Monaghan's timestep routine is more efficient, in
part because it takes shorter timesteps when shocks are strong (that
is, when there are large velocity differentials between neighboring
particles).  The agreement between the 3D and 1D calculations for the
internal
energy $U$ and entropy $S$ was strongly dependent on the AV parameters
$\alpha$ and $\beta$ (see \S 7.4), but only
weakly dependent on the Courant number $C_N$ or timestep routine.

Such calculations are a useful and realistic way to calibrate spurious
transport, since the test problem, which includes shocks and
significant fluid motion, has many of the same properties as real
astrophysical problems.  In fact, the recoil shocks in stellar
collisions do tend to be nearly planar, so that even the 1-D geometry
of the shock fronts is realistic.  The periodic boundary conditions
play the role of gravity in the sense that they prevent the gas from
expanding to infinity.

\subsection{Shear Flows}
 
To test the various AV forms in the presence of a shear flow, we impose
the so-called slipping boundary conditions on a periodic box, as is
commonly done in molecular dynamics (see, e.g., [\cite{Naitoh-Ono1976}]).
The resulting ``stationary Couette flow'' has a velocity field close to
$(v_x,v_y,v_z)=(v_0y/L,0,0)$ and allows us to measure the numerical
viscosity of the particles.  As in the shock-tube tests, we also
examine spurious mixing in the direction perpendicular to the fluid
flow.  These shear tests therefore allow us to further investigate the
accuracy of our SPH code as a function of the AV parameters and
scheme.  We find that both the Hernquist \& Katz AV [eq.~(\ref{pi2})]
and the Balsara AV [eq.~(\ref{piDB})] exhibit less viscosity than the
classical AV [eq.~(\ref{pi})].  While the HK AV produces the smallest
numerical viscosity for these pure shear flows, it also has the largest
spurious diffusion coefficient (see Table IV).  The product $\eta D$ is
smallest for the HK AV, indicating that this form is well suited for
keeping spurious mixing at a manageable level during calculations
involving shear flows.  For all three forms of the AV, increasing
$\alpha$ and $\beta$ tends to damp out the noise and consequently
decrease spurious mixing, but it also increases the spurious shear
viscosity.

Rotation plays an important role in many hydrodynamic processes.  For
instance, a collision between stars can yield a rapidly and
differentially rotating merger remnant.  Even in the absence of shocks,
AV tends to damp away differential rotation due to the relative
velocity of neighboring particles at slightly different radii, and an
initially differentially rotating system will tend towards rigid
rotation on the viscous dissipation timescale. In systems best modeled
with a perfect fluid, ideally with a viscous timescale $\tau=\infty$,
any such angular momentum transport introduced by the SPH scheme is
spurious.

As a concrete example, we consider an axisymmetric equilibrium
configuration differentially rotating with an angular velocity profile
$\Omega(\varpi) \propto \varpi^{-\lambda}$, where $\varpi$ is the
distance from the rotation axis and $\lambda$ is a constant of order
unity.  We then analytically estimate the viscous dissipation timescale
for each of the three AVs discussed in \S 2.  These analytic estimates
are found to closely match numerically measured values of the
timescale.  Both the Hernquist \& Katz AV [eq.~(\ref{pi2})] and the
Balsara AV [eq.~(\ref{piDB})] yield longer viscous timescales than the
classical AV [eq.~(\ref{pi})], and hence are better at maintaining the
angular velocity profile.  The Balsara AV does best in this regard.

\subsection{Artificial Viscosity Forms and Parameters}
\label{subsection:artificial}

When choosing values of AV parameters, one must weigh the relative
importance of shocks, shear, and fluid mixing.  For this reason, it is
an application-dependent, somewhat subjective matter to specify
``optimal values'' of $\alpha$ and $\beta$.  Here, however, we roughly
delineate the boundaries of the region in parameter space that gives
acceptable results.

Our shock-tube tests of \S 5 are all done with periodic cells each
containing mass $M=0.625$.  We find that the quantity
$(\Delta(U/E)_{max})^2+((\gamma-1)\Delta S_{max}/M)^2$ is a convenient
measure of how well a calculation matches the 1-D code's results for
both internal energy and entropy (note that $(\gamma-1)\Delta S_{max}/M
\sim \Delta A_{max}/A$ for small $\Delta S_{max}$).  Values of
$\Delta(U/E)_{max}$ and $\Delta S_{max}$ are listed in Tables 1 through
4.

Examination of the final three columns in Table I leads us to the
following acceptable ranges for $\alpha$ in our $\gamma=5/3$ low Mach
number shock-tube tests: $0.2\lesssim\alpha\lesssim 1$ for the
classical AV, $0.1\lesssim\alpha\lesssim 0.5$ for the HK AV, and
$0.2\lesssim2\alpha/\gamma\lesssim 1$ for the Balsara AV.  If spurious
diffusion is not a concern, these ranges for $\alpha$ can all be
extended down to a lower limit of $\alpha=0$.  For a given value of
$\alpha$, the acceptable range of $\beta$ is approximately given by
$0.8 \lesssim 2\alpha+\beta\lesssim 3.3$ for the classical AV, and $0.6
\lesssim 2\alpha+\beta \lesssim 1.3$ for the HK AV, and $0.8 \lesssim
(2\alpha+\beta)2/\gamma\lesssim 3.3$ for the Balsara AV.  For
parameters in these ranges, all three AVs handle the low Mach number
shocks with roughly the same level of accuracy.  When Monaghan's
timestep routine is used with $C_N=0.3$, values of $\alpha$ and $\beta$
which worked particularly well in our low Mach calculations included
$\alpha=0.2$, $\beta=1$ for the classical AV, $\alpha=0.3$, $\beta=0.5$
for the HK AV, and $\alpha=0.5\times\gamma/2$, $\beta=\gamma/2$ for the
Balsara AV.

For our high Mach number tests, inspection of Tables 3 and 4 leads to
the following acceptable ranges for the AV parameters:
$1.3\lesssim\alpha+\beta\lesssim 3.5$ for the classical AV, $1
\lesssim\alpha+\beta\lesssim 1.6$ for the HK AV, and
$1.9\lesssim(\alpha+\beta)2/\gamma\lesssim 4$ for the Balsara AV.  The
Balsara AV seems capable of handling these high Mach number shocks
marginally better than the classical AV, and both are more accurate
than the HK AV.  Values of $\alpha$ and $\beta$ which worked
particularly well in both of our $\gamma=5/3$ and $\gamma=3$ high Mach
calculations included $\alpha=1$, $\beta=1.5$ for the classical AV, and
$\alpha=2\times\gamma/2$, $\beta=\gamma/2$ for the Balsara AV.  With
the HK AV, $\alpha=0.5$, $\beta=1$ worked quite well for $\gamma=5/3$,
as did $\alpha=0.5$, $\beta=0.5$ for $\gamma=3$.  By performing these
high Mach calculations for two different values of $\gamma$, we have
determined that the ranges of acceptable AV parameters are only weakly
dependent on the equation of state for both the classical AV and the HK
AV.  For the Balsara AV, we find that the AV parameters should be
scaled with $\gamma$, so that softer equations of state require larger
AV parameters.

Our shear tests of \S 6 allow us to further examine the accuracy of our
SPH code as a function of the AV parameters.  Not surprisingly,
increasing the strength of the AV tends to increase the measured
viscosity $\eta$ and decrease the measured spurious diffusion
coefficient $D$.  The product of the viscosity and the diffusion
coefficient provides a convenient (but somewhat arbitrary) measure of a
calculation's accuracy.  We find that increasing $\alpha$ typically
tends to increase the product $\eta D$ in our shear tests, and we
consequently choose as the ``optimal'' value of $\alpha$ a relatively
small value for which the shock-tube tests (both low and high Mach
number) give acceptable results.

The combined results of our shock-tube and shear tests therefore
suggest a single set of AV parameters which are appropriate in a large
number of situations:  $\alpha\approx 0.5$, $\beta\approx 1$ for the
classical AV, $\alpha\approx\beta\approx0.5$ for the Hernquist \& Katz
AV, and $\alpha\approx\beta\approx \gamma/2$ for the Balsara AV.  We
will refer to these parameters as ``optimal''; however, these choices
should be modified depending on the particular application.  For
instance, if spurious mixing is not a concern and if only weak shocks
(${\cal M}\lo 2$) are expected during a calculation, then a smaller
value of $\alpha$ is appropriate.  Likewise, if strong shocks are
expected (${\cal M}\go$ a few) and shear viscosity is not a concern,
then a stronger AV is justified.

The above recommended values for $\alpha$ and $\beta$ correspond to a
somewhat weaker AV than is typically suggested in the literature (e.g.
$\alpha\approx 1$, $\beta\approx 2$ for the classical AV).  While
larger AV parameters are appropriate in extreme cases (${\cal M}\go
10$), we feel our recommended values are slightly more accurate for
most situations. Furthermore, since errors do not change significantly
when the energy equation is integrated instead of the entropy equation
(the only major difference being a larger $\Delta S_{max}$ for the
energy equation, by a roughly constant amount, see Table II), we
conclude that these ``optimal'' parameters are insensitive to the means
by which the thermodynamics is treated.  However, we have not tested
the dependence of the optimal AV parameters on the neighbor number
$N_N$ in detail, nor have we performed test calculations in which both
shear flows and shocks are {\it simultaneously} occurring.

Morris \& Monaghan [\cite{Morris-Monaghan1997}] have recently tested the classical AV of
equation (\ref{mu}) with a {\it time-varying} viscosity parameter
$\alpha$, and with $\beta=2\alpha$.  The evolution of $\alpha$ is
determined for each particle by a source and decay equation, causing
the AV to be significantly active only in the immediate vicinity of a
shock.  The results of their tests are encouraging, and their idea of
time-varying AV coefficients could be applied to any AV form.

Our results concerning the various AV forms can be summarized as follows.
We find that the AVs defined by equations (\ref{pi}) and (\ref{piDB})
do equally well both in their handling of shocks and in their
controlling of spurious mixing, and do slightly better than equation
(\ref{pi2}).  Furthermore, both equations (\ref{pi2}) and (\ref{piDB}) do
introduce less numerical viscosity than equation (\ref{pi}).
Since use of equation (\ref{piDB}), Balsara's form of AV, does indeed
significantly decrease the amount of shear viscosity without
sacrificing accuracy in the treatment of shocks, we conclude that it is
an appropriate choice for a broad range of problems.  This is
consistent with the successful use of Balsara's AV reported by Navarro
\& Steinmetz [\cite{Navarro-Steinmetz1997}] in their models of rotating
galaxies.

Balsara's viscosity was constructed to be quite similar to the
classical AV in form; the main difference is that Balsara's form
contains a ``switch'' which suppresses the AV in regions of large
vorticity.  It is a simple matter to generate more sensitive switches
than the one in equation (\ref{piDB}).  For instance, instead of
$(f_i+f_j)/2$ one could use $f_if_j$ [or more generally $(f_if_j)^k$,
with $k\go 1$).  Alternatively, in place of the form function $f_i$
defined by equation (\ref{fi}), one could use
\begin{equation}
g_i={({\bf \nabla}\cdot {\bf v})_i^2 \over ({\bf \nabla}\cdot {\bf v})_i^2
+({\bf \nabla}\times {\bf v})_i^2 + \eta' c_i^2/h_i^2
}. \label{gi}
\end{equation}
As expected from scaling analyses such as in \S 6.2, the viscous
dissipation timescale can be increased by adopting more sensitive
switches such as these.  However, such switches also tend to allow a
faster rate of spurious particle diffusion.  We have performed a
handful of tests which suggest that such generalizations of Balsara's
AV may also handle shocks well, although more tests are necessary.

\acknowledgments 

J.C.L.\ is supported in part by NSF AST 93-15375 and by a New York
Space Grant Fellowship.  A.S.\ is supported in part by the Natural
Sciences and Engineering Research Council of Canada.  F.A.R.\ is
supported in part by NSF Grant AST-9618116 and by a Sloan Research
Fellowship.  S.L.S.\ is supported in part by NSF Grant AST 96-18524 and
NASA Grant NAG5-7152 at the University of Illinois at
Urbana-Champaign.  Some computations were performed at the Cornell
Theory Center.  This work was also partially supported by the National
Computational Science Alliance under Grant AST970022N and utilized the
NCSA SGI/CRAY POWER CHALLENGEarray and the NCSA SGI/CRAY Origin2000.

\begin{table}\scriptsize

\centering

{\sc Table I: Low Mach Number Shock-Tube Cases with $\gamma=5/3$}
\vspace{.04in}
\ 

\begin{tabular}{crrccrrcccl} \hline\hline\vspace{-0.13in}
\ \\
\vspace{0.04in}
  &&&     &  & steps  & steps  & $\Delta E/E $ & $\delta_y^2+\delta_z^2$ & &\\
AV&&& $dt$&  &  to    &  to    &  at           &  $[n^{-2/3}]$           & &\\
routine & $\alpha$ & $\beta$ & routine & $C_N$ & $t=1$ & $t=4$ & $t=4$ & at $t=4$ & $\Delta(U/E)_{max}$ & $ \Delta S_{max}$  \\ \hline
\vspace{-0.13in}\\
None&   &     & S & 0.1 & 436&1664& 0.06\% &115.7&0.111&0.14\\ 
 C & 0 & 0.1 & S & 0.1 & 413&1402& 0.04\% &25.25&0.044&0.051\\ %
 C & 0 & 1   & S & 0.8 &  38& 140& 4.56\% & 3.16&0.014&0.025\\ 
 C & 0 & 2.5 & S & 0.1 & 295&1121& 0.04\% & 1.92&0.014&0.018\\ %
 C & 0 & 10  & S & 0.1 & 281&1078& 0.04\% & 0.88&0.030&0.030\\ %
 C & 0 & 100 & S & 0.1 & 307&1072& 0.05\% & 0.40&0.064&0.064\\ %
 C &0.1& 1   & M & 0.3 & 163& 572& 0.13\% & 2.51&0.012&0.019\\ 
 C &0.2&0.5  & M & 0.3 & 145& 523& 0.25\% & 2.43&0.013&0.019\\ 
 C &0.2& 1   & M & 0.3 & 167& 585& 0.11\% & 1.81&0.010&0.020\\ 
 C &0.2& 1   & M & 0.8 &  63& 218& 1.31\% & 1.81&0.013&0.011\\ %
 C &0.2& 1.25& M & 0.3 & 175& 612& 0.07\% & 1.72&0.011&0.020\\ 
 C &0.3& 1   & M & 0.3 & 170& 604& 0.09\% & 1.54&0.012&0.020\\ 
 C &0.5& 1   & M & 0.3 & 180& 653& 0.08\% & 1.10&0.017&0.019\\ 
 C &0.5& 1   & M & 0.8 &  68& 245& 0.78\% & 1.09&0.016&0.015\\ 
 C & 1 & 0   & S & 0.8 &  36& 134& 1.41\% & 0.78&0.021&0.018\\ %
 C & 1 & 1   & S & 0.8 &  39& 164& 1.25\% & 0.76&0.025&0.020\\ %
 C & 1 & 1   & M & 0.8 &  81& 307& 0.41\% & 0.74&0.025&0.023\\ %
 C & 1 & 1.25& M & 0.3 & 221& 832& 0.03\% & 0.76&0.026&0.025\\ 

 C & 1 & 2   & S & 0.8 &  42& 171& 0.92\% & 0.72&0.028&0.025\\ 
 C & 2 & 0   & S & 0.1 & 278&1063& 0.02\% & 0.51&0.035&0.034\\ %
 C & 2 & 1   & S & 0.8 &  56& 231& 0.72\% & 0.52&0.040&0.049\\ %
 C & 3 & 0   & S & 0.1 & 275&1053& 0.01\% & 0.41&0.043&0.042\\ %
 C & 3 & 1   & S & 0.8 &  79& 329& 2.18\% & 0.40&0.047&0.066\\ %
 C &10 & 0   & S & 0.1 & 265&1035& 0.24\% & 0.11&0.071&0.068\\ %

HK & 0 & 1.25& M & 0.3 & 116& 449& 0.28\% & 7.40&0.016&0.015\\ 
HK &0.1& 0.5 & M & 0.3 & 111& 440& 0.40\% & 8.97&0.018&0.016\\ 
HK &0.1& 0.5 & M & 0.8 &  42& 161& 2.79\% & 6.95&0.014&0.025\\ 
HK &0.1& 1   & M & 0.8 &  45& 171& 2.00\% & 4.64&0.018&0.014\\ %
HK &0.1& 2   & M & 0.8 &  52& 191& 0.98\% & 3.19&0.025&0.025\\ 
HK &0.2& 0.5 & M & 0.3 & 117& 463& 0.31\% & 4.63&0.013&0.012\\ 
HK &0.2& 0.75& M & 0.3 & 119& 467& 0.24\% & 3.95&0.016&0.016\\ 
HK &0.3& 0.5 & M & 0.3 & 125& 493& 0.22\% & 3.05&0.016&0.017\\ 
HK &0.4& 0.5 & M & 0.3 & 135& 534& 0.15\% & 2.45&0.020&0.022\\ 
HK &0.5& 0.5 & M & 0.3 & 145& 572& 0.11\% & 1.97&0.025&0.027\\ 
HK &0.5& 1   & M & 0.3 & 148& 579& 0.06\% & 1.78&0.029&0.032\\ 
HK &0.5& 1   & M & 0.8 &  56& 218& 0.43\% & 1.80&0.030&0.031\\ 
HK & 1 & 0   & M & 0.3 & 196& 768& 0.01\% & 1.21&0.039&0.040\\ 
HK & 1 & 1   & M & 0.3 & 198& 772& 0.02\% & 1.13&0.044&0.045\\ 
HK & 1 & 1   & M & 0.8 &  75& 291& 0.32\% & 1.16&0.044&0.046\\ %

HK & 1 & 1   & S & 0.8 &  39& 151& 4.70\% & 1.36&0.043&0.080\\ %

 B & 0                 & 2.5$\times\gamma/2$ & M & 0.3 & 192& 687& 0.05\% & 5.11&0.012&0.011\\ 
 B &0.2$\times\gamma/2$& 0.5$\times\gamma/2$ & M & 0.3 & 144& 534& 0.28\% & 6.86&0.021&0.018\\ 
 B &0.5$\times\gamma/2$& 1$\times\gamma/2$   & M & 0.3 & 173& 637& 0.12\% & 1.98&0.010&0.019\\ 

 B & 1$\times\gamma/2$ & 0.75$\times\gamma/2$& M & 0.3 & 206& 800& 0.09\% & 1.14&0.018&0.019\\ 
 B & 1$\times\gamma/2$ & 1$\times\gamma/2$   & M & 0.3 & 211& 811& 0.07\% & 1.13&0.019&0.020\\ 
 B & 1$\times\gamma/2$ & 1$\times\gamma/2$   & M & 0.8 &  79& 304& 0.54\% & 1.08&0.020&0.018\\ 
 B & 1$\times\gamma/2$ & 1.25$\times\gamma/2$& M & 0.3 & 216& 819& 0.05\% & 1.12&0.020&0.020\\ 
 B & 1$\times\gamma/2$ & 2$\times\gamma/2$   & M & 0.3 & 305&1195& 0.05\% & 0.74&0.031&0.031\\ 
 B & 2$\times\gamma/2$ & 0                   & M & 0.3 & 233& 855& 0.02\% & 1.07&0.022&0.023\\ 
 B & 2$\times\gamma/2$ & 1$\times\gamma/2$   & M & 0.3 & 309&1212& 0.04\% & 0.70&0.032&0.031\\ 
 B & 2$\times\gamma/2$ & 1.25$\times\gamma/2$& M & 0.3 & 311&1213& 0.03\% & 0.71&0.032&0.032\\ 

\end{tabular}
\end{table}

\begin{table}\scriptsize
\centering

{\sc Table II: Low Mach Number Shock-Tube Cases with $\gamma=5/3$\\
(Classical AV, Simple timestep routine, $C_N=0.1$)}
\vspace{.04in}
\ 

\begin{tabular}{rclccccccl} \hline\hline\vspace{-0.13in}
\ \\
\vspace{0.04in}
          &         &          &            & steps  & steps  & $\Delta E/E $ & $\delta_y^2+\delta_z^2$ & &\\
          &         &          &evolution   &  to    &  to    &  at           &  $[n^{-2/3}]$           & &\\
 $\alpha$ & $\beta$ & $\eta^2$ &equation    & $t=1$ & $t=4$ & $t=4$ & at $t=4$ & $\Delta(U/E)_{max}$ & $ \Delta S_{max}$  \\ \hline
\vspace{-0.13in}\\
 0 & 1   & 0.01 &entropy  & 313&1172& 0.04\% & 3.9&0.015&0.016\\ 
 0 & 1   & 0.01 &energy   & 335&1295& 0.04\% & 4.1&0.016&0.021\\ 

 1 & 0   & 0.01 &entropy  & 285&1082& 0.01\% & 0.8&0.021&0.023\\ 
 1 & 0   & 0.01 &energy   & 309&1222& 0.01\% & 0.8&0.022&0.030\\ 

 1 & 1   & 0.002&entropy  & 283&1076& 0.01\% & 0.7&0.025&0.025\\ 
 1 & 1   & 0.002&energy   & 306&1215& 0.01\% & 0.8&0.024&0.032\\ 

 1 & 1   & 0.01 &entropy & 283&1076& 0.01\% & 0.8&0.025&0.025\\ 
 1 & 1   & 0.01 &energy  & 306&1210& 0.01\% & 0.8&0.024&0.032\\ 

 1 & 1   & 0.05 &entropy & 283&1079& 0.01\% & 0.8&0.024&0.025\\ 
 1 & 1   & 0.05 &energy  & 307&1219& 0.02\% & 0.8&0.024&0.031\\ 

 2 & 1   & 0.01 &entropy & 278&1061& 0.02\% & 0.5&0.036&0.035\\ 
 2 & 1   & 0.01 &energy  & 304&1197& 0.00\% & 0.5&0.035&0.042\\ 

 3 & 1   & 0.01 &entropy & 275&1053& 0.01\% & 0.4&0.044&0.044\\ 
 3 & 1   & 0.01 &energy  & 304&1198& 0.00\% & 0.4&0.044&0.050\\ 

\end{tabular}
\end{table}

\clearpage

\begin{table}\scriptsize
\centering

{\sc Table III: High Mach Number Shock-Tube Cases with $\gamma=5/3$}
\vspace{.04in}
\ 

\begin{tabular}{crrccrrcccl} \hline\hline\vspace{-0.13in}
\ \\
\vspace{0.04in}
  &&&     &  & steps  & steps  & $\Delta E/E $ & $\delta_y^2+\delta_z^2$ & &\\
AV&&& $dt$&  &  to    &  to    &  at           & $[n^{-2/3}]$            & &\\
routine & $\alpha$ & $\beta$ & routine & $C_N$ & $t=1$ & $t=4$ & $t=4$ & at $t=4$ & $\Delta(U/E)_{max}$ & $ \Delta S_{max}$  \\ \hline
\vspace{-0.13in}\\
None&  &     & M & 0.3 &  97& 376& 0.04\% &300.1&0.207&0.85\\ 
 C & 0 & 1   & M & 0.1 & 411&1512& 0.02\% & 6.02&0.028&0.11\\ 
 C & 0 & 1   & M & 0.8 &  83& 247& 1.49\% & 5.46&0.026&0.12\\ 
 C & 0 & 5   & M & 0.8 & 124& 371& 1.82\% & 1.39&0.065&0.14\\ 

 C &0.1& 1   & M & 0.3 & 227& 687& 0.07\% & 3.84&0.024&0.096\\ %
 C &0.2& 0.5 & M & 0.3 & 206& 682& 0.05\% & 4.57&0.027&0.14\\ 
 C &0.2& 1   & M & 0.3 & 231& 715& 0.06\% & 2.85&0.024&0.089\\ %
 C &0.2&1.25 & M & 0.3 & 238& 730& 0.12\% & 2.27&0.027&0.084\\ 
 C &0.3& 1   & M & 0.3 & 233& 746& 0.06\% & 2.13&0.025&0.085\\ 
 C &0.3& 1.25& M & 0.3 & 243& 763& 0.12\% & 1.76&0.028&0.081\\ 
 C &0.5& 1   & M & 0.3 & 245& 827& 0.05\% & 1.38&0.027&0.079\\ 
 C &0.5& 1.25& M & 0.3 & 252& 830& 0.10\% & 1.26&0.031&0.075\\ 
 C &0.5& 2.5 & M & 0.3 & 283& 896& 0.16\% & 1.06&0.046&0.063\\ 
 C &0.7& 1.5 & M & 0.3 & 268& 936& 0.08\% & 0.94&0.037&0.068\\ 

 C & 1 & 0   & M & 0.8 &  97& 386& 0.71\% & 1.16&0.027&0.127\\ %
 C & 1 & 1   & M & 0.8 & 106& 389& 0.27\% & 0.85&0.033&0.076\\ %
 C & 1 & 1.5 & M & 0.3 & 292&1058& 0.05\% & 0.82&0.042&0.062\\ 
 C & 1 & 2   & M & 0.3 & 299&1057& 0.08\% & 0.79&0.048&0.063\\ 
 C & 1 & 2   & M & 0.8 & 112& 397& 0.02\% & 0.80&0.045&0.069\\ %
 C & 2 & 2   & M & 0.8 & 146& 557& 0.04\% & 0.56&0.059&0.089\\ %
 C & 5 & 0   & M & 0.8 & 258&1039& 0.08\% & 0.27&0.077&0.12\\ 

HK & 0 & 1.25& M & 0.3 & 131& 494& 0.28\% & 9.71&0.053&0.072\\ 
HK &0.2& 0.5 & M & 0.3 & 144& 555& 0.33\% &13.41&0.043&0.086\\ 
HK &0.5& 0.5 & M & 0.3 & 186& 727& 0.11\% & 4.04&0.041&0.080\\ 
HK &0.5& 1   & M & 0.3 & 180& 698& 0.08\% & 2.78&0.060&0.066\\ 
HK & 1 & 0   & M & 0.3 & 249& 976& 0.08\% & 2.90&0.029&0.11\\ 
HK & 1 & 0.25& M & 0.3 & 251& 979& 0.04\% & 2.26&0.046&0.082\\ 
HK & 1 & 1   & S & 0.8 &  44& 163& 3.20\% & 1.62&0.069&0.093\\ %
HK & 1 & 1   & M & 0.3 & 238& 941& 0.02\% & 1.57&0.073&0.083\\ 
HK & 1 & 1   & M & 0.8 &  89& 350& 0.08\% & 1.55&0.068&0.077\\ %

 B & 0                 & 2.5$\times\gamma/2$ & M & 0.3 & 279& 834& 0.31\% & 8.25&0.030&0.077\\ 
 B &0.2$\times\gamma/2$& 0.5$\times\gamma/2$ & M & 0.3 & 194& 664& 0.10\% &16.99&0.055&0.19\\ 
 B &0.5$\times\gamma/2$& 0.75$\times\gamma/2$& M & 0.3 & 243& 854& 0.02\% & 5.35&0.029&0.13\\ 
 B &0.5$\times\gamma/2$& 1$\times\gamma/2$   & M & 0.3 & 254& 857& 0.02\% & 4.35&0.025&0.11\\ 

 B & 1$\times\gamma/2$ & 0.75$\times\gamma/2$& M & 0.3 & 293&1076& 0.02\% & 1.88&0.024&0.089\\ 
 B & 1$\times\gamma/2$ & 1$\times\gamma/2$   & M & 0.3 & 300&1106& 0.02\% & 1.57&0.026&0.074\\ 
 B & 1$\times\gamma/2$ & 1$\times\gamma/2$   & M & 0.8 & 112& 413& 0.33\% & 1.62&0.024&0.080\\ %
 B & 1$\times\gamma/2$ & 1.25$\times\gamma/2$& M & 0.3 & 301&1077& 0.03\% & 1.45&0.028&0.068\\ 
 B & 1$\times\gamma/2$ & 1.5 $\times\gamma/2$& M & 0.3 & 306&1080& 0.05\% & 1.40&0.031&0.066\\ 
 B & 1$\times\gamma/2$ & 2$\times\gamma/2$   & M & 0.3 & 316&1100& 0.09\% & 1.29&0.037&0.064\\ 
 B & 2$\times\gamma/2$ & 0                   & M & 0.3 & 403&1617& 0.03\% & 0.91&0.030&0.065\\ 
 B & 2$\times\gamma/2$ & 1$\times\gamma/2$   & M & 0.3 & 405&1577& 0.00\% & 0.79&0.041&0.058\\ 
 B & 2$\times\gamma/2$ & 1.25$\times\gamma/2$& M & 0.3 & 406&1562& 0.01\% & 0.81&0.043&0.063\\ 

\end{tabular}
\end{table}

\begin{table}\scriptsize
\centering

{\sc Table IV: High Mach Number Shock-Tube Cases with $\gamma=3$}
\vspace{.04in}
\ 

\begin{tabular}{crrrrcccl} \hline\hline\vspace{-0.13in}
\ \\
\vspace{0.04in}
  &&& steps  & steps  & $\Delta E/E $ & $\delta_y^2+\delta_z^2$ & &\\
AV&&&  to    &  to    &  at           & $[n^{-2/3}]$            & &\\
routine & $\alpha$ & $\beta$ & $t=1$ & $t=4$ & $t=4$ & at $t=4$ & $\Delta(U/E)_{max}$ & $ \Delta S_{max}$  \\ \hline
\vspace{-0.13in}\\

C  & 0.2  & 0.5  & 238& 849& 1.10\% & 1.90&0.140&0.073\\
C  & 0.28 & 0.56 & 240& 867& 1.01\% & 1.27&0.114&0.065\\
C  & 0.3  & 1.0  & 248& 877& 0.63\% & 1.03&0.081&0.049\\
C  & 0.5  & 1.0  & 261& 938& 0.49\% & 0.79&0.045&0.039\\
C  & 0.5  & 1.25 & 264& 939& 0.37\% & 0.81&0.037&0.034\\
C  & 0.7  & 1.5  & 284&1013& 0.26\% & 0.68&0.040&0.024\\
C  & 0.9  & 1.8  & 307&1106& 0.20\% & 0.59&0.047&0.022\\
C  & 1.0  & 1.5  & 313&1147& 0.24\% & 0.58&0.049&0.021\\
HK & 0.2  & 0.5  & 184& 708& 1.30\% & 3.48&0.076&0.042\\
HK & 0.28 & 0.28 & 223& 880& 0.54\% & 1.28&0.061&0.026\\
HK & 0.5  & 0.5  & 216& 844& 0.62\% & 1.43&0.048&0.027\\
HK & 0.5  & 1.0  & 214& 836& 0.48\% & 1.21&0.077&0.023\\
HK & 0.7  & 0.5  & 243& 955& 0.40\% & 1.15&0.076&0.025\\ 
HK & 0.9  & 0.9  & 269&1063& 0.18\% & 0.83&0.115&0.037\\
B  & 0.5$\times\gamma/2$  & 1.0$\times\gamma/2$  & 271&1014& 0.79\% & 1.35&0.094&0.060\\
B  & 0.56$\times\gamma/2$ & 0.56$\times\gamma/2$ & 286&1100& 0.85\% & 1.41&0.103&0.068\\
B  & 1.0$\times\gamma/2$  & 0.75$\times\gamma/2$ & 329&1269& 0.47\% & 0.83&0.039&0.042\\ 
B  & 1.0$\times\gamma/2$  & 1.0$\times\gamma/2$  & 326&1240& 0.45\% & 0.82&0.031&0.038\\
B  & 1.0$\times\gamma/2$  & 1.25$\times\gamma/2$ & 324&1226& 0.39\% & 0.81&0.033&0.035\\
B  & 1.8$\times\gamma/2$  & 1.8$\times\gamma/2$  & 421&1610& 0.22\% & 0.60&0.066&0.018\\
B  & 2.0$\times\gamma/2$  & 1.0$\times\gamma/2$  & 446&1722& 0.24\% & 0.56&0.066&0.018\\

\end{tabular}
\end{table}

\begin{table}\scriptsize
\centering

{\sc Table V: $N=1000, N_N=64$, $v_0/c_s=0.1\gamma^{-1/2}$, $\gamma=5/3$, $dt=0.01$ Shear tests}
\vspace{.04in}
\ 

\begin{tabular}{ccrcrll} \hline\hline\vspace{-0.13in}
\ \\
\vspace{0.04in}
AV routine & $\alpha$ & $\beta$ & $\langle{v_y^2+v_z^2 \over c_s^2}\rangle^{1/2}$ & $\eta~~[M c_s n^{2/3}]$ & $D~~[c_s n^{-1/3}]$ & $\eta D~~[M c_s^2 n^{1/3}]$\\ \hline
\vspace{-0.13in}\\

 None&   &     & 0.337&	3.0(1)$\times 10^{-4}$ & 2.85(6) & 8.4(4)$\times 10^{-4}$\\
   C &0.0& 1.00& 0.020&	1.332(1)$\times 10^{-3}$ & 4.59(5)$\times 10^{-3}$ & 6.13(7)$\times 10^{-6}$\\
   C &0.0& 2.50& 0.016&	2.763(5)$\times 10^{-3}$ & 3.89(4)$\times 10^{-3}$ & 1.07(1)$\times 10^{-5}$\\
   C &0.0&10.00& 0.012&	8.71(3)$\times 10^{-3}$ & 3.73(7)$\times 10^{-3}$ & 3.25(7)$\times 10^{-5}$\\
   C &0.3& 0.50& 0.013&	5.60(4)$\times 10^{-3}$ & 3.91(5)$\times 10^{-3}$ & 2.19(3)$\times 10^{-5}$\\
   C ($\eta^2=0.002$) &0.3& 1.00& 0.013&	6.16(6)$\times 10^{-3}$ & 3.51(3)$\times 10^{-3}$ & 2.16(3)$\times 10^{-5}$\\
   C                  &0.3& 1.00& 0.013&	6.05(5)$\times 10^{-3}$ & 3.53(7)$\times 10^{-3}$ & 2.14(5)$\times 10^{-5}$\\
   C ($\eta^2=0.05$)  &0.3& 1.00& 0.013&	5.71(4)$\times 10^{-3}$ & 3.95(4)$\times 10^{-3}$ & 2.26(3)$\times 10^{-5}$\\
   C &0.5& 1.00& 0.012&	9.09(10)$\times 10^{-3}$ & 3.53(7)$\times 10^{-3}$ & 3.21(7)$\times 10^{-5}$\\
   C &0.8& 1.25& 0.012&	1.37(3)$\times 10^{-2}$ & 3.78(6)$\times 10^{-3}$ & 5.2(1)$\times 10^{-5}$\\
   C &1.0& 1.00& 0.012&	1.64(4)$\times 10^{-2}$ & 3.68(4)$\times 10^{-3}$ & 6.0(1)$\times 10^{-5}$\\
   C &1.0& 1.25& 0.012&	1.66(3)$\times 10^{-2}$ & 3.44(9)$\times 10^{-3}$ & 5.7(2)$\times 10^{-5}$\\
   C &2.0& 0.00& 0.010&	3.1(1)$\times 10^{-2}$ & 3.7(1)$\times 10^{-3}$ & 1.12(5)$\times 10^{-4}$\\
   C &3.0& 0.00& 0.009&	4.8(2)$\times 10^{-2}$ & 3.57(3)$\times 10^{-3}$ & 1.71(8)$\times 10^{-4}$\\
  HK &0.0& 1.25& 0.082&	1.72(5)$\times 10^{-4}$ & 0.17(4) & 3.0(7)$\times 10^{-5}$\\
  HK &0.0&10.00& 0.038&	5.08(4)$\times 10^{-4}$ & 2.1(5)$\times 10^{-2}$ & 1.1(3)$\times 10^{-5}$\\
  HK &0.1& 0.50& 0.066&	4.15(2)$\times 10^{-4}$ & 0.11(4) & 4.5(17)$\times 10^{-5}$\\
  HK &0.5& 0.50& 0.024&	1.34(1)$\times 10^{-3}$ & 5.4(1)$\times 10^{-3}$ & 7.3(2)$\times 10^{-6}$\\
  HK &0.5& 1.00& 0.025&	1.39(3)$\times 10^{-3}$ & 5.32(15)$\times 10^{-3}$ & 7.4(3)$\times 10^{-6}$\\
  HK &1.0& 1.00& 0.022&	2.64(3)$\times 10^{-3}$ & 5.05(13)$\times 10^{-3}$ & 1.33(4)$\times 10^{-5}$\\
   B &$0.0\times\gamma/2$ & $ 1.00\times\gamma/2$ &0.026&	2.72(1)$\times 10^{-4}$ & 1.16(3)$\times 10^{-2}$ & 3.13(7)$\times 10^{-6}$\\
   B &$0.0\times\gamma/2$ & $ 2.50\times\gamma/2$ &0.023&	4.53(1)$\times 10^{-4}$ & 7.0(1)$\times 10^{-3}$ & 3.20(7)$\times 10^{-6}$\\
   B &$0.0\times\gamma/2$ & $10.00\times\gamma/2$ &0.020&	1.055(3)$\times 10^{-3}$ & 5.8(1)$\times 10^{-3}$ & 6.1(1)$\times 10^{-6}$\\
   B ($\eta^2=0.002$)&$0.5\times\gamma/2$ & $ 0.50\times\gamma/2$ &0.013&	2.33(2)$\times 10^{-3}$ & 4.0(2)$\times 10^{-3}$ & 9.3(5)$\times 10^{-6}$\\
   B &$0.5\times\gamma/2$ & $ 1.00\times\gamma/2$ &0.018&	2.25(1)$\times 10^{-3}$ & 3.42(35)$\times 10^{-3}$ & 7.7(8)$\times 10^{-6}$\\
   B &$1.0\times\gamma/2$ & $ 0.00\times\gamma/2$ &0.015&	4.22(3)$\times 10^{-3}$ & 3.15(17)$\times 10^{-3}$ & 1.33(7)$\times 10^{-5}$\\
   B &$1.0\times\gamma/2$ & $ 1.00\times\gamma/2$ &0.015&	4.33(2)$\times 10^{-3}$ & 3.93(5)$\times 10^{-3}$ & 1.70(3)$\times 10^{-5}$\\
   B &$1.0\times\gamma/2$ & $ 2.00\times\gamma/2$ &0.014&	4.444(7)$\times 10^{-3}$ & 4.15(8)$\times 10^{-3}$ & 1.84(3)$\times 10^{-5}$\\
   B &$2.0\times\gamma/2$ & $ 0.00\times\gamma/2$ &0.013&	8.5(1)$\times 10^{-3}$ & 3.69(5)$\times 10^{-3}$ & 3.13(5)$\times 10^{-5}$\\
   B &$3.0\times\gamma/2$ & $ 0.00\times\gamma/2$ &0.012&	1.584(25)$\times 10^{-2}$ & 3.82(6)$\times 10^{-3}$ & 6.0(1)$\times 10^{-5}$\\

\end{tabular}
\end{table}

\begin{table}\scriptsize
\centering
{\sc Table VI: $N=1000, N_N=64$, $v_0/c_s=0.5\gamma^{-1/2}$, $\gamma=5/3$, $dt=0.01$ Shear tests}
\vspace{.04in}
\ 

\begin{tabular}{ccrcrll} \hline\hline\vspace{-0.13in}
\ \\
\vspace{0.04in}
AV routine & $\alpha$ & $\beta$ & $\langle{v_y^2+v_z^2 \over c_s^2}\rangle^{1/2}$ & $\eta~~[M c_s n^{2/3}]$ & $D~~[c_s n^{-1/3}]$ & $\eta D~~[M c_s^2 n^{1/3}]$\\ \hline
\vspace{-0.13in}\\

 None&   &     & 0.128	  &  4.8(4)$\times 10^{-5}$ & 0.38(7) &  1.8(4)$\times 10^{-5}$\\
   C &0.2& 0.75& 0.029	  &  9.3(6)$\times 10^{-3}$ &  2.2(2)$\times 10^{-2}$ &  2.1(2)$\times 10^{-4}$\\
   C ($\eta^2=0.002$)&0.3& 1.00& 0.026	  &  1.5(1)$\times 10^{-2}$ &  2.0(1)$\times 10^{-2}$ &  3.0(3)$\times 10^{-4}$\\
   C &0.3& 1.00& 0.026&  1.5(1)$\times 10^{-2}$ &  2.0(1)$\times 10^{-2}$ &  2.9(3)$\times 10^{-4}$\\
   C ($\eta^2=0.05$) &0.3& 1.00& 0.026	  &  1.3(1)$\times 10^{-2}$ &  2.1(2)$\times 10^{-2}$ &  2.9(3)$\times 10^{-4}$\\
   C &0.3& 0.50& 0.028& 1.19(10)$\times 10^{-2}$ &  2.1(1)$\times 10^{-2}$ &  2.4(3)$\times 10^{-4}$\\
   C &0.4& 0.50& 0.026&  1.6(2)$\times 10^{-2}$ &  1.9(2)$\times 10^{-2}$ &  3.1(4)$\times 10^{-4}$\\
   C &0.5& 0.50& 0.024&  2.2(2)$\times 10^{-2}$ &  2.0(1)$\times 10^{-2}$ &  4.3(5)$\times 10^{-4}$\\
   C &0.5& 1.00& 0.023&  2.5(2)$\times 10^{-2}$ &  1.8(1)$\times 10^{-2}$ &  4.5(6)$\times 10^{-4}$\\
   C &0.8& 1.25& 0.019&  4.5(5)$\times 10^{-2}$ &  1.7(1)$\times 10^{-2}$ & 7.5(10)$\times 10^{-4}$\\
   C &1.0& 0.25& 0.019&  5.4(7)$\times 10^{-2}$ & 1.59(7)$\times 10^{-2}$ & 8.5(12)$\times 10^{-4}$\\
  HK &0.0& 1.25& 0.079& 2.66(5)$\times 10^{-4}$ & 0.15(1) &  4.0(3)$\times 10^{-5}$\\
  HK &0.0&10.00& 0.063& 1.65(1)$\times 10^{-3}$ &  7.3(4)$\times 10^{-2}$ & 1.21(7)$\times 10^{-4}$\\
  HK &0.1& 0.50& 0.069& 3.91(5)$\times 10^{-4}$ & 0.106(4) &  4.1(2)$\times 10^{-5}$\\
  HK &0.2& 0.50& 0.062& 6.69(7)$\times 10^{-4}$ &  7.3(3)$\times 10^{-2}$ &  4.9(2)$\times 10^{-5}$\\
  HK &0.2& 0.75& 0.061& 7.11(8)$\times 10^{-4}$ &  6.8(3)$\times 10^{-2}$ &  4.9(2)$\times 10^{-5}$\\
  HK &0.3& 0.50& 0.059&  9.7(2)$\times 10^{-4}$ &  5.8(3)$\times 10^{-2}$ &  5.6(4)$\times 10^{-5}$\\
  HK &0.4& 0.50& 0.056& 1.28(3)$\times 10^{-3}$ &  5.4(5)$\times 10^{-2}$ &  6.9(7)$\times 10^{-5}$\\
  HK &0.5& 0.50& 0.055& 1.66(6)$\times 10^{-3}$ &  5.5(4)$\times 10^{-2}$ &  9.2(8)$\times 10^{-5}$\\
  HK &0.8& 1.25& 0.052&  2.8(1)$\times 10^{-3}$ &  4.6(7)$\times 10^{-2}$ &  1.3(2)$\times 10^{-4}$\\
  HK &1.0& 0.25& 0.051&  3.7(2)$\times 10^{-3}$ &  4.4(6)$\times 10^{-2}$ &  1.6(2)$\times 10^{-4}$\\
   B &$0.0\times\gamma/2$ & $ 1.00\times\gamma/2$ &0.054	  & 5.90(4)$\times 10^{-4}$ &  5.3(3)$\times 10^{-2}$ &  3.1(2)$\times 10^{-5}$\\
   B &$0.0\times\gamma/2$ & $ 2.50\times\gamma/2$ &0.045	  &1.245(9)$\times 10^{-3}$ &  3.1(3)$\times 10^{-2}$ &  3.8(4)$\times 10^{-5}$\\
   B &$0.0\times\gamma/2$ & $10.00\times\gamma/2$ &0.036	  & 4.10(2)$\times 10^{-3}$ & 2.87(6)$\times 10^{-2}$ & 1.18(3)$\times 10^{-4}$\\
   B ($\eta^2=0.002$) &$0.5\times\gamma/2$ & $ 0.50\times\gamma/2$ &0.036	  &  3.8(2)$\times 10^{-3}$ &  2.3(3)$\times 10^{-2}$ & 8.7(10)$\times 10^{-5}$\\
   B ($\eta^2=0.05$) &$0.5\times\gamma/2$ & $ 0.50\times\gamma/2$ &0.037	  &  3.6(2)$\times 10^{-3}$ &  2.2(3)$\times 10^{-2}$ & 7.8(13)$\times 10^{-5}$\\
   B &$0.5\times\gamma/2$ & $ 1.00\times\gamma/2$ &0.036	  &  4.1(2)$\times 10^{-3}$ &  2.4(2)$\times 10^{-2}$ & 9.6(11)$\times 10^{-5}$\\
   B &$0.8\times\gamma/2$ & $ 1.25\times\gamma/2$ &0.032	  &  7.1(4)$\times 10^{-3}$ &  2.3(2)$\times 10^{-2}$ &  1.6(2)$\times 10^{-4}$\\
   B &$1.0\times\gamma/2$ & $ 0.00\times\gamma/2$ &0.031	  &  8.9(6)$\times 10^{-3}$ &  2.2(2)$\times 10^{-2}$ &  2.0(2)$\times 10^{-4}$\\
   B &$1.0\times\gamma/2$ & $ 0.75\times\gamma/2$ &0.030	  &  9.2(6)$\times 10^{-3}$ &  2.1(2)$\times 10^{-2}$ &  1.9(2)$\times 10^{-4}$\\
   B &$1.0\times\gamma/2$ & $ 1.00\times\gamma/2$ &0.030	  &  9.6(7)$\times 10^{-3}$ &  2.0(3)$\times 10^{-2}$ &  2.0(3)$\times 10^{-4}$\\
   B &$1.0\times\gamma/2$ & $ 1.25\times\gamma/2$ &0.030	  &  9.5(6)$\times 10^{-3}$ &  2.2(2)$\times 10^{-2}$ &  2.1(3)$\times 10^{-4}$\\
   B &$1.0\times\gamma/2$ & $ 2.00\times\gamma/2$ &0.030	  &  9.9(6)$\times 10^{-3}$ &  2.2(2)$\times 10^{-2}$ &  2.2(2)$\times 10^{-4}$\\
   B &$2.0\times\gamma/2$ & $ 0.00\times\gamma/2$ &0.024	  &  2.5(2)$\times 10^{-2}$ &  2.0(2)$\times 10^{-2}$ &  4.9(7)$\times 10^{-4}$\\
\end{tabular}
\end{table}

\clearpage

\begin{table}\scriptsize
\centering
{\sc Table VII: $N=1000, v_0/c_s=0.1\gamma^{-1/2}, \gamma=5/3, dt=0.01$ Shear tests}
\vspace{.04in}
\ 

\begin{tabular}{ccrccrll} \hline\hline\vspace{-0.13in}
\ \\
\vspace{0.04in}
AV routine & $\alpha$ & $\beta$ & $N_N$ & $\langle{v_y^2+v_z^2 \over c_s^2}\rangle^{1/2}$ & $\eta~~[M c_s n^{2/3}]$ & $D~~[c_s n^{-1/3}]$ & $\eta D~~[M c_s^2 n^{1/3}]$\\ \hline
\vspace{-0.13in}\\

   B &$0.0\times\gamma/2$ & $ 1.00\times\gamma/2$ & 20 & 0.060	  & 6.63(7)$\times 10^{-4}$ & 7.0(3)$\times 10^{-3}$ & 4.7(2)$\times 10^{-6}$\\
   B &$0.0\times\gamma/2$ & $ 1.00\times\gamma/2$ & 32 & 0.037	  & 2.98(2)$\times 10^{-4}$ & 6.7(2)$\times 10^{-3}$ & 2.00(7)$\times 10^{-6}$\\
   B &$0.0\times\gamma/2$ & $ 1.00\times\gamma/2$ & 64 & 0.026	  & 2.72(1)$\times 10^{-4}$ & 1.16(3)$\times 10^{-2}$ & 3.13(7)$\times 10^{-6}$\\
   B &$1.0\times\gamma/2$ & $ 0.00\times\gamma/2$ & 20 & 0.027	  & 4.85(3)$\times 10^{-3}$ & 5.5(2)$\times 10^{-3}$ & 2.67(10)$\times 10^{-5}$\\
   B &$1.0\times\gamma/2$ & $ 0.00\times\gamma/2$ & 48 & 0.017	  & 4.48(2)$\times 10^{-3}$ & 3.85(8)$\times 10^{-3}$ & 1.72(4)$\times 10^{-5}$\\
   B &$1.0\times\gamma/2$ & $ 0.00\times\gamma/2$ & 64 & 0.015	  & 4.22(3)$\times 10^{-3}$ & 3.2(2)$\times 10^{-3}$ & 1.33(7)$\times 10^{-5}$\\
   B &$1.0\times\gamma/2$ & $ 1.00\times\gamma/2$ & 20 & 0.026	  & 4.92(4)$\times 10^{-3}$ & 5.16(8)$\times 10^{-3}$ & 2.54(5)$\times 10^{-5}$\\
   B &$1.0\times\gamma/2$ & $ 1.00\times\gamma/2$ & 64 & 0.015	  & 4.33(2)$\times 10^{-3}$ & 3.93(5)$\times 10^{-3}$ & 1.70(3)$\times 10^{-5}$\\

\end{tabular}
\end{table}

\clearpage

\begin{figure}
\caption{
This sequence of cross-sectional slabs, each of thickness $\Delta
z=1.02 n^{-1/3}$, in a periodic box of dimension $19n^{-1/3}\times
19n^{-1/3}\times 19n^{-1/3}$ demonstrates the instability of a simple
cubic lattice.  (a) At $t=0$ the $N=19^3$ equal mass SPH particles,
each with $N_N\approx 32$ neighbors, are initially motionless with only
minuscule deviations (due to numerical roundoff errors) from the unstable
equilibrium positions of a
simple cubic lattice.  (b) At $t=190 n^{-1/3} c_s^{-1}$ the particles
are in the process of shifting their positions.  (c) By $t=380 n^{-1/3}
c_s^{-1}$ the particles have settled into a new, stable lattice
structure.
} \label{crystal}
\end{figure}

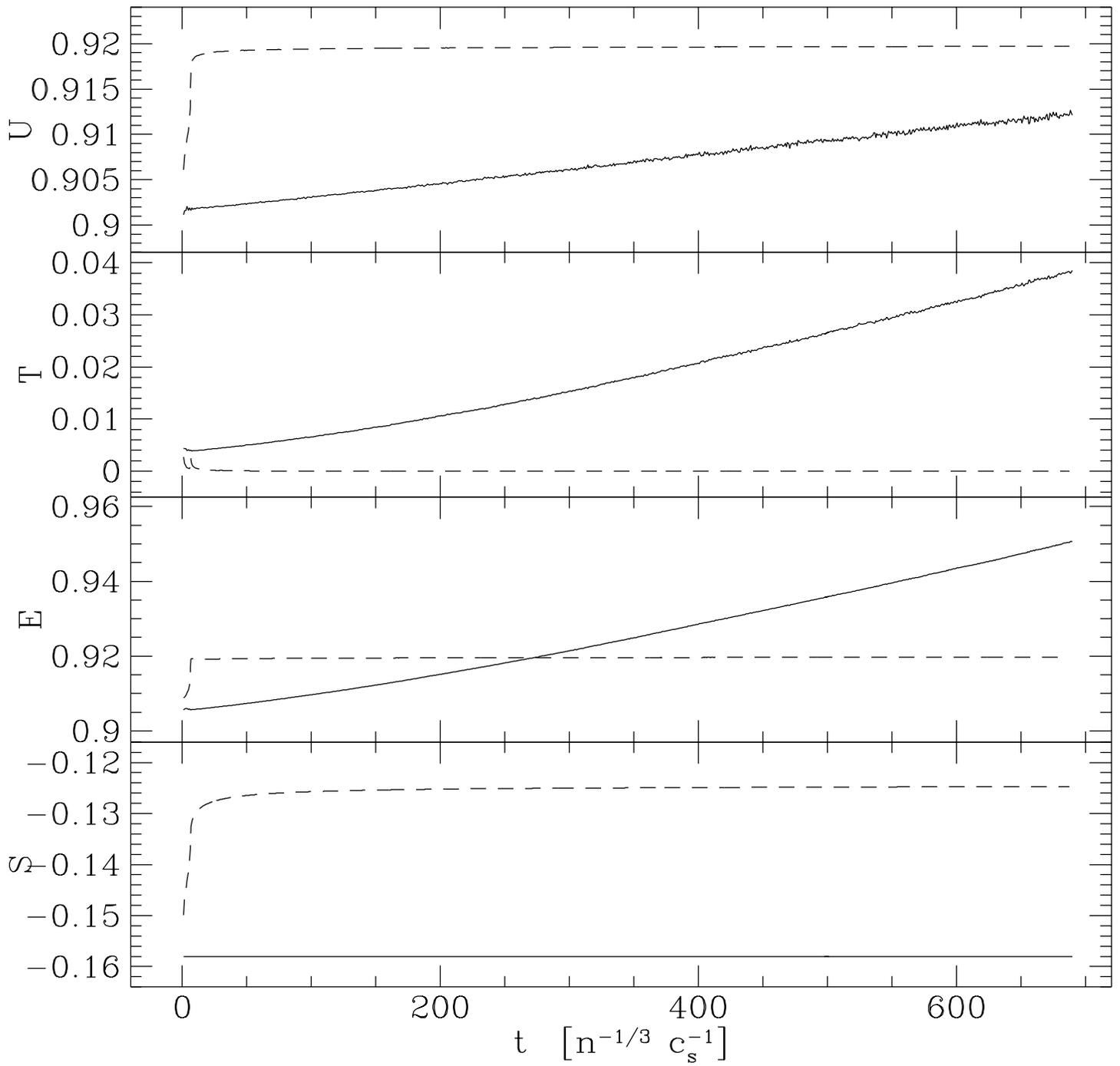
\begin{figure}
\caption{
The internal energy $U$, kinetic energy $T$, total
energy $E=U+T$, and entropy $S$ of the $N=20^3$ equal mass
particles interacting with $N_N\approx 64$ neighbors for a calculation
without AV (solid curve) and a calculation with
AV (dashed curve).  In contrast to the previous simple box
tests, the smoothing lengths $h_i$ are allowed to vary.   The particles
began on a simple cubic lattice with a Maxwell-Boltzmann velocity
distribution.
} \label{utes}
\end{figure}

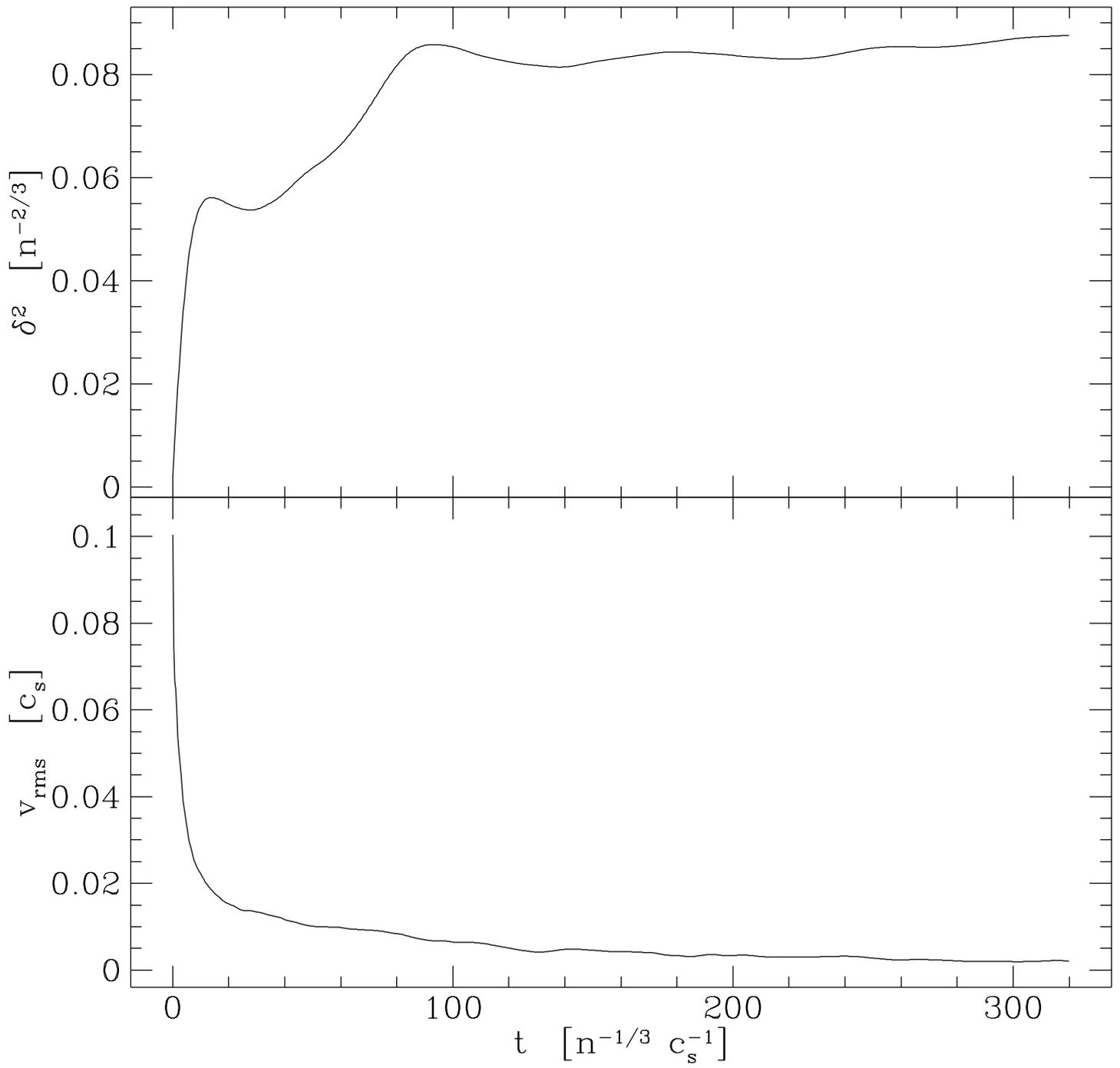
\begin{figure}
\caption{
The mean square deviation $\delta^2$ and root mean square velocity
$v_{rms}$ as a function of time for $N=16^3$ equal mass SPH particles
with $N_N\approx32$ in a typical simple box test.  Here the AV is given
by equation (\ref{pi}) with $\alpha=1$, $\beta=2$ and $\eta^2=0.01$.  The
particles begin in a simple cubic lattice with a Maxwell-Boltzmann
velocity distribution.  The AV drives $v_{rms}$ to
zero, so that the mean square deviation $\delta^2$ approaches a
constant and the diffusion coefficient $D=d\delta^2/dt$ becomes zero.
} \label{d2vtAV}
\end{figure}

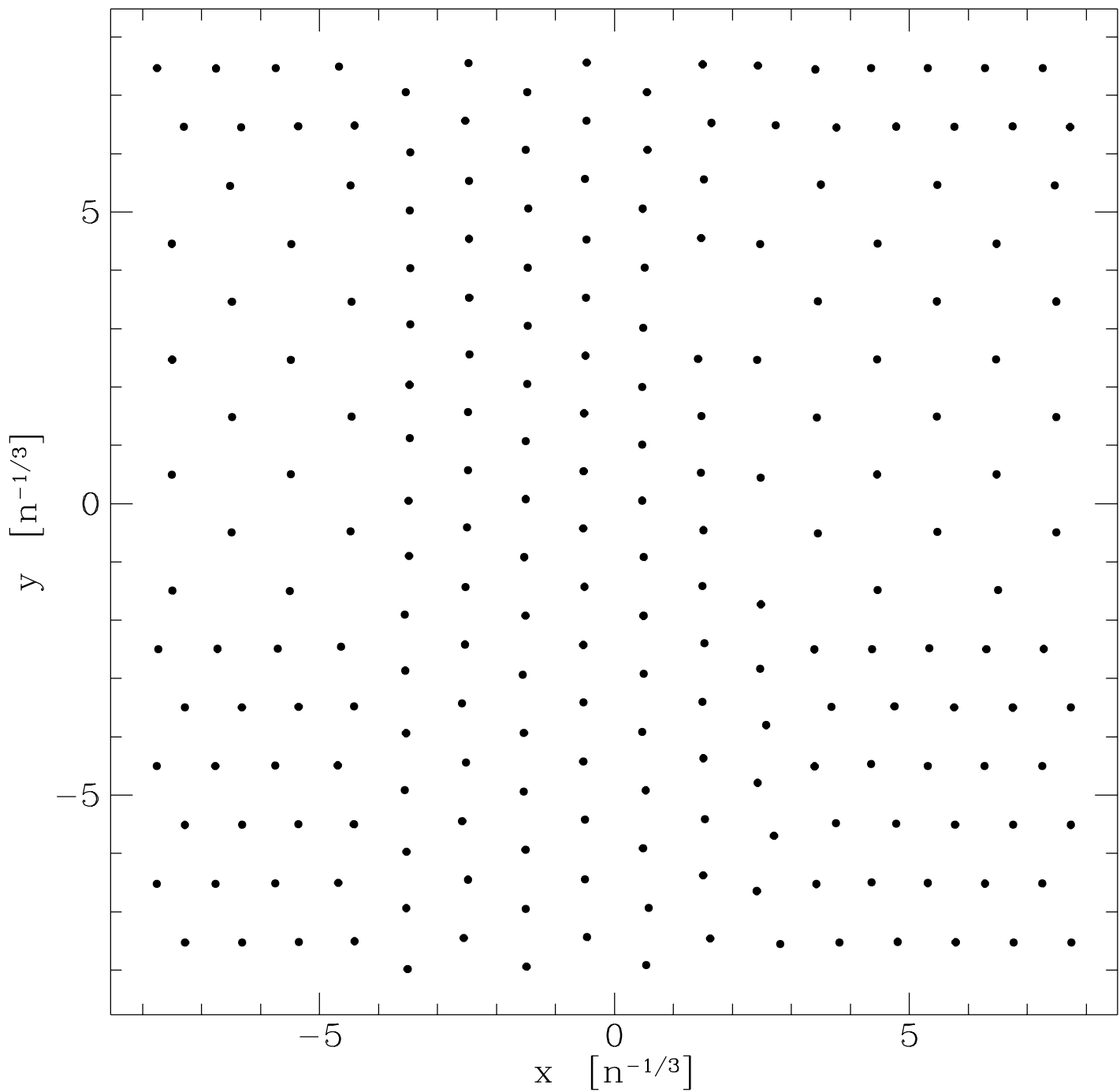
\begin{figure}
\caption{
A cross-sectional slab of thickness $\Delta z=0.6n^{-1/3}$ of the final
particle configuration for the simple box test presented in
Fig.~\ref{d2vtAV}.  There are clear dislocations separating the different
lattice orientations.  The initially noisy system has been quenched, or
``frozen," into a crystal by the AV so quickly that
the SPH particles did not have opportunity to settle into a single
orientation.
} \label{quench}
\end{figure}

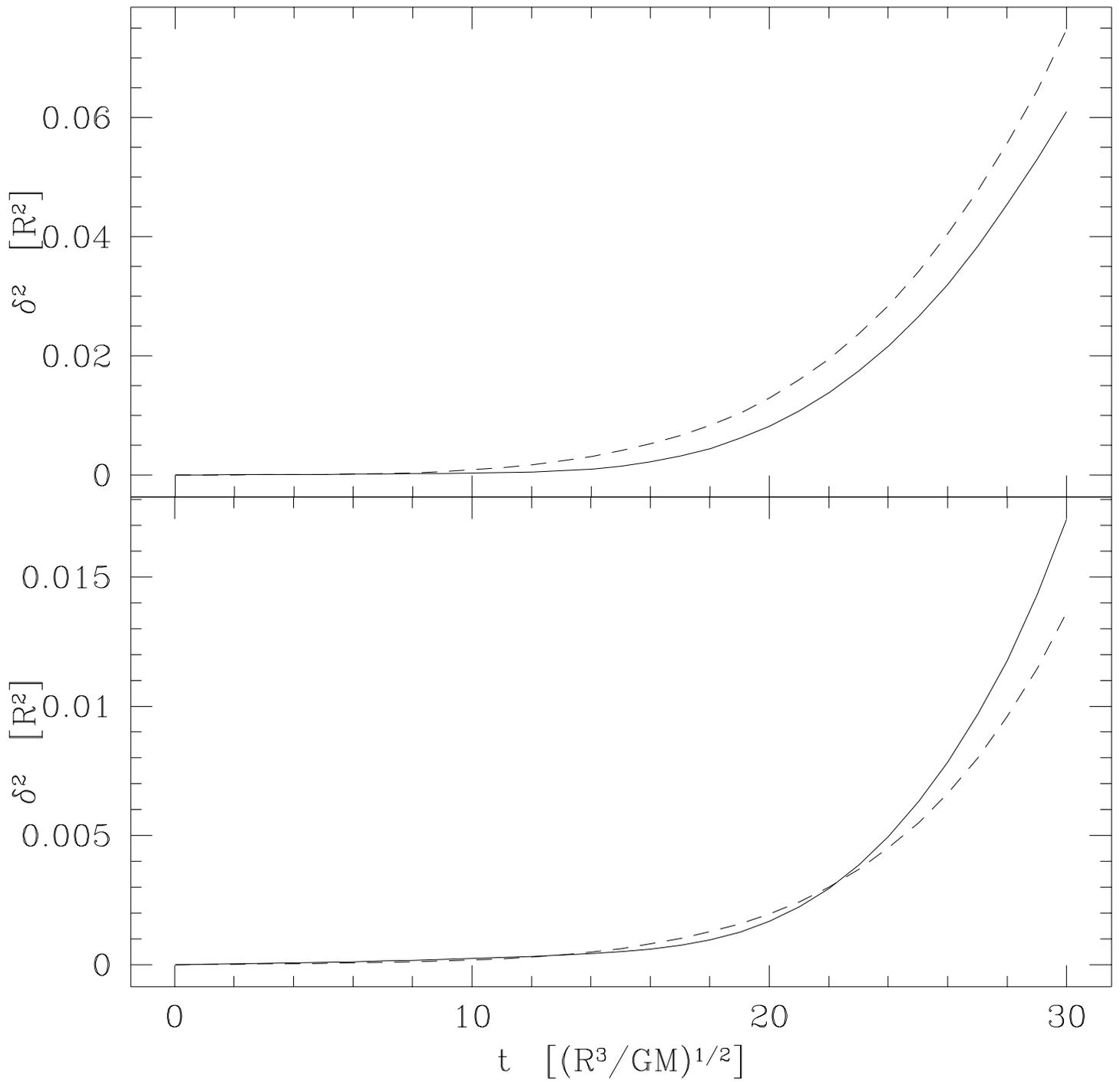
\begin{figure}
\caption{
The ``predicted'' (dashed curve) and actual mean square displacement
(solid curve) for the innermost 6400 particles in an equilibrium
$n=1.5$ polytrope of mass $M$ and radius $R$ modeled with $N=13949$
equal mass particles and $N_N\approx64$.  For the top frame
equation~(\ref{rdot}) is used to update particle positions, while in
the bottom frame equation~(\ref{XSPH}), the XSPH method, is
implemented.  
} \label{d2vtpoly}
\end{figure}

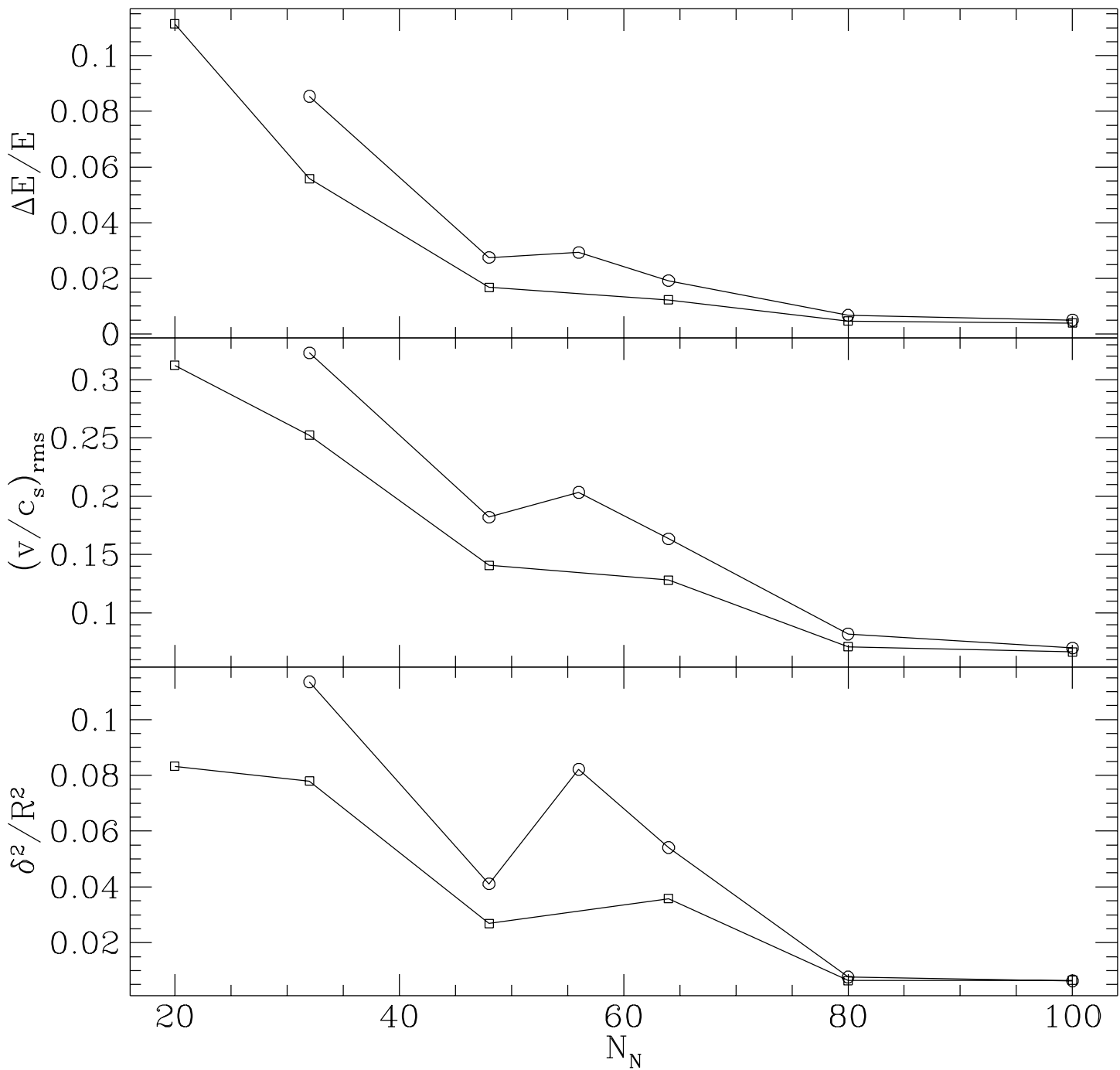
\begin{figure}
\caption{
The fractional spurious change in total energy $\Delta E/E$,
$(v/c_s)_{rms}$ and the mean square diffusion distance $\delta^2$ as a
function of $N_N$ evaluated at a time $t=25 (R^3/GM)^{1/2}$ during
calculations of an equilibrium $n=1.5$ polytrope.
Circular data points correspond to a polytrope modeled with $N=30000$
particles, while square data points correspond to those with $N=13949$
particles.
} \label{nn}
\end{figure}

%

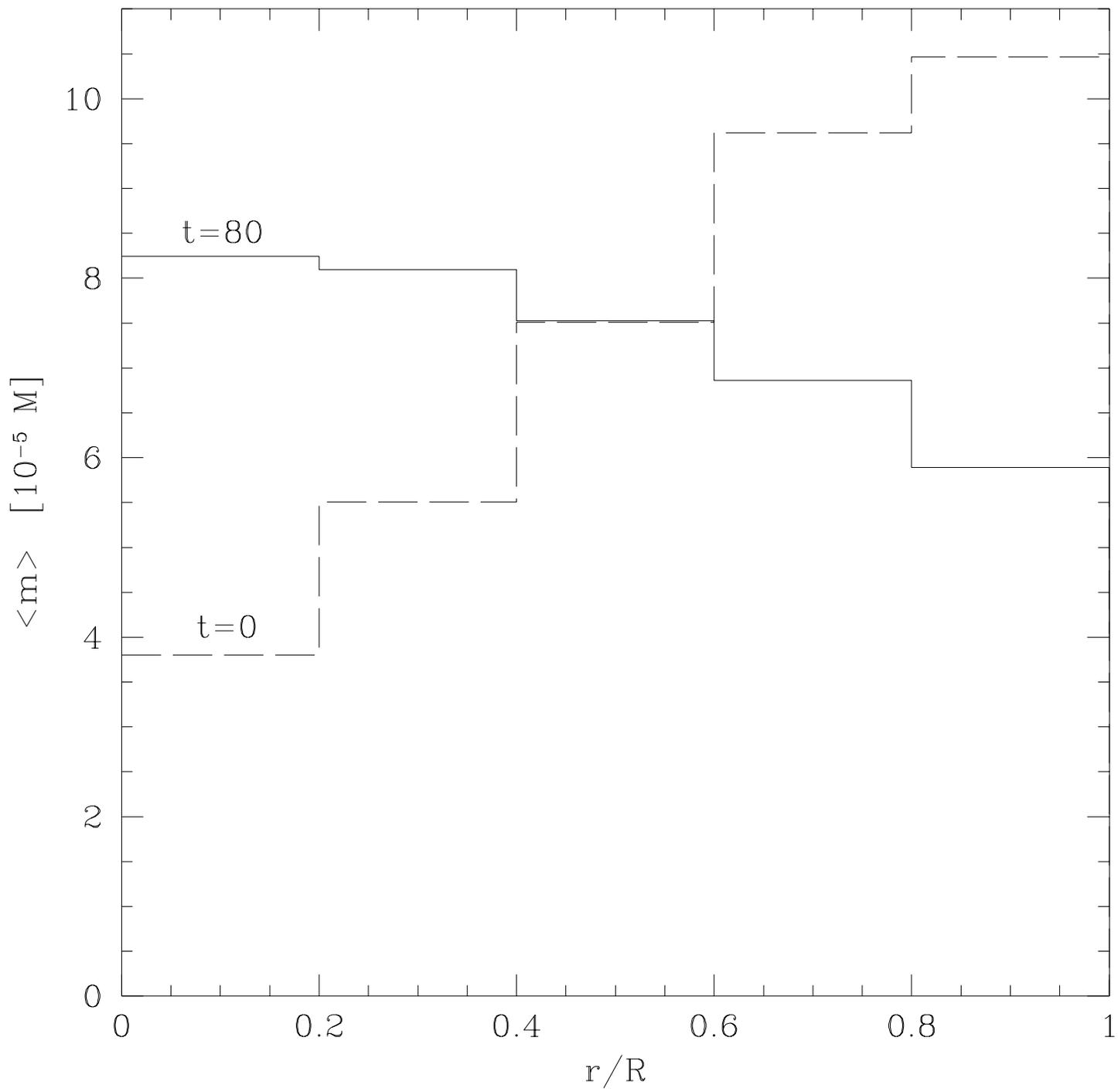
\begin{figure}
\caption{
Histogram of the average SPH particle mass $\langle m \rangle$ in five
radial bins for the initial configuration (dashed curve) and $t=80
(R^3/GM)^{1/2}$ configuration (solid curve) during the evolution of an
equilibrium $n=1.5$ polytrope of mass $M$ and radius $R$.  This
calculation employs $N=13949$ particles with $N_N\approx 64$, $C_N=0.8$,
the simple timestep routine, and no AV.
} \label{bhist}
\end{figure}

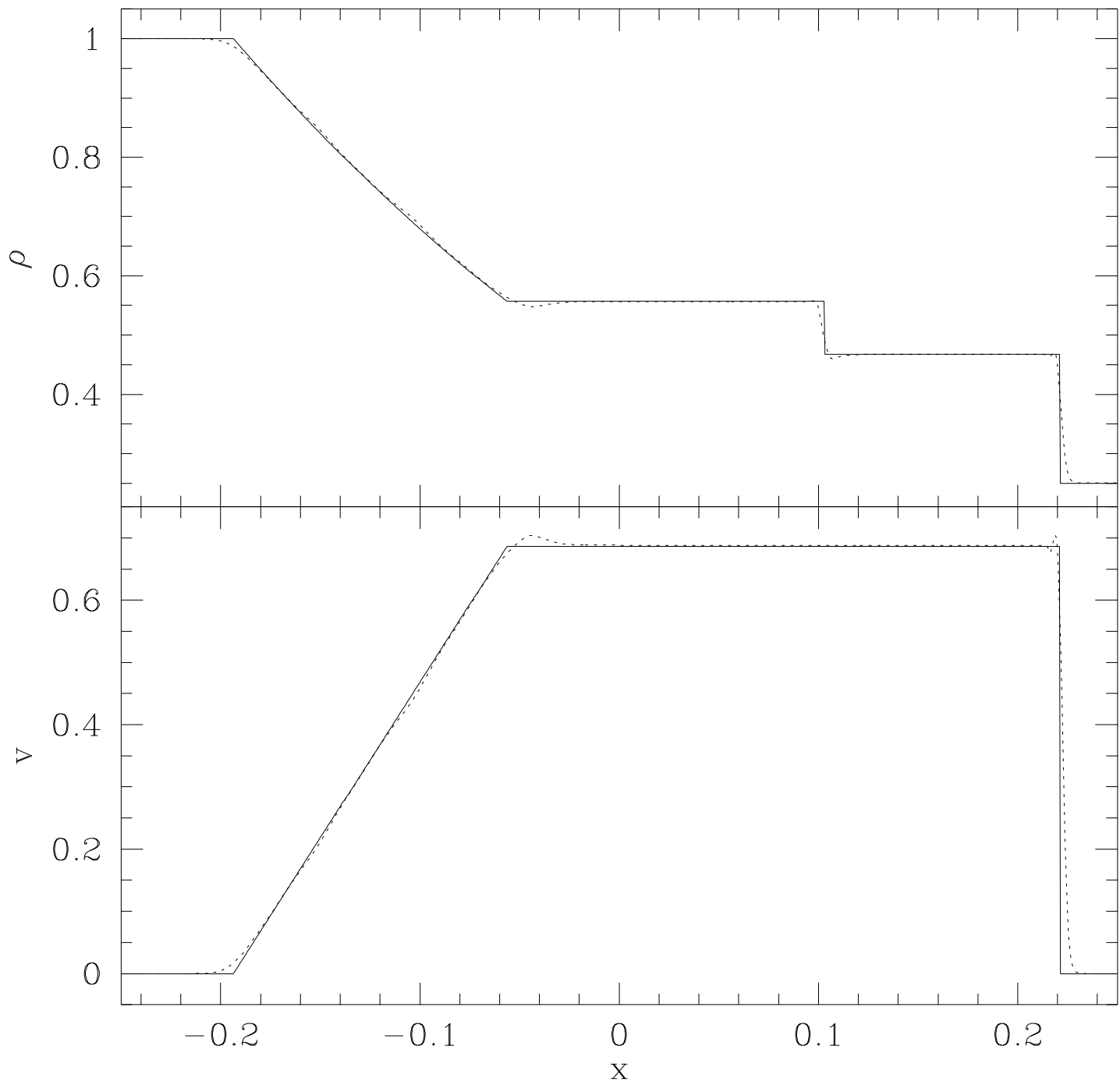
\begin{figure}
\caption{
Density and velocity profiles in a shock-tube test with Mach number
${\cal M}\approx 1.6$ as given by the quasi-analytic solution (solid
curve) and our 1-dimensional SPH code (dotted curve) at a time
$t=0.15$.  An adiabatic equation of state is used with $\gamma=5/3$.
} \label{rhox}
\end{figure}

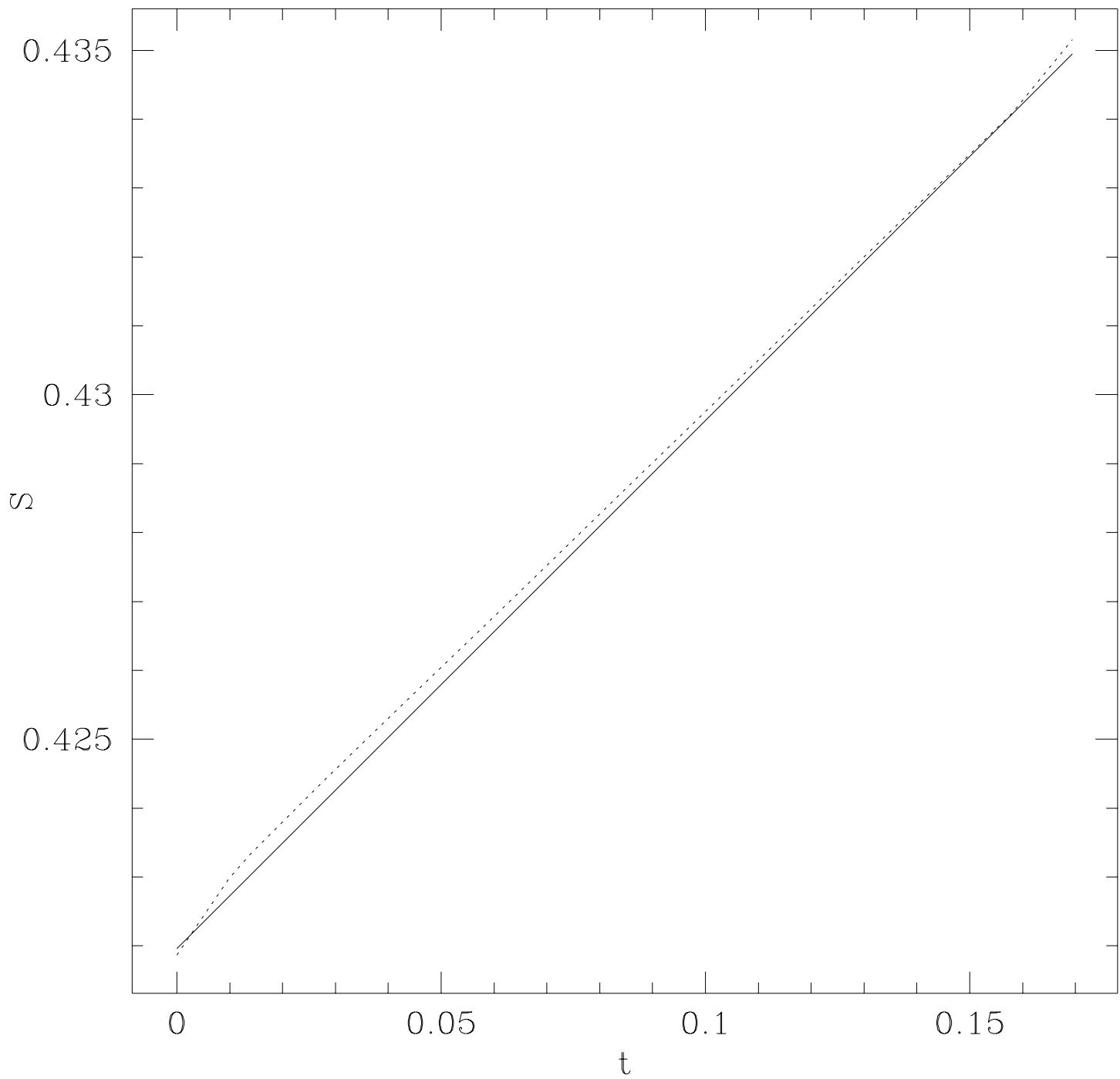
\begin{figure}
\caption{
Entropy $S$ in a shock-tube at early times $t$, as
given by the quasi-analytic solution (solid line) and our 1D SPH code
(dotted curve), for the same calculation presented in Fig.~\ref{rhox}.
} \label{svst}
\end{figure}

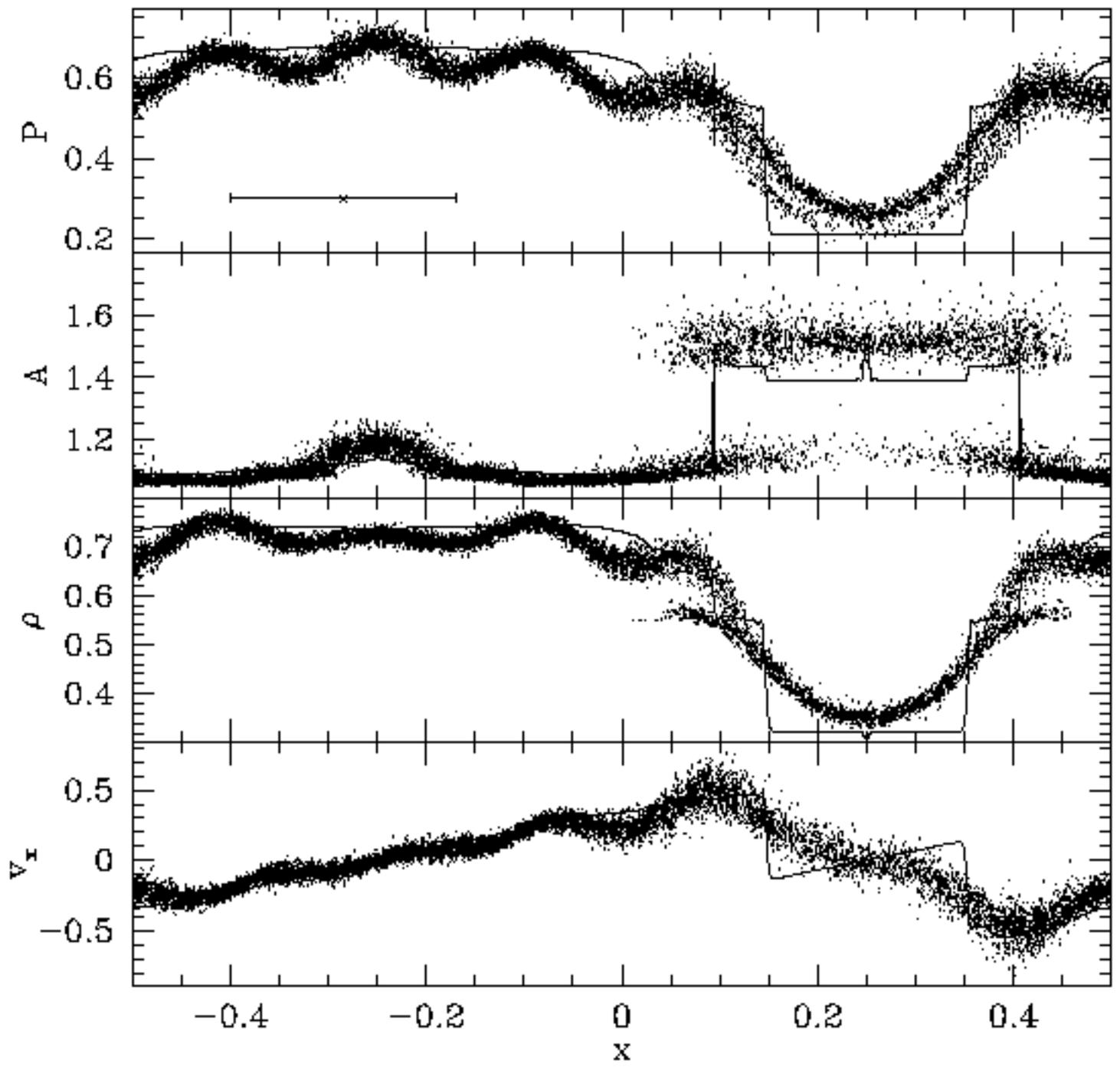
\begin{figure}
\caption{
The pressure $P$, entropy variable $A$, density $\rho$ and velocity
component $v_x$ as given by our 1D code (solid curve) and by one of our
3D calculations (dots) at the relatively late time $t=1$, for the same
shock-tube test shown in Figures \ref{rhox} and \ref{svst}.  The bar in
the lower left corner of the uppermost frame has a total length of
$4\langle h\rangle$, where $\langle h\rangle=0.058$ is the average
smoothing length in the 3D calculation.
} \label{latetmost}
\end{figure}

\begin{figure}
\caption{
Dependence of the results of shock-tube calculations on the AV
parameters $\alpha$ and $\beta$ for the classical AV with our
3D SPH code:  (a) $\alpha=0$; $\beta=1$ (short), $2.5$ (long), $10$
(dot short), (b) $\beta=0$; $\alpha=1$ (short), $2$ (long), $3$ (dot
short), $10$ (dot long), (c) $\beta=1$; $\alpha=0$ (dot), $1$ (short
dash), $2$ (long dash), $3$ (dot dash).  In all cases $\eta^2=0.01$.
The solid line in the bottom two frames corresponds to our benchmark 1D
calculation.
} \label{varyAV}
\end{figure}

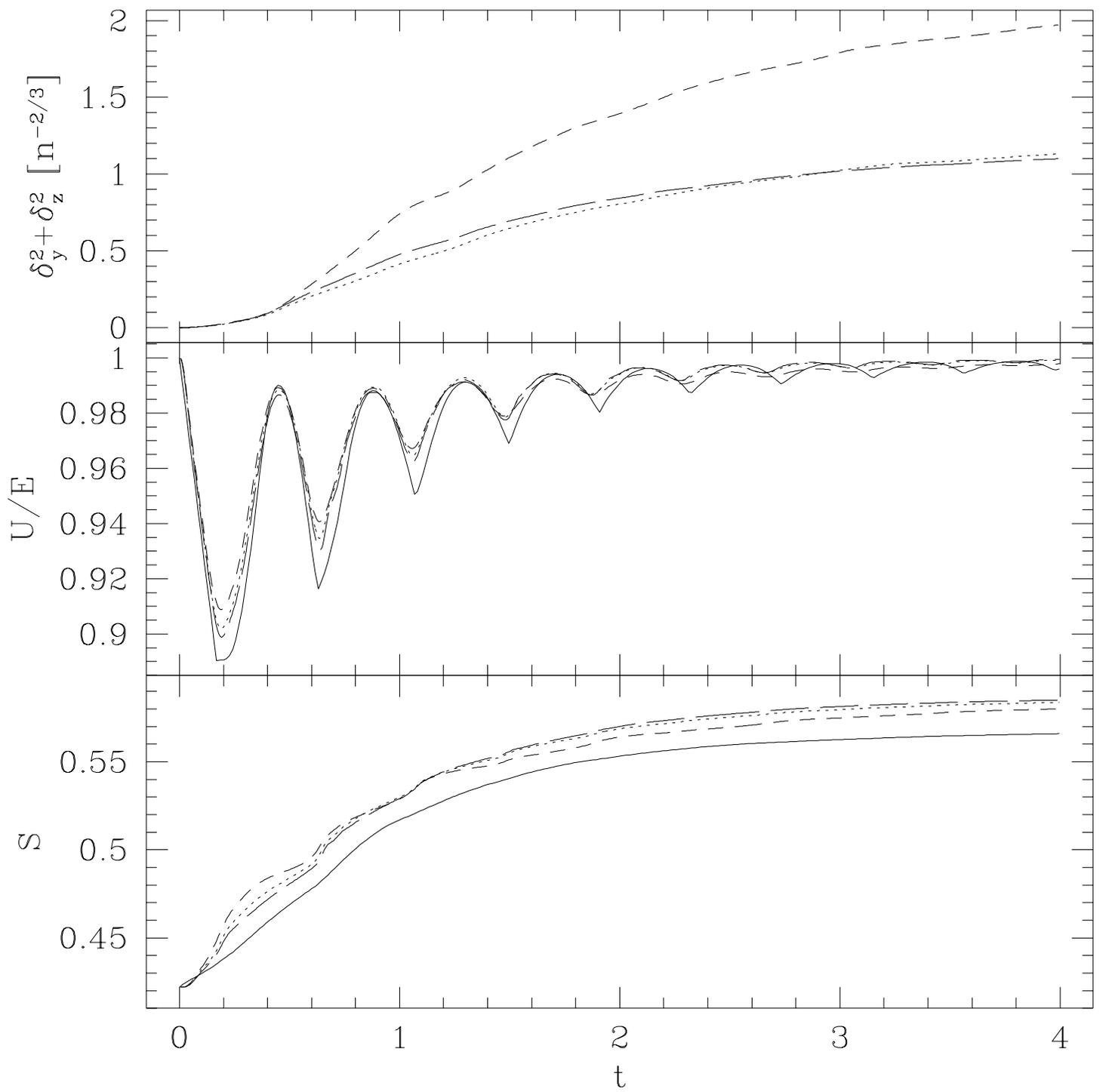
\begin{figure}
\caption{
The average square displacement perpendicular to the bulk fluid flow
$\delta_y^2+\delta_z^2$, the ratio of internal to total energy $U/E$,
and the entropy $S$ for three $\gamma=5/3$ shock-tube calculations with
different forms of AV:  the classical AV with $\alpha=0.5$, $\beta=1$
(long dashed curve), the HK AV with $\alpha=\beta=0.5$ (short dashed
curve), and the Balsara AV with $\alpha=\beta=\gamma/2$
(dotted curve).  The solid curve in the bottom two frames results from
our 1D SPH code.}
\label{compareav}
\end{figure}

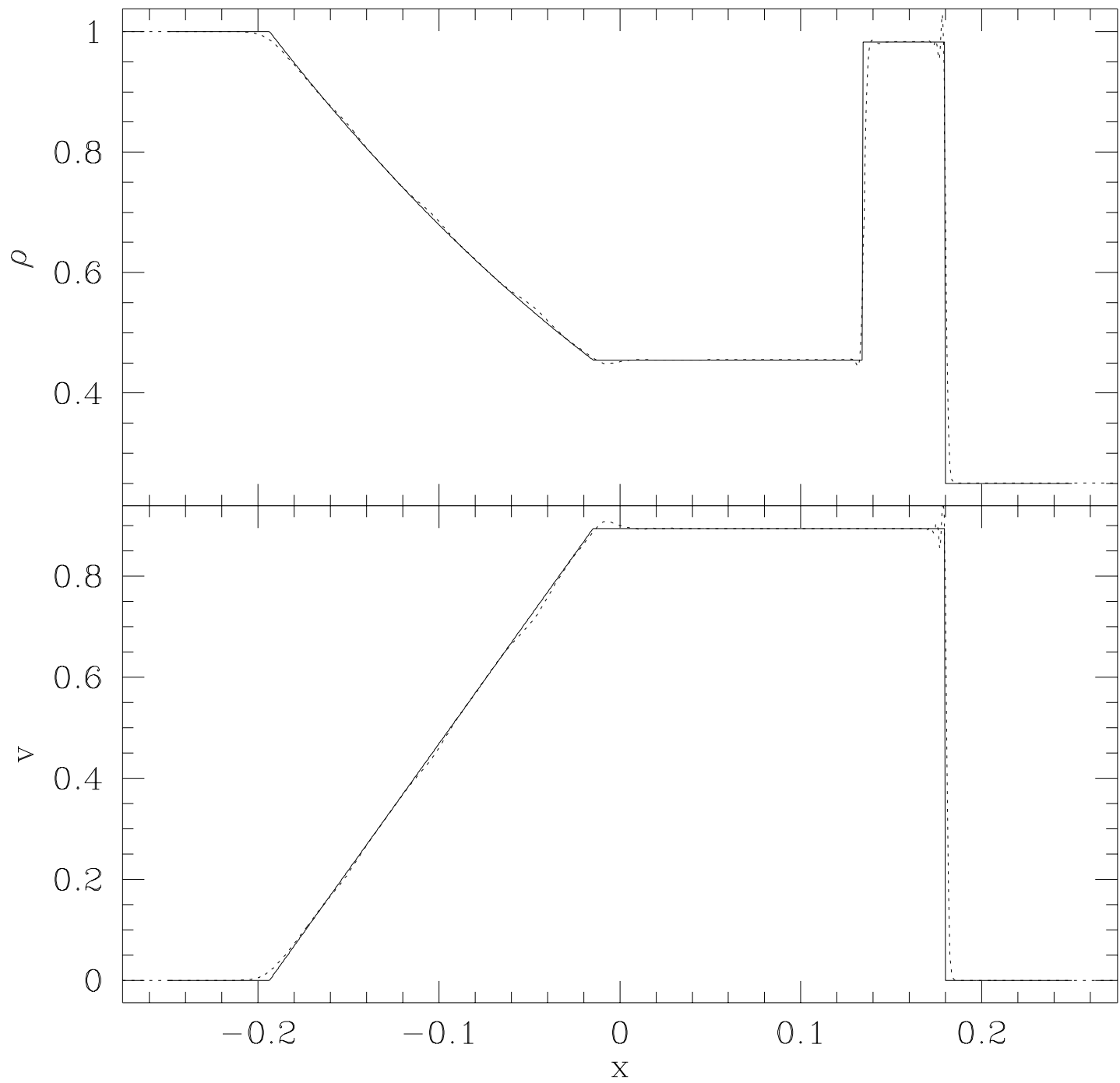
\begin{figure}
\caption{
As Fig.~\ref{rhox}, but for a higher Mach number (${\cal M}\approx13.2$)
shock-tube test. The solid line is the quasi-analytic solution, while
the dotted line is the result of our 1D SPH code.
} \label{1dhighMach}
\end{figure}

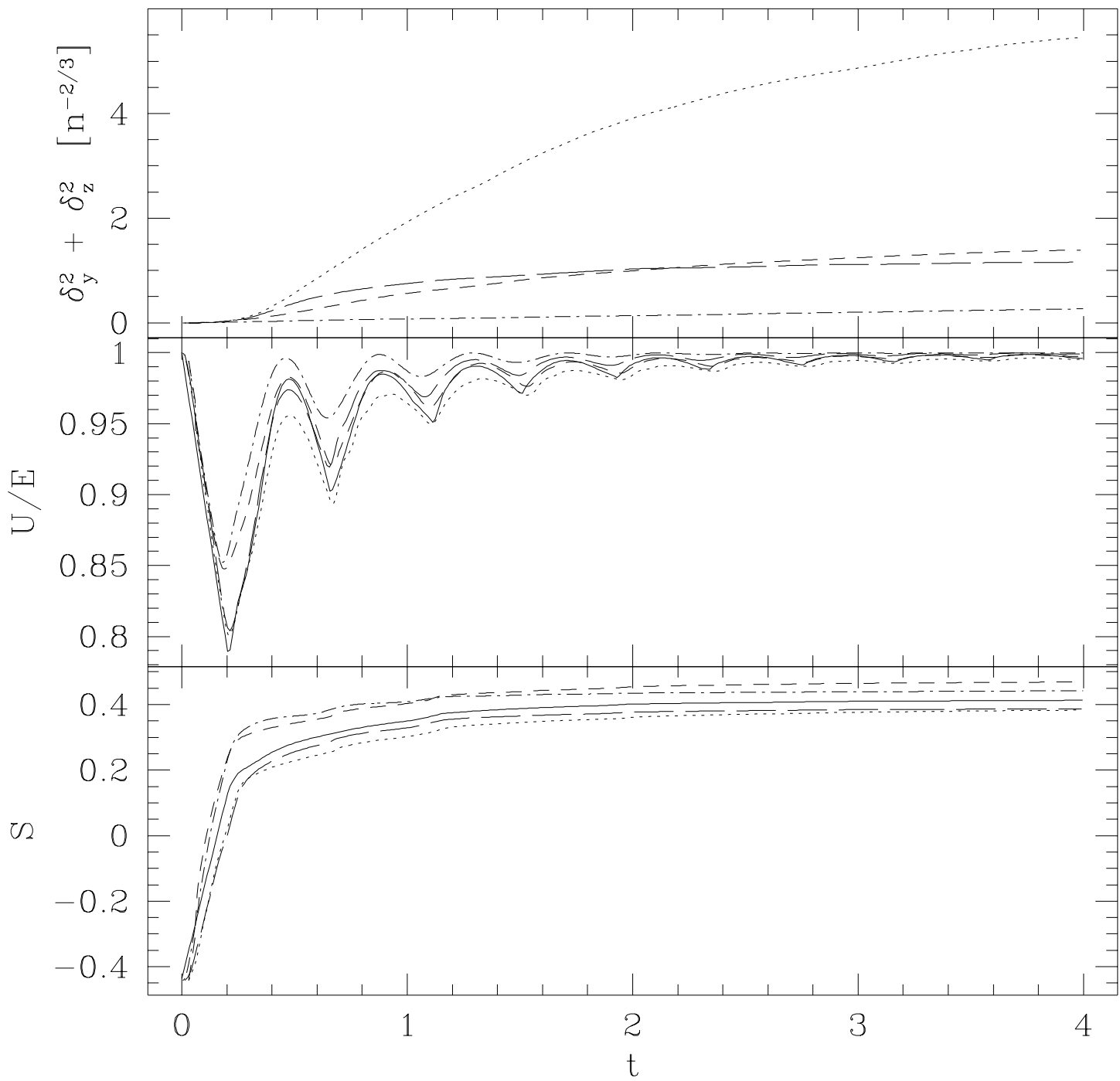
\begin{figure}
\caption{
Dependence of the high Mach number shock-tube calculations on the
AV parameters $\alpha$ and $\beta$ for the classical AV
and our 3D SPH code. The different lines represent different values, as
follows:
$\alpha=0, \beta=1$ (dotted curve);
$\alpha=0, \beta=5$ (short dashed curve);
$\alpha=1, \beta=0$ (long dashed curve);
$\alpha=5, \beta=0$ (dot-dash).
The solid curve is the result of the 1D calculation presented in
Fig.~\ref{1dhighMach}.
} \label{highMachs}
\end{figure}

\clearpage

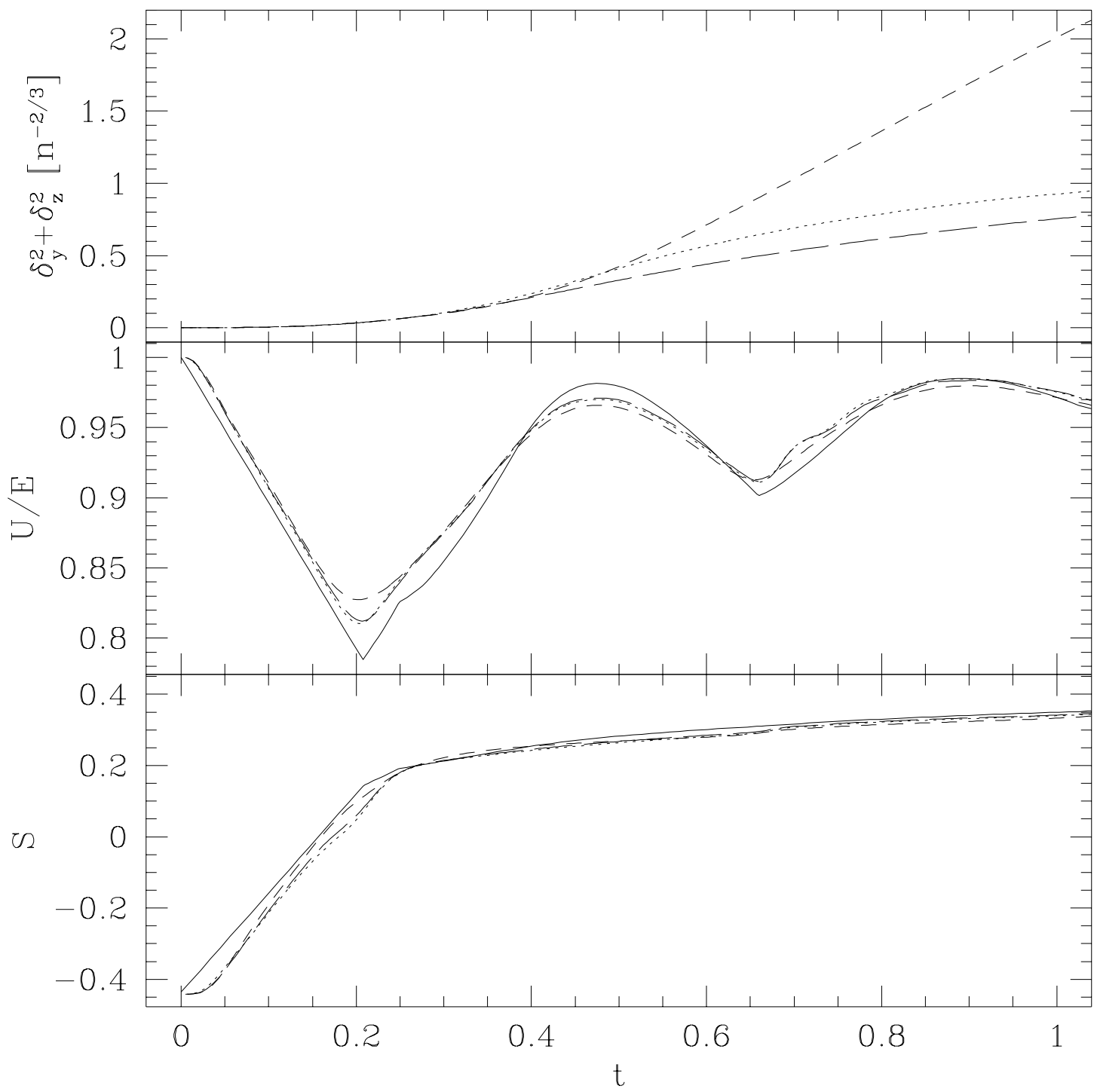
\begin{figure}
\caption{
As Fig.~\ref{compareav}, but for our high Mach number shock-tube test.
} \label{compareavhighm}
\end{figure}

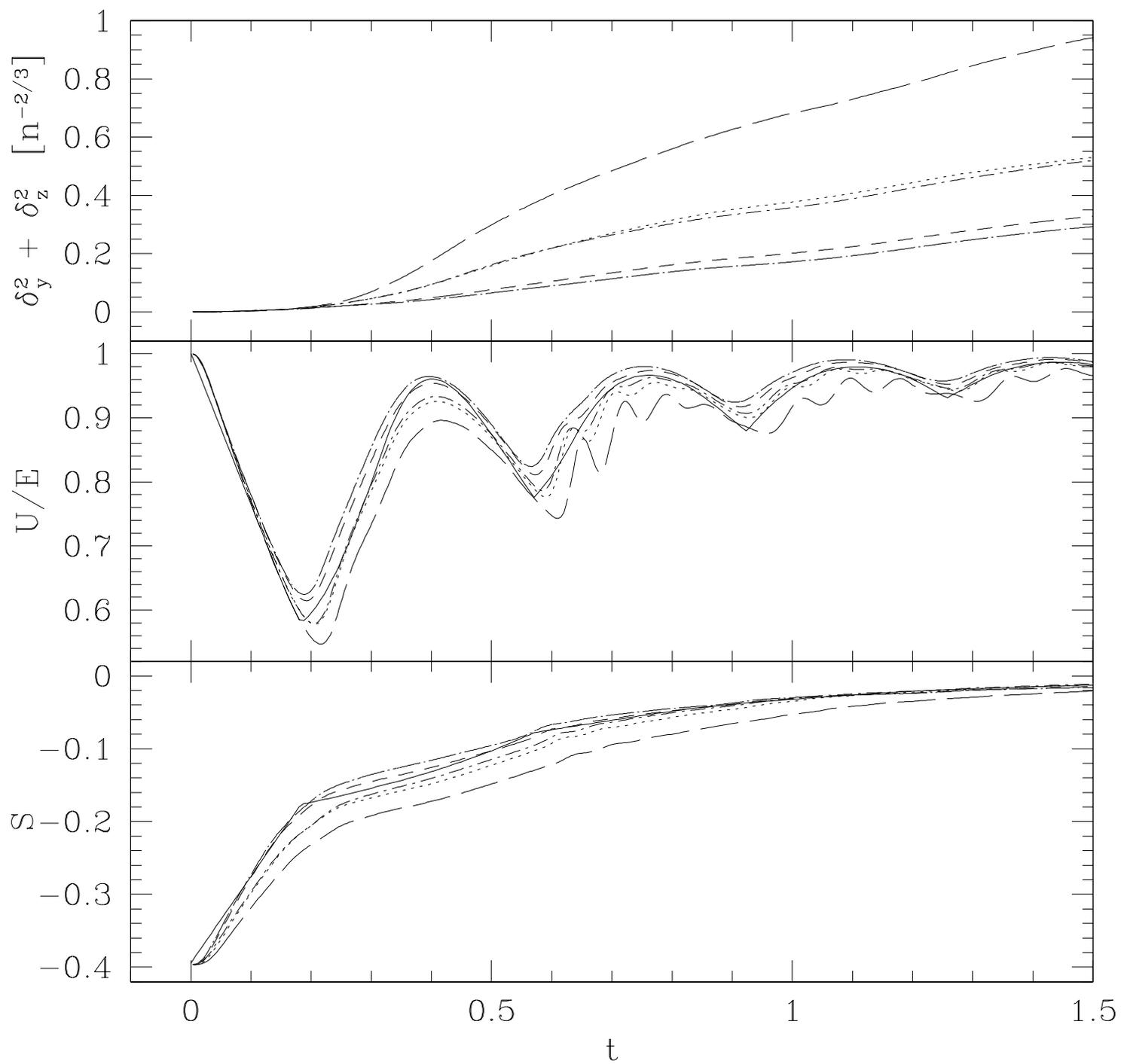
\begin{figure}
\caption{
As Fig.~\ref{compareav}, but for the high Mach number shock-tube test
with $\gamma=3$. The solid line is for the 1D calculation, and the
others are the results of the 3D calculations with the following AV
schemes and parameters: dotted, $\alpha=0.5, \beta=1.0$, classical AV;
short dash, $\alpha=0.9, \beta=1.8$, classical AV; long dash,
$\alpha=0.28, \beta=0.56$, classical AV; dot-short
dash, $\alpha=1.0\times \gamma/2, \beta=1.0\times \gamma/2$, Balsara AV;
dot-long dash, $\alpha=1.8\times \gamma/2, \beta=1.8\times \gamma/2$,
Balsara AV; short dash-long dash, $\alpha=0.56\times \gamma/2,
\beta=0.56\times \gamma/2$, Balsara AV.  
} \label{compareavgam3}
\end{figure}

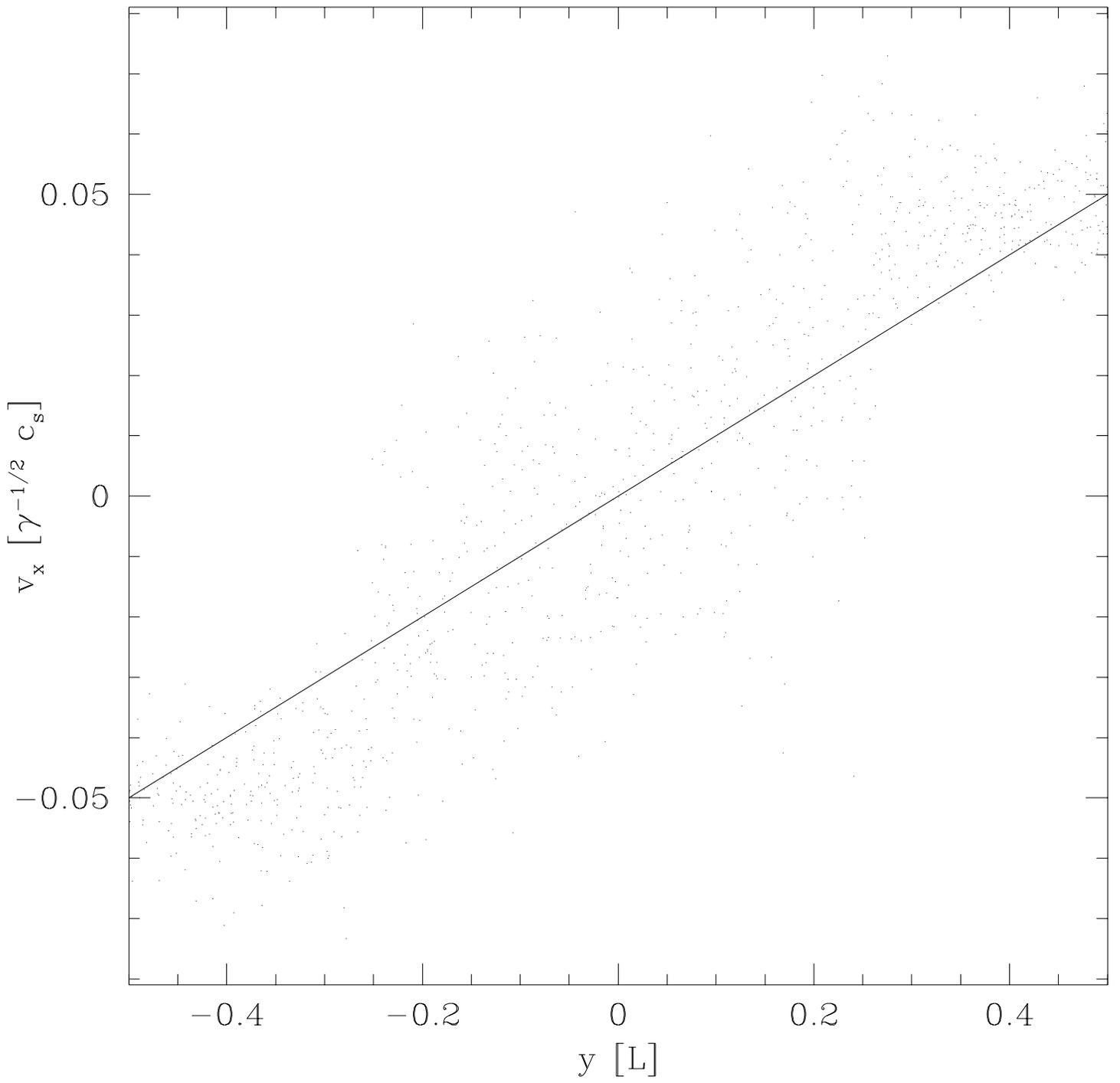
\begin{figure}
\caption{
Particle velocities in the $x$-direction plotted against their $y$
coordinates.  Slipping boundary conditions at $y=\pm {1\over 2}$ are
used to maintain the shear flow.  The system was relaxed without AV for
the first 10 time units towards a configuration with $v_x=v_0y/L$, $v_y=v_z=0$
(the solid line), and then allowed to evolve with AV for another 10 time
units to the state shown in this figure.  Here
$v_0=0.1c_s\gamma^{-1/2}$, and $L=1$ is the unit of length.   We used
$N=1000$ particles each with $N_N=64$ neighbors on average and the
classical AV with $\alpha=0.5$ and $\beta=1$.
} \label{couette}
\end{figure}

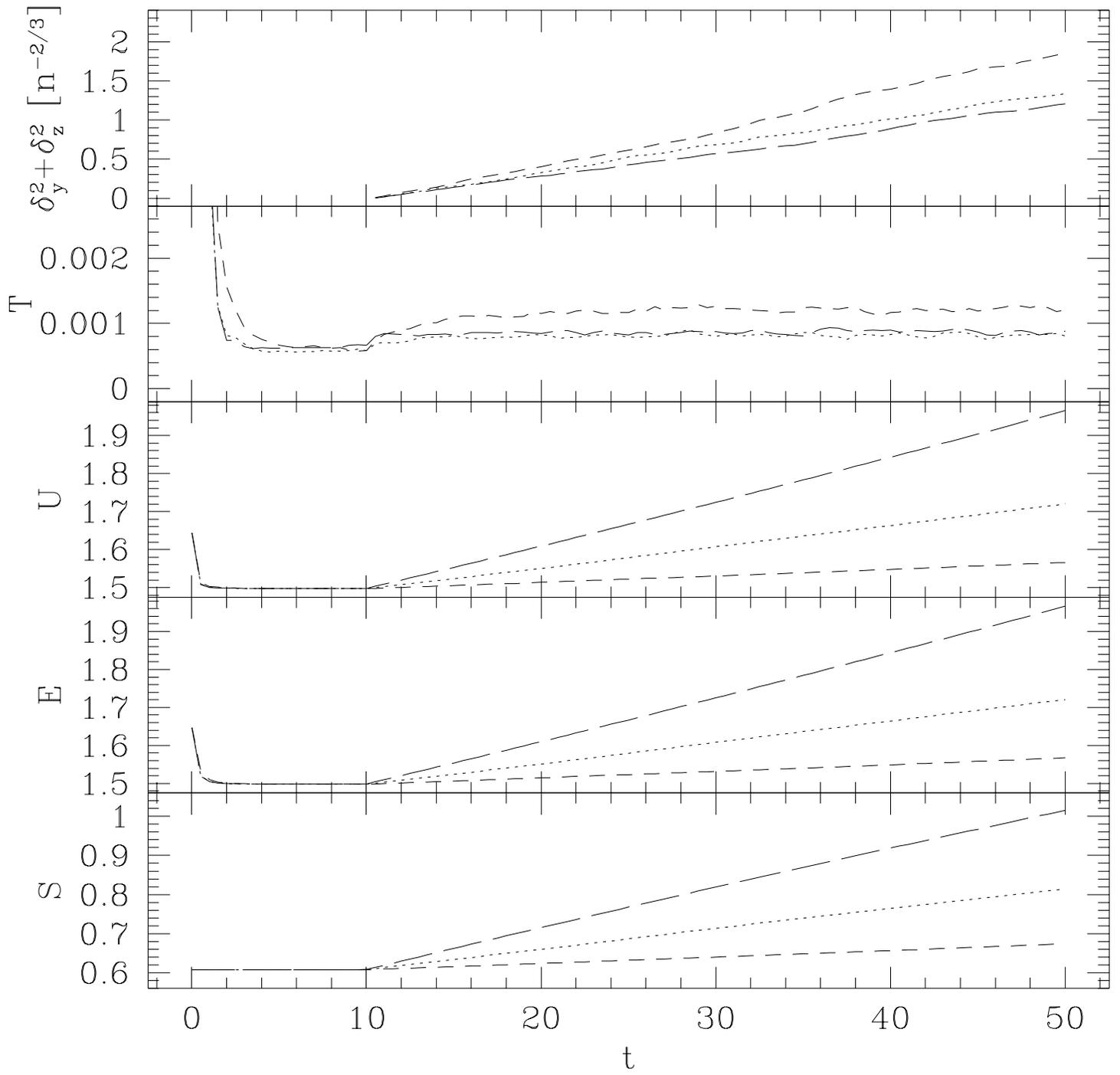
\begin{figure}
\caption{
The spurious square displacement in the direction perpendicular to the
fluid flow, energies, and entropy as a function of time in three
calculations of a shear flow using $N=1000$ and $N_N=64$ with different
forms of AV:
the classical AV with $\alpha=0.5$, $\beta=1$ (long dashed curve), the
HK AV with $\alpha=\beta=0.5$ (short dashed curve), and the Balsara AV
with $\alpha=\beta=1 \times \gamma/2$ (dotted curve).  The system was
relaxed for the first 10 time units towards a situation with
$(v_x,v_y,v_z)=(0.1c_s\gamma^{-1/2}y/L,0,0)$, while from $t=10$ to 50 the
system freely evolves with slipping boundary conditions.
} \label{shear}
\end{figure}

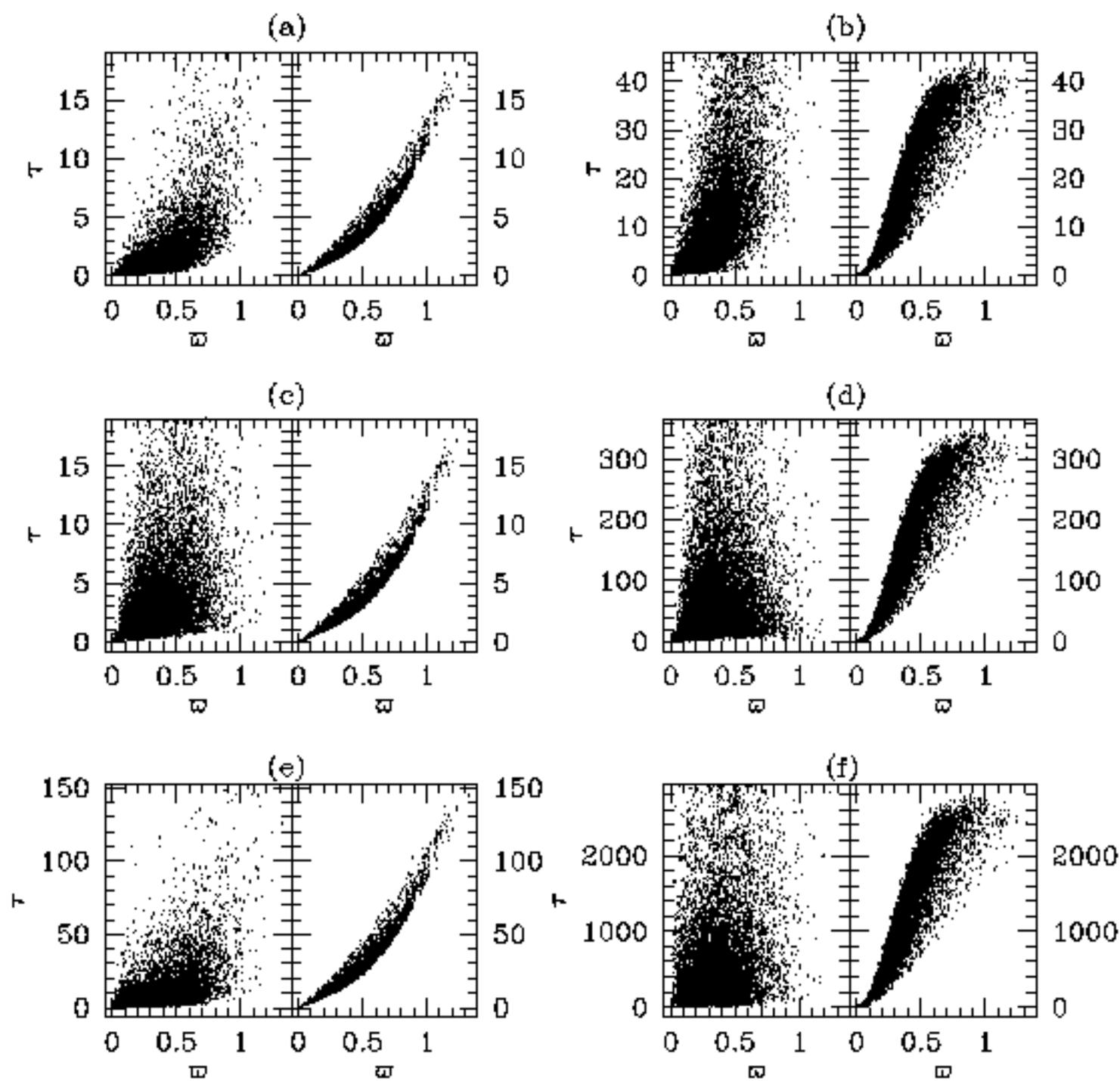
\begin{figure}
\caption{
The viscous timescale as a function of the distance $\varpi$ from the
rotation axis for various artificial viscosities in a system which has
been relaxed into a rapidly, differentially rotating configuration:
(a) $\alpha=2$ and $\beta=0$ in equation~(\ref{pi}),
(b) $\alpha=0$ and $\beta=2$ in equation~(\ref{pi}),
(c) $\alpha=2$ and $\beta=0$ in equation~(\ref{pi2}),
(d) $\alpha=0$ and $\beta=2$ in equation~(\ref{pi2}),
(e) $\alpha=2\times \gamma/2$ and $\beta=0$ in equation~(\ref{piDB}), and
(f) $\alpha=0$ and $\beta=2\times\gamma/2$ in equation~(\ref{piDB}).
Both the actual timescale $\tau_i=v_i/|-\sum_j m_j
\Pi_{ij}{\bf \nabla}_i W_{ij}|$ computed directly from the SPH code
(left frame) and the analytic estimate (right frame) are shown.
Estimates are computed
from equation~(\ref{tau}) with $k_1=k_2=1$ used as an approximation for (a)
and (b), from equation~(\ref{tau2}) with $k_1'=k_2'=1$ for (c) and (d), and
from equation~(\ref{tauDB}) with $k_1^{\prime\prime}=k_2^{\prime\prime}=1$ for
(e) and (f).
} \label{tsall}
\end{figure}

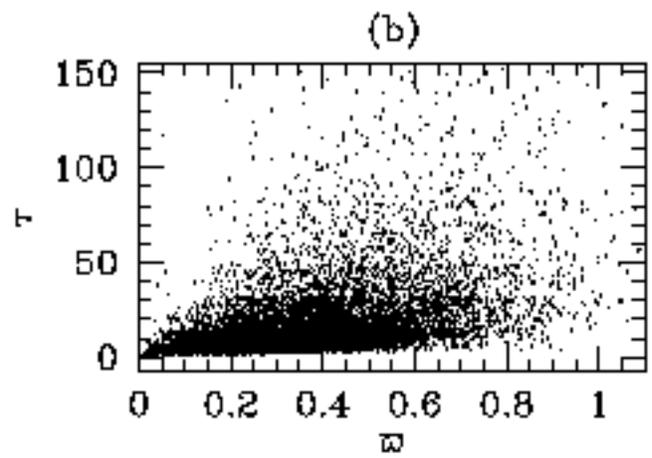
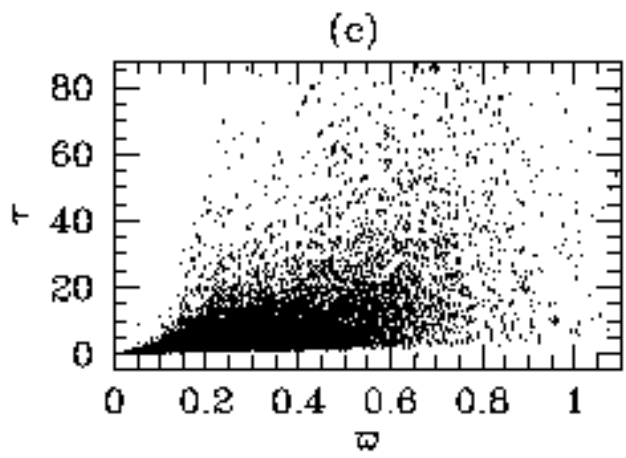
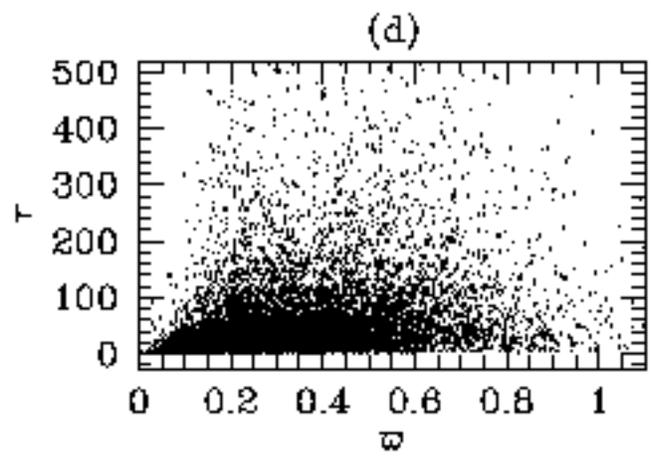
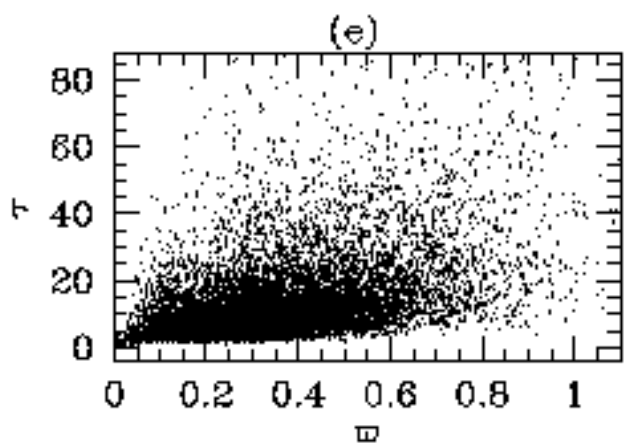
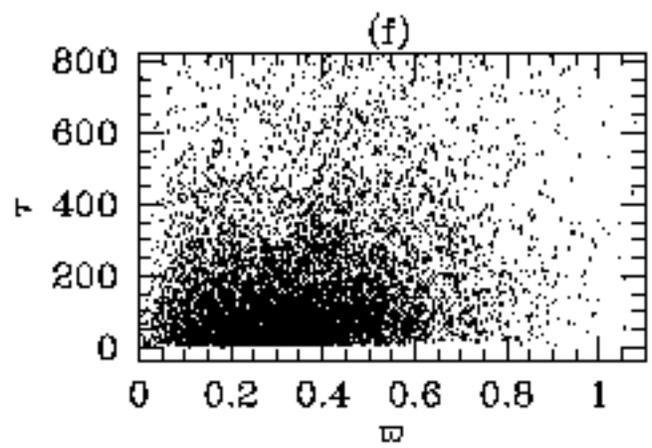
\begin{figure}
\caption{
The timescale $\tau_i=v_i/|-\sum_j m_j
\Pi_{ij}{\bf \nabla}_i W_{ij}|$ computed directly from the SPH code
for various artificial
viscosities after 1 time unit of free evolution.  The various AV schemes and
parameters $\alpha$ and $\beta$ are the same as in Fig.~\ref{tsall}.
} \label{tsall_noisy}
\end{figure}

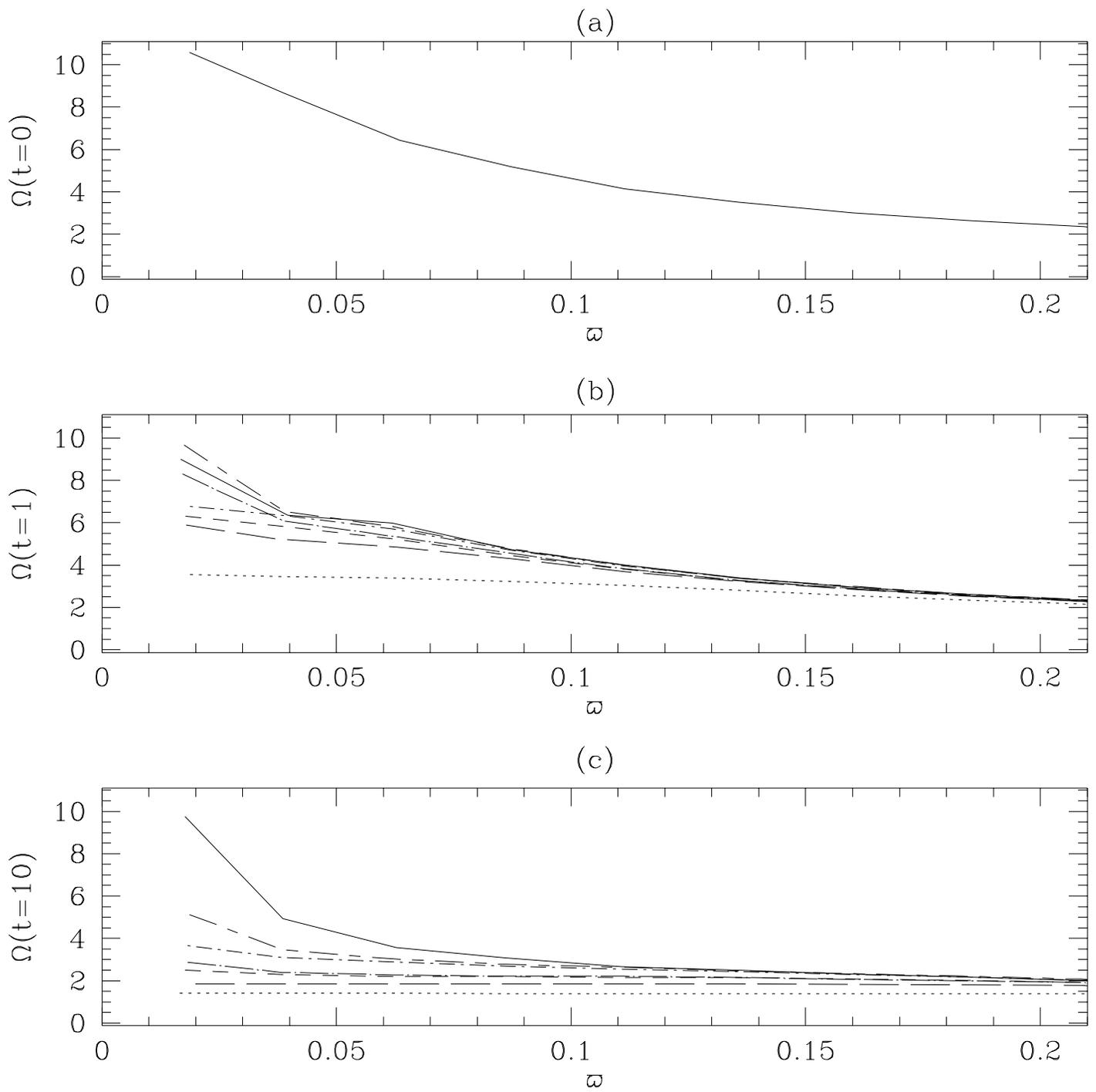
\begin{figure}
\caption{
The angular velocity $\Omega$ as a function of cylindrical radius
$\varpi$ at times (a) $t=0$, (b) $t=1$ and (c) $t=10$ in seven
different calculations which began with the same initial conditions but
implemented different artificial viscosities, namely, from top to
bottom in (c):  no AV (solid curve),
$\alpha=0$ and $\beta=2\times\gamma/2$ in equation~(\ref{piDB}) (short dash - long dash),
$\alpha=0$ and $\beta=2$ in equation~(\ref{pi2}) (dot - short dash),
$\alpha=2\times\gamma/2$ and $\beta=0$ in equation~(\ref{piDB}) (dot - long dash),
$\alpha=0$ and $\beta=2$ in equation~(\ref{pi}) (short dash),
$\alpha=2$ and $\beta=0$ in equation~(\ref{pi2}) (long dash), and
$\alpha=2$ and $\beta=0$ in equation~(\ref{pi}) (dotted).
} \label{oall}
\end{figure}

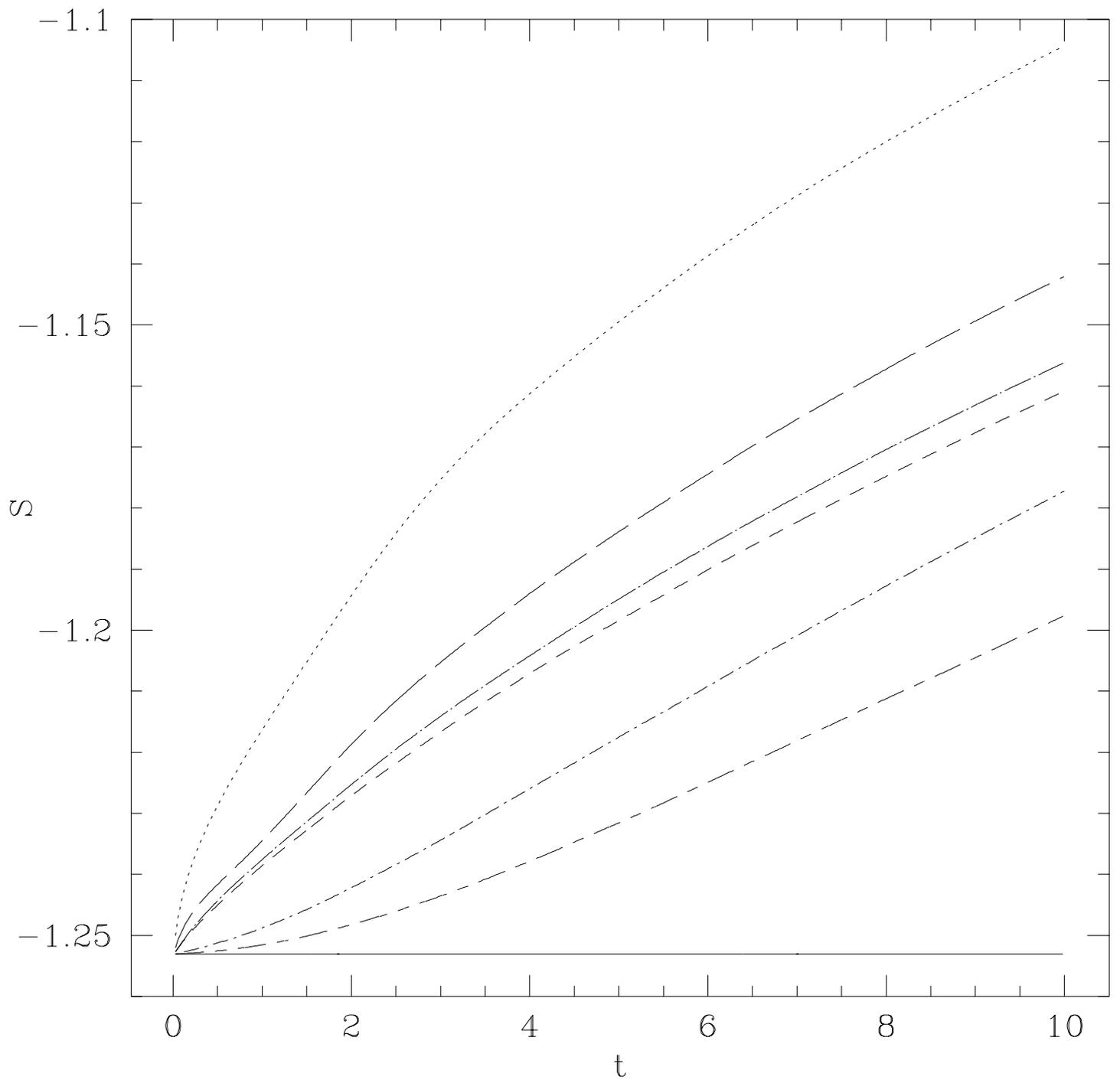
\begin{figure}
\caption{
Entropy $S$ as a function of time for the seven calculations presented
in Fig.~\ref{oall}.  The various line types are as in Fig.~\ref{oall}.
} \label{svstshear}
\end{figure}

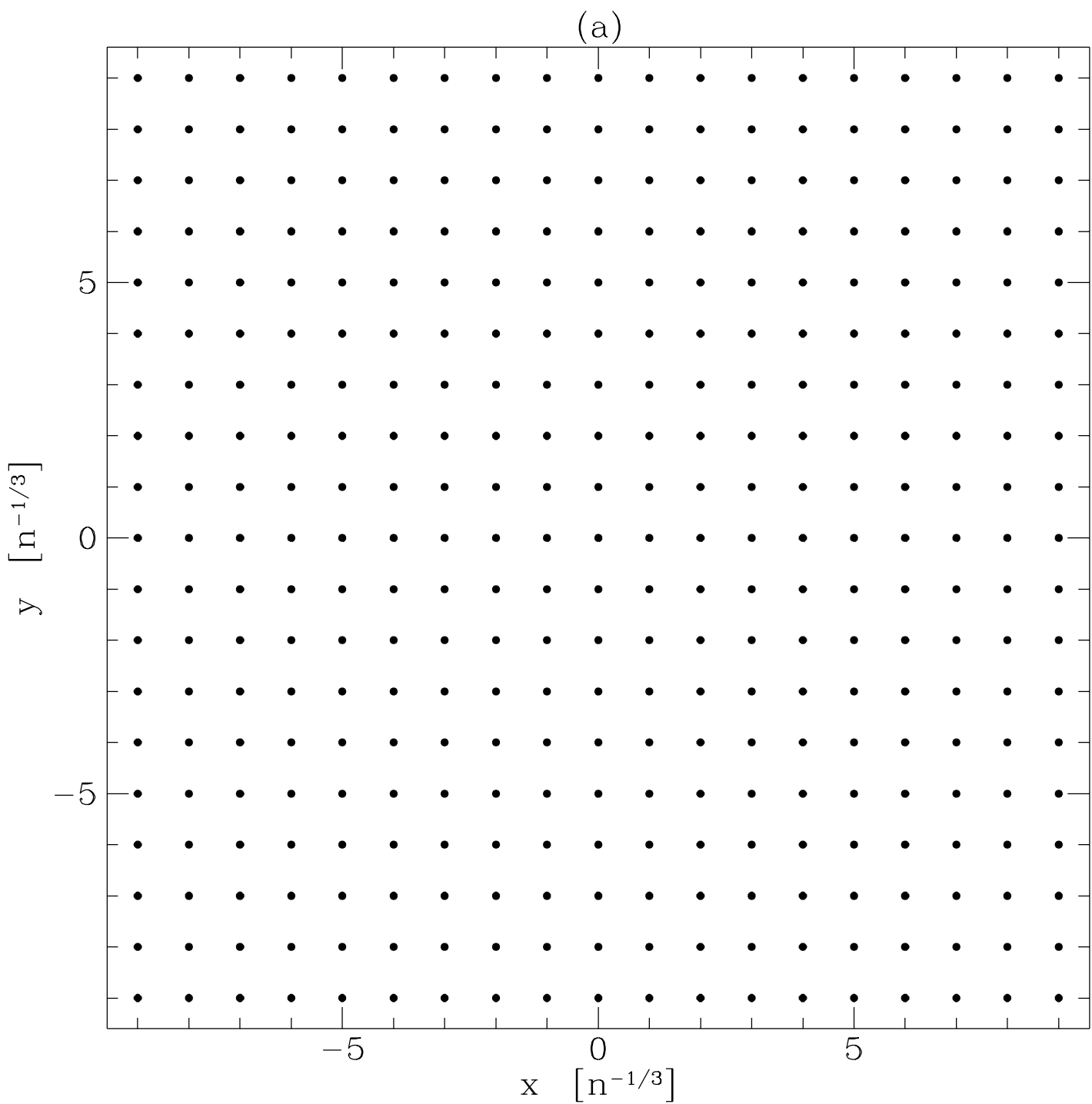
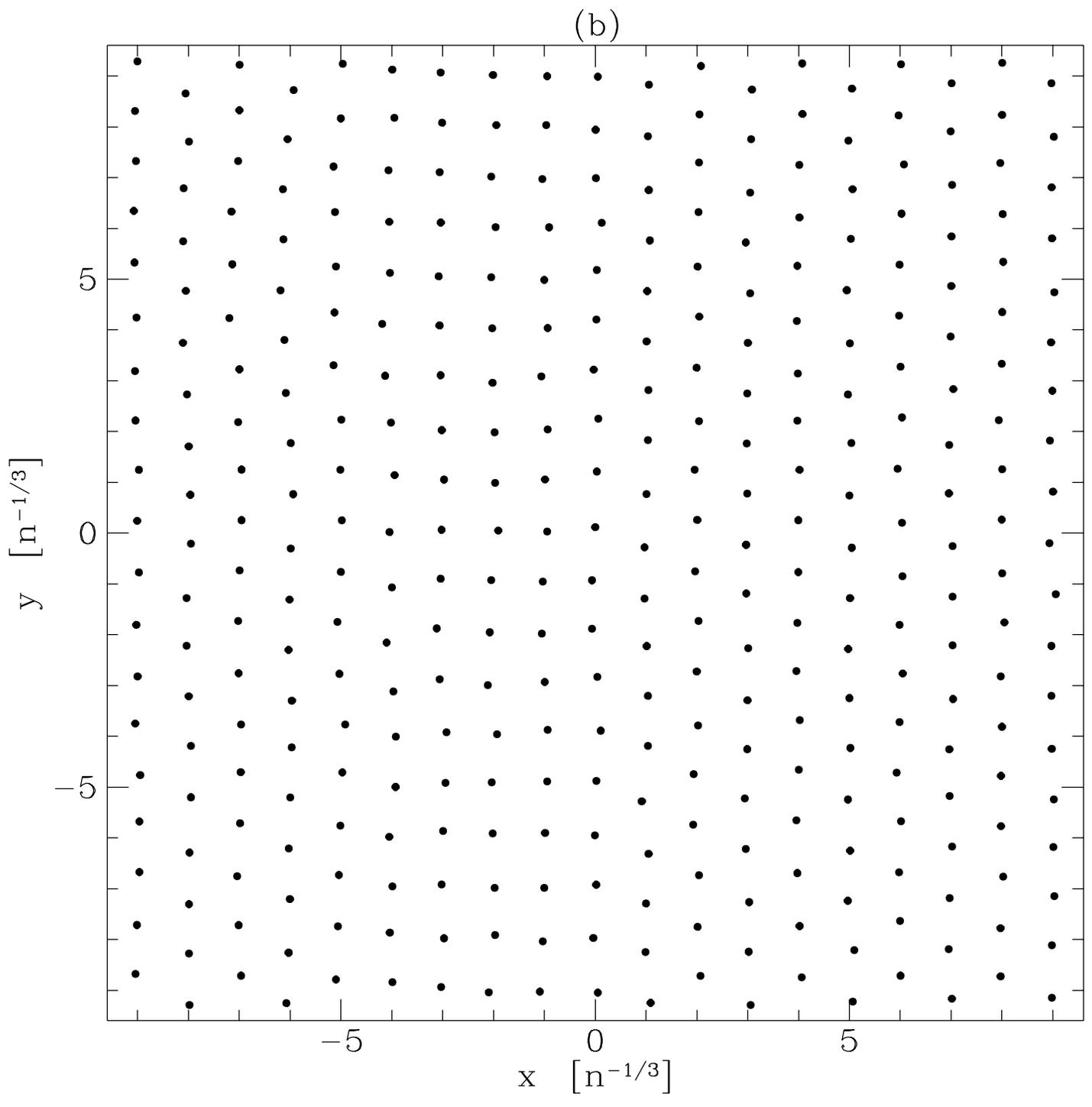
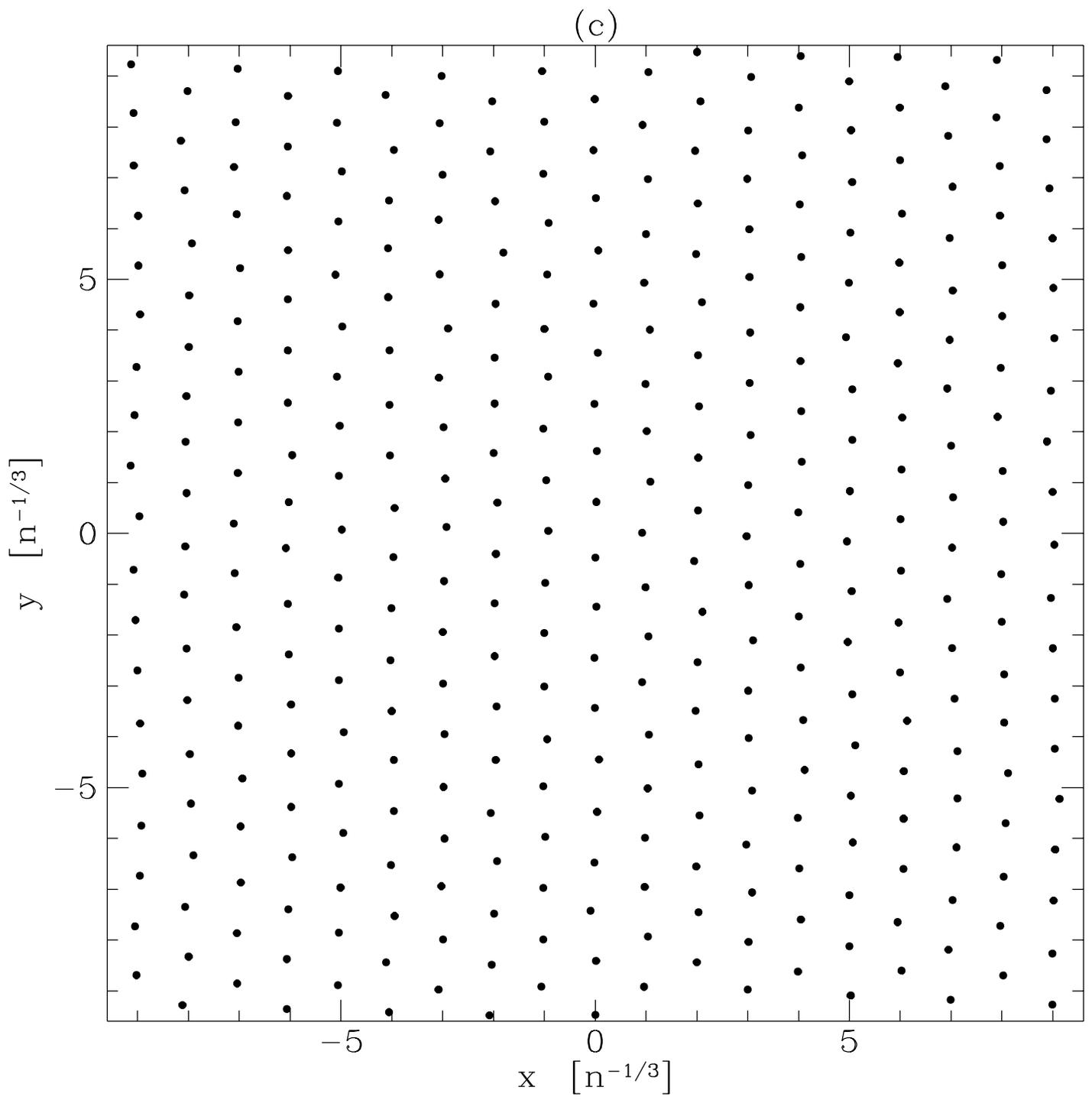
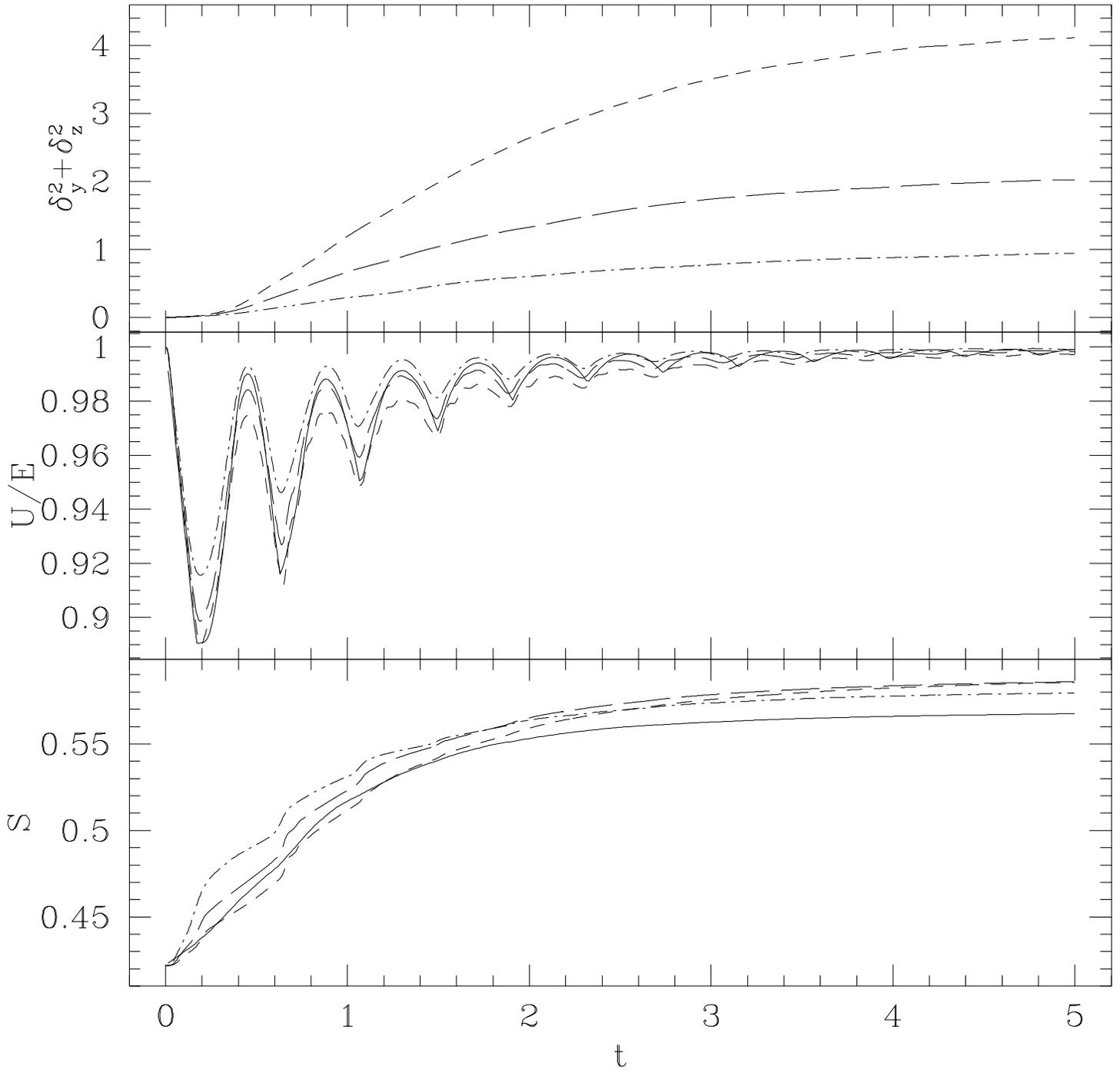
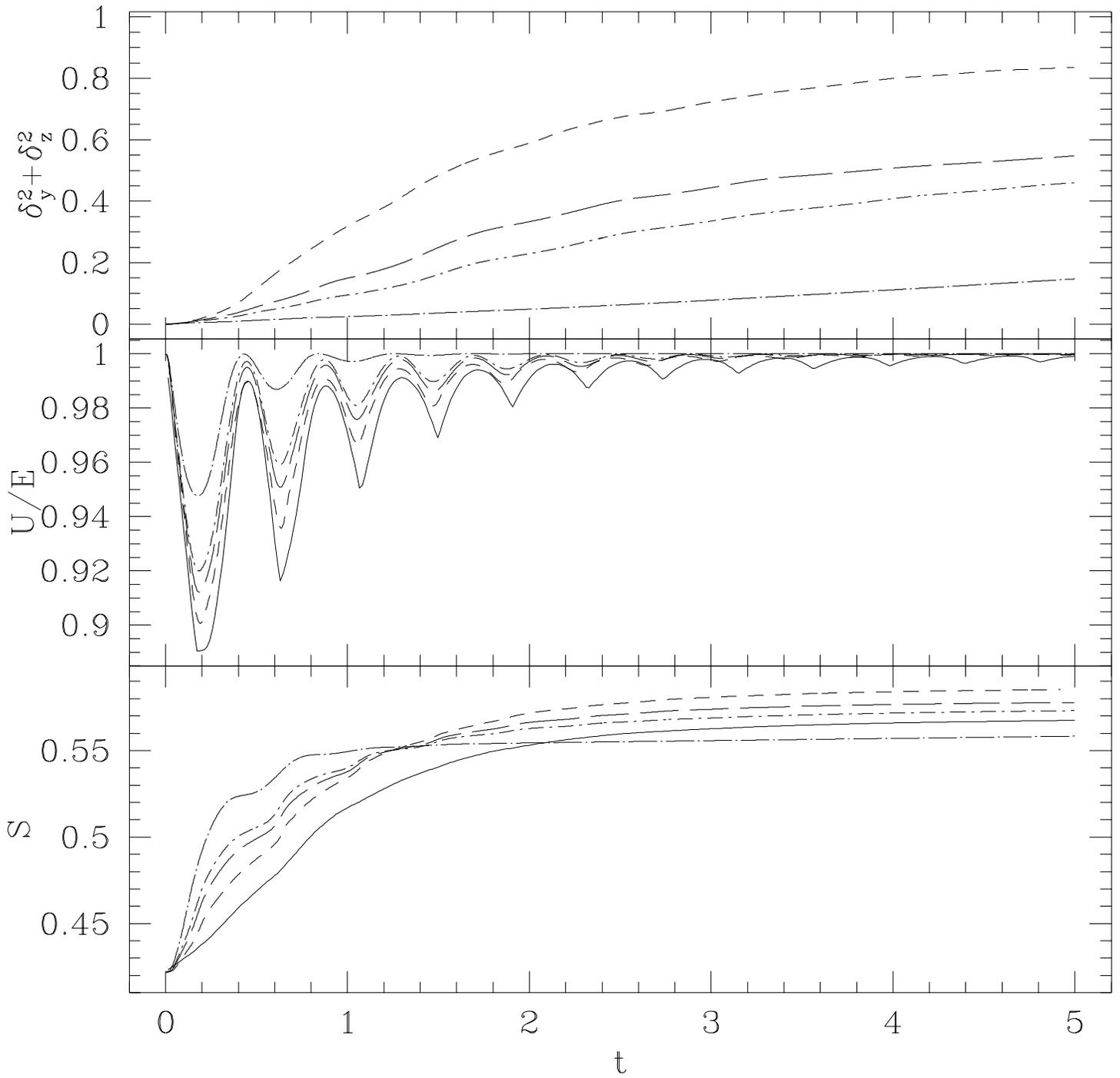
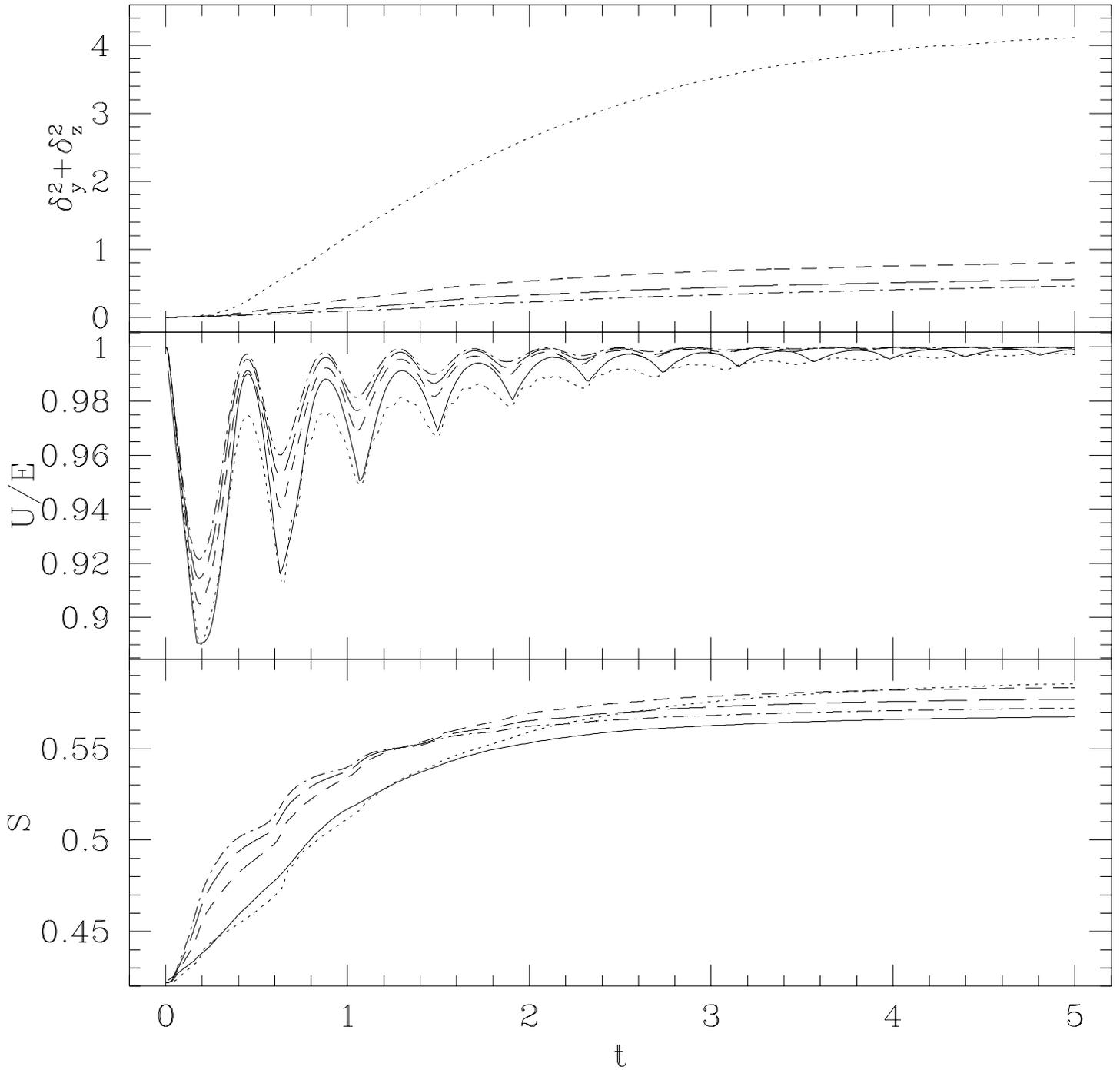
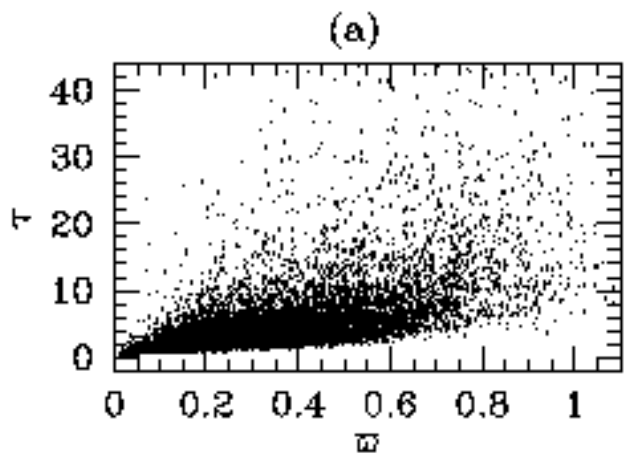
\end{document}